\documentclass[fleqn,a4paper]{aa}
\usepackage{graphicx}
\usepackage[varg]{txfonts}    
\usepackage[sort]{natbib}           

\bibpunct{(}{)}{;}{a}{}{,}
\defcitealias{Peretal.04}{Paper~I}

\newcommand{\kms}{km\,s$^{-1}$}
\newcommand{\Msun}{M$_{\sun}$}
\newcommand{\msun}{M_{\sun}}
\newcommand{\Lsun}{L$_{\sun}$}
\newcommand{\lsun}{L_{\sun}}
\newcommand{\Mdot}{M$_{\sun}$\,yr$^{-1}$}
\newcommand{\mdot}{M_{\sun}\,yr^{-1}}
\newcommand{\etal}{et al.}

\newcommand{\heii}{He\,{\sc ii}}
\newcommand{\teff}{T_\mathrm{eff}}
\newcommand{\ccm}{cm$^{-3}$}
\newcommand{\ergs}{erg\,s$^{-1}$}

\newcommand{\oiii}{[O\,{\sc iii}]}

\newcommand{\Tx}{$T_\mathrm{X}$}
\newcommand{\Lwind}{L_\mathrm{wind}}
\newcommand{\Vwind}{v_\mathrm{wind}}
\newcommand{\dotMw}{\dot{M}_\mathrm{wind}}
\newcommand{\dotMb}{\dot{M}_\mathrm{hb}}
\newcommand{\pdx}[2]{\frac{\partial #1}{\partial #2}}
\newcommand{\pdf}[2]{\frac{\partial}{\partial #2}\left( #1 \right)}

\begin{document}

\title{The evolution of planetary nebulae}
\subtitle{V. The diffuse X-ray emission%
             \thanks{Dedicated to the memory of M. Perinotto, a dear friend and
	             esteemed colleague who died unexpectedly and much too
		     early on August 15, 2007.}}
\titlerunning{The evolution of planetary nebulae. V.}

\author{M. Steffen \and D. Sch\"onberner \and A. Warmuth}

\institute{Astrophysikalisches Institut Potsdam, An der Sternwarte 16, 14482 Potsdam,
           Germany\\
\email{msteffen@aip.de, deschoenberner@aip.de, awarmuth@aip.de}}

\offprints{M. Steffen, \email{msteffen@aip.de}}

\date{Received \today / Accepted }

\abstract{Observations with space-borne X-ray telescopes revealed the existence of
          soft, diffuse X-ray emission from the inner regions of planetary nebulae.
          Although the existing images support the idea that this emission arises
          from the hot shocked central-star wind which fills the inner cavity of a
          planetary nebula, existing models have difficulties to explain the
	  observations consistently.
}
{We investigate how the inclusion of thermal conduction changes the physical 
 parameters of the hot shocked wind gas and the amount of X-ray emission
 predicted by time-dependent hydrodynamical models of planetary nebulae 
 {with central stars of normal, hydrogen-rich surface composition.}
}
{We upgraded our 1D hydrodynamics code NEBEL by to account for
 energy transfer due to heat conduction, which is of importance
 at the interface separating the hot shocked wind gas (`hot bubble') 
 from the much cooler nebular material. With this new version of NEBEL we 
 recomputed a selection of our already existing hydrodynamical sequences and 
 obtained synthetic X-ray spectra for representative models along the evolutionary 
 tracks by means of the freely available CHIANTI package.
}
{Heat conduction leads to lower temperatures and higher densities
 within a bubble and brings the physical properties of the X-ray emitting domain 
 into close agreement with the values derived from observations.
 The amount of X-rays emitted during the course of evolution depends on the energy
 dumped into the bubble by the fast stellar wind, on the efficiency of `evaporating'
 cool nebular gas via heat conduction, and on the bubble's expansion rate. 
 We find {from our models} that the X-ray luminosity of a planetary nebula 
 increases during its 
 evolution across the HR diagram until stellar luminosity and wind power decline.
 Depending on the central-star mass and the evolutionary phase, our models
 predict X-ray [0.45--2.5 keV] luminosities between $10^{-8}$ and $10^{-4}$ of 
 the stellar bolometric luminosities, in good agreement with the observations.
 Less than 1\% of the wind power is radiated away in this X-ray band.
 Although temperature, density, and also the mass of
 the hot bubble is significantly altered by heat conduction, the dynamics of 
 the whole system remains practically the same.
}
{Heat conduction allows the construction of nebular models which predict the correct
 amount of X-ray emission and at the same time are fully consistent
 with the observed mass-loss rate \emph{and} wind speed.
 Thermal conduction must be considered as a viable
 physical process for explaining the diffuse X-ray emission from planetary nebulae
 with closed inner cavities.
 Magnetic fields must then be absent or extremely weak.
}

\maketitle

\section{Introduction}
  The modern, very successful concept for the formation and evolution of the main
  structures of Planetary Nebulae (PNe) is based on the dynamical effects caused by
  (i) the interaction of a rapidly varying central-star stellar wind with the slow
  AGB wind ejected earlier, and (ii) the heating of this circumstellar material
  by photoionization.    The wind from the central star is very fast, exceeding
  in most cases 1000~\kms, and passes through a shock before making contact with
  the dense slow AGB wind.  A tenuous but very hot
  ($\ga\! 10^7~\mbox{K}$) `bubble' is formed since $9/16$ of the kinetic wind energy 
  is converted into internal energy behind a strong adiabatic shock.
  The bubble is separated from the outer, much cooler nebular gas by a contact
  discontinuity (or contact surface). The bubble's thermal expansion accelerates 
  the inner layers of the PN.
  For recent reviews see \citet{SchSt.03} and \citet{StSch.06}.

  The shocked wind gas including the thin transition regime towards the PN proper 
  are the likely sites of diffuse X-ray emission, mainly by
  thermal bremsstrahlung and line emission \citep{VK.85}.
  Any positive detection of diffuse X-ray emission 
  coming from the inner cavities enclosed by nebular shells
  is a direct confirmation of the wind interaction scenario as sketched above.
  Thus there was considerable interest to observe selected PNe with X-ray
  satellites once they became available.  Since the surface brightness in X-rays
  is expected to be quite small \citep[see][]{VK.85}, only very few positive
  detections have been reported to date.   Summaries of X-ray
  observations of PNe conducted to date with the different satellites are
  presented in \citet{Chuetal.03}, \citet{G.06}, and \citet{Kast.07}. 

  Although the observational confirmation of the existence of shocked wind material
  was very gratifying,
  the properties of the X-ray emitting gas were disturbing.  It turned out that
  the X-ray spectra are rather soft, indicating temperatures of only
  $\approx\! (1\ldots 3)\times10^6~\mbox{K}$.  The electron densities of the emitting
  volumes vary between about 20 and 200~\ccm.  These results are in sharp contrast
  with theoretical expectations since the \emph{observed} wind velocities 
  and mass-loss rates demand bubble temperatures of 
  $\approx\!\!10^7\dots 10^8~\mbox{K}$, and electron densities far \emph{below} the 
  observed ones.

  \citet{ASB.06} and \citet{ASBB.07} proposed that the
  remains of the much slower wind blown during the early PN evolution (`early wind')
  when heat conduction is unimportant are responsible for the properties of the 
  observed X-ray emission.  Also collimated outflows (jets) can play a role.
  So it appears that several physical mechanisms exist (heat conduction 
  and/or mixing, early wind, jets) which all may contribute to the 
  X-ray emission, as pointed out by \citet{SoKa.03}. The jet-wind interaction
  has recently been studied in detail by \citet{AMS.08}.

  In two fundamental papers on the properties of interstellar bubbles by
  \citet{CCW.75} and \citet{Wetal.77} it was demonstrated that heat conduction by
  electrons across the contact surface which separates the hot shocked wind gas from
  the much cooler swept-up matter is a natural mechanism to account for the observed
  X-ray and EUV line emissions.
  Heat conduction enforces `evaporation' of cool gas into the hot bubble,
  leading to a shell of matter at the bubble's outer edge with properties just
  ideal for explaining the observed X-ray and EUV line emission.

  Since planetary nebulae are virtually scaled down versions of the interstellar
  bubbles (or \textsc{H\,ii} regions), heat conduction should be important for
  them as well.  To our knowledge, \citet{So.94} was the first to look into
  the consequences of heat conduction in PNe.  He concluded that the observed
  X-ray emission must come from the heat conduction front, and not from the hot
  bubble as a whole. Heat conduction is also capable of explaining the unusually 
  strong UV lines (e.g. of \ion{O}{vi}) seen in some PNe, which cannot be produced 
  by regular photoionization.
  \citet{So.94} noted also that weak tangential magnetic fields would suppress
  heat conduction very effectively, and hence also the X-ray emission.
  But even without magnetic fields X-rays from old, large PNe may escape
  detection because the X-ray surface brightness falls below the detection 
  limit of existing X-ray satellites.
  We note in passing that the same considerations apply to wind-blown bubbles
  around massive stars as well \citep[cf.][]{WWW.94}.

  Despite of their potential
  importance, quantitative theoretical calculations of X-ray and EUV 
  emissions from PNe based on the concept of heat conduction are rare. 
  \citet{ZP.96,ZP.98} succeeded in developing analytical solutions based on the 
  concept of \citet{Wetal.77} and were able to present quantitative predictions for 
  a few specified cases.  Using  analytical solutions of heat conduction, \cite{GChuG.04}
  modelled the transition layers between the hot bubble and the nebular regime for
  \object{NGC~6543} in order to compute the \textsc{O\,vi} emission lines seen in
  the FUSE spectra.

  Two hydrodynamical studies addressed recently the issue of X-ray emission from
  planetary nebulae \citep{SS.06,ASBB.07}.  The basic philosophy of both studies
  is to avoid thermal conduction and to select the properties of the wind such that
  temperature and density of the X-ray emitting gas agree with the observations.
  No relation between stellar parameters and wind properties is considered, and
  the feedback of radiation to the hydrodynamics is neglected.
  To explain the observed X-ray luminosities,
  these authors came up with wind parameters which disagree strongly with both
  the predictions of the theory of radiation-driven
  winds \citep[e.g.][]{Pauletal.88} \emph{and} with the directly observed mass-loss
  rates and wind speeds of the objects in question.
  More precisely, the necessary mass-loss rates are too large
  and the corresponding wind speeds too low -- by as much as a factor 5.%
  \footnote{The only exception is the central star of NGC~2392 whose
  wind speed is much too low for its position in the Hertzsprung-Russell
  diagram.}

  The results of \citet{ASBB.07} are particularly interesting.  Consistence with
  existing observations in terms of X-ray luminosity and temperature of the X-ray
  emitting gas could be achieved only for rapidly decaying central-star winds:
  while the wind speed increases linearly with time, the mass-loss rate  must decline
  quite rapidly to very small values \citep[see][Figs. 6 and 8 therein]{ASBB.07}.
  For the best choice of \citet{ASBB.07}, the central-star wind reaches
  $10^{-10}$ \Mdot\ already at a modest wind speed of only 750 \kms.

  We iterate that combinations of mass-loss rates and outflow speeds
  as they are found in \citet{ASBB.07} are in severe conflict with both
  current theories of radiation-driven winds from hot stars 
  (see Fig.~\ref{mod.prop}
  in Sect. \ref{bubb.struc} for details) \emph{and} with the observations
  (see Table~\ref{tab.xray} in Sect. \ref{xray.lum}), with the apparent 
  exception of NGC~2392. In our opinion the existing
  studies of the X-ray emission from planetary nebulae which neglect thermal
  conduction by electrons are therefore far from being convincing.
  After all, thermal conduction is a physical process 
  inherent to all hydrodynamical systems and becomes important in rarefied 
  plasmas wherever the mean free path of the electrons is large enough. 
  Thermal conduction
  can \emph{only} be modified or even suppressed by the existence of magnetic
  fields \citep[cf.][]{BBF.90}.

  Since no self-consistent radiation-hydrodynamics computations including heat
  conduction have been performed to date\footnote{
  \cite{MF.95} computed the X-ray emission of their 2D hydrodynamics models, but heat 
  conduction was not considered.}, and
  urged by the fact that the new observations which became available by the
  Chandra and XMM-Newton satellites lack a convincing interpretation by detailed
  modeling, we decided to update our 1D radiation-hydrodynamics code NEBEL by
  incorporating a thermal conduction module.  We recomputed then some of our sequences
  from \citet[Paper~I hereinafter]{Peretal.04} with heat conduction
  self-consistently included.  Based on the new thermal structure of the bubble we
  modelled the X-ray and EUV emissions at selected positions along the evolutionary
  sequences by using the well-documented CHIANTI code.

  We emphasize that in our study (i) wind and stellar evolution are consistently
  connected within the framework of the theory of radiation-driven winds, and (ii)
  no fit to the data by manipulating the models is done.  It is our belief that
  only with such an approach it is possible to interprete the data appropriately
  and to deduce meaningful constraints regarding the relevant physics.  
  Obviously, our approach is only applicable to objects which do not depart 
  too much from sphericity, have closed shells where a bubble of shocked,
  hot wind gas is able to exist, and have central stars of normal, 
  hydrogen-rich surface composition.

  We avoid to discuss in the present work objects with 
  [WC] central stars for two reasons: (i) Their evolution is at present not known, and
  (ii) the physics of heat conduction in hydrogen-free and carbon-rich plasmas has
  still to be worked out.  We note that the X-ray emission from a hydrogen-free
  plasma is expected to be stronger, making it more likely to detect 
  extended X-ray emission from PNe with [WC] central stars. The present number of 
  positive detections is, however, too small as to make any definitive statements 
  concerning the ratio of PNe X-ray sources with normal and [WC] central stars, 
  respectively.

  The computational details of how the thermal conduction is treated are described
  in the Sect.~\ref{computate}, followed by a discussion of the resulting bubble
  structures in Sect.~\ref{bubb.struc}. Section~\ref{x-rays} is devoted to the
  X-ray emission emerging from the conduction front and how it develops with time.
  In Sect.~\ref{comp.obs} we discuss extensively how our models compare with the
  existing observations.
  The paper concludes with Sect.~\ref{diss.concl}.
  Part of the results presented here can be found in \citet{SSW.06}.

\section{The computations}                                   \label{computate}

\subsection{The hydrodynamical models}
\label{models}

  The basic idea behind our modeling is to couple a spherical circumstellar
  envelope, being the relic of the strong wind on the asymptotic giant branch
  (AGB), to a post-AGB star model of certain mass and to follow numerically
  the hydrodynamical evolution of the envelope across the Hertzsprung-Russell 
  diagram towards the white-dwarf cooling path. This is achieved by applying 
  our 1D radiation-hydrodynamics code NEBEL as
  described in more detail in \citet{Peretal.98}. The inner boundary condition
  in terms of density and velocity is provided by the variable central star wind.
  For the ionizing photon flux the star is assumed to radiate as a black body for
  each given effective temperature.

  The radiation part of our hydrodynamics code, CORONA, is described in
  \citet{MS.97}. We point out that this code is designed
  to compute ionization, recombination, radiative heating and line cooling fully
  time-dependently. At each volume element, the cooling function is composed
  of the contributions of all the ions considered and computed according to the
  actual plasma parameters. For each chemical element listed in
  Table~\ref{tab.element}, up to 12 ionization stages are taken into account, 
  amounting to a total of 76 ions. 

\begin{table}[t]           
\caption{Elemental abundances, $\epsilon_\mathrm{i}$, used in the computations of
         our hydrodynamical models, in (logarithmic) number fractions relative to
         hydrogen,
         $\log {\epsilon}_\mathrm{i} = \log (n_\mathrm{i}/n_\mathrm{H}) + 12$.}
\label{tab.element}
\centering
\begin{tabular}{ccccccccc}
\hline\hline\noalign{\smallskip}
  H    &    He  &   C    &   N   &    O   &   Ne  &   S   &   Cl   &   Ar  \\
\noalign{\smallskip}\hline\noalign{\smallskip}
 12.00 &  11.04 &  8.89  &  8.39 &  8.65  &  8.01 &  7.04 &  5.32  &  6.46  \\
\noalign{\smallskip}\hline
\end{tabular}
\end{table}

  In \citetalias{Peretal.04} a very detailed description of how the nebular
  properties depend on the chosen initial envelope configurations and central-star
  models is given.  For the present study we selected representative sequences and
  recomputed them with and without heat conduction included.
  The new simulations consider 9 chemical elements instead of 6, with abundances
  as listed in Table~\ref{tab.element}.
  A compilation of the sequences investigated here is given in Table~\ref{tab.mod}.
  The 0.595~\Msun\ post-AGB model has been introduced
  in \citet{Schetal.05}.  Sequence No.~10a is a variant of sequence No.~10 of
  \citetalias{Peretal.04} with the AGB mass-loss rate doubled. Note that
  the luminosities listed in Table~\ref{tab.mod} correspond to the early part of
  the post-AGB evolution. They decrease slowly during the evolution across
  the Hertzsprung-Russell diagram.

\begin{table}[t]
\centering
\caption{Hydrodynamical sequences of model planetary nebulae
         used in this work. The sequence numbers correspond to Table\,1 in
         \citetalias{Peretal.04} and Table\,2 in \citet{Schetal.05}, but the
         additional notation `HC' and `HC2' indicates that heat conduction is 
         included according to method~1 and 2, respectively
	 (see Sect.~\ref{heat.cond}).  The stellar luminosities refer to
	 $\teff $\,=\,$ 30\,000$~K, and the peak mass-loss rate of the AGB 
         hydrodynamical simulation (6, 6a) is about $1\times10^{-4}$~\Mdot.
         {\sc Type} indicates the structure adopted for the AGB envelope:
         `A' means constant mass loss rate, $\rho$$\,\simeq$$\,r^{-2}$, `B' 
         means structure from hydrodynamical simulation (see 
         \citetalias{Peretal.04} for details). 
}
\label{tab.mod}
\begin{tabular}{lcrccc}
\hline\hline\noalign{\smallskip}
No. & $M $    & $L$\quad\quad        & $\dot{M}_{\rm agb}$~~~& $v_{\rm agb}$ & {\sc Type}\\[0.5mm]
    & (\Msun) & (\Lsun)    & (\Mdot)          &        (\kms)          &       \\
\noalign{\smallskip}
\hline\noalign{\smallskip}
22        & 0.565& 3\,981 & $3\times10^{-5}$          &  10            & {\sc A}\\
22-HC2    & 0.565& 3\,981 & $3\times10^{-5}$          &  10            & {\sc A}\\
\noalign{\medskip}
6a        & 0.595& 5\,593 & Hydro.\ sim.           &\hspace*{-2.5mm}$\simeq$12 & {\sc C}\\
6a-HC     & 0.595& 5\,593 & Hydro.\ sim.           &\hspace*{-2.5mm}$\simeq$12 & {\sc C}\\
6a-HC2    & 0.595& 5\,593 & Hydro.\ sim.           &\hspace*{-2.5mm}$\simeq$12 & {\sc C}\\
\noalign{\medskip}
6         & 0.605& 6\,280 & Hydro.\ sim.          &\hspace*{-2.5mm}$\simeq$12 & {\sc C}\\
6-HC      & 0.605& 6\,280 & Hydro.\ sim.          &\hspace*{-2.5mm}$\simeq$12 & {\sc C}\\
6-HC2     & 0.605& 6\,280 & Hydro.\ sim.          &\hspace*{-2.5mm}$\simeq$12 & {\sc C}\\
\noalign{\medskip}
10        & 0.696&11\,615 & $ 1\times10^{-4} $         &  15           & {\sc A}\\
10-HC2    & 0.696&11\,615 & $ 1\times10^{-4} $         &  15           & {\sc A}\\
10a-HC    & 0.696&11\,615 & $ 2\times10^{-4} $         &  15           & {\sc A}\\
10a-HC2   & 0.696&11\,615 & $ 2\times10^{-4} $         &  15           & {\sc A}\\
\noalign{\smallskip}\hline
\end{tabular}
\end{table}

\subsection{The treatment of electron heat conduction}
\label{heat.cond}
\subsubsection{Physical description}
  Electron heat conduction is described as a diffusion process, with the
  heat flux $\vec{q}$ given by
  \begin{equation}
  \vec{q} = -D \; \vec{\nabla} T_{\rm e} .
  \label{E1}
  \end{equation}
  Following \citet{Sp.62} and \citet{CMK.77}, the electron mean free
  path $\lambda$ is a function of electron temperature, $ T_{\rm e}$, and 
  electron number density, $n_{\rm e}$, and can be written as
  \begin{equation}
  \lambda = 2.625 \times 10^{5} \; T_{\rm e}^{2} / n_{\rm e} / \ln \Lambda \;\;\;\;
  \mathrm{[cm]},
  \label{E2}
  \end{equation}
  where the Coulomb Logarithm, $\ln \Lambda$, can be approximated as
  \begin{equation}
  \ln \Lambda = \left\{\begin{array}{ll}
             9.425 + 3/2\; \ln T_{\rm e} - 1/2\; \ln n_{\rm e}, &
             \;\; T_{\rm e} \le  4.2 \times 10^{5}\;\mathrm{K}, \\
             22.37 + \;\;\;\;\;\;\; \ln T_{\rm e} -1/2\; \ln n_{\rm e}, &
             \;\; T_{\rm e} >  4.2 \times 10^{5}\;\mathrm{K},
             \end{array} \right .
  \label{E3}
  \end{equation}
  for a pure hydrogen plasma. The diffusion coefficient $D$ is then
  given by
  \begin{equation}
   D = 7.04 \times 10^{-11} \; \lambda \; n_{\rm e} \; T_{\rm e}^{1/2} \;\;\;\;
  [{\rm erg}\,{\rm s}^{-1}\,{\rm K}^{-1}\,{\rm cm}^{-1}].
  \label{E4}
  \end{equation}
  At high $T_{\rm e}$ and low $n_{\rm e}$, the electron mean free path $\lambda$ becomes
  very large according to Eq.\,(\ref{E2}) (actually it can become much larger
  than the dimensions of the `hot bubble'), and the diffusion approximation 
  is no longer valid. Rather, the heat flux cannot exceed the saturation limit
  \begin{equation}
  \vec{q}_{\rm sat} = 1.72 \times 10^{-11} \; T_{\rm e}^{3/2} n_{\rm e} \;\;\;\;
  \mathrm{[erg\,cm}^{-2}\,\mathrm{s}^{-1}].
  \label{E5}
  \end{equation}
  This is an approximate upper
  limit expressing the fact that the heat flux cannot be larger than
  the heat content, $3/2\, n_{\rm e} k T_{\rm e}$, times a characteristic electron
  transport velocity, $v_{\rm char}$. For a more detailed derivation of
  Eq.\,(\ref{E5}) see \citet{CMK.77}.

  \subsubsection{Numerical Treatment}
   Each time step $\Delta t$ of the hydrodynamical simulations is divided
   into 3 successive steps for updating the state vector $\vec{Q}$
   (operator splitting):\\
   (a)~~~ $\vec{Q}(t) \hspace*{9mm} \Rightarrow$
          ~~~~~{\bf advection}~~~~~ $\Rightarrow \vec{Q_1}(t+\Delta t)$; \\
   (b)~~~ $\vec{Q_1}(t+\Delta t) \Rightarrow $ {\bf heat conduction}
          $\Rightarrow \vec{Q_2}(t+\Delta t)$ \\
          \hspace*{7mm} (1st energy update at constant mass density $\rho$); \\
   (c)~~~ $\vec{Q_2}(t+\Delta t) \Rightarrow $ ~~~~~~{\bf radiation}~~~~~~
          $\Rightarrow \vec{Q_{\mbox{ }}}(t+\Delta t)$ \\
          \hspace*{7mm} (2nd energy update at constant mass density $\rho$). \\[2mm]
  Ionization is frozen during step (b), being updated subsequently in step (c).
   For step (b), we solve the diffusion equation in spherical coordinates,
  \begin{equation}
  \pdx{E}{t} = \rho c_v\; \pdx{T_{\rm e}}{t} = 
               \frac{1}{r^2}\; \pdf{r^2\; D \; \pdx{T_{\rm e}}{r}}{r},
  \label{E6}
  \end{equation}
  ($E$: internal energy per unit volume, $c_v$: specific heat at constant 
  volume, $r$: radial coordinate) with a fully implicit, 
  standard numerical method. No additional
  constraints are imposed on the time step by this implicit energy update.
  
  The diffusion coefficient $D$ is evaluated at the \emph{cell
  boundaries} from the local physical conditions according to
  Eq.\,(\ref{E4}), with $\lambda$ replaced by $\overline{\lambda}$ (see below)
  to take into account the effect of saturation. In order to take care
  of the fact that the diffusion coefficient itself depends on
  temperature, we have adopted the following procedure to obtain the
  temperature update from initial temperature $T_1$ to final temperature 
  $T_2$ due to electron heat conduction acting over the time interval 
  $\Delta t\,$: (i) we calculate the initial diffusion coefficient
  $D_0 = D(T_1)$; (ii) starting from $T_1$, we solve the diffusion 
  equation Eq.\,(\ref{E6}) for time step $\Delta t / 2$ with $D = D_0$ 
  taken to be constant over time,
  resulting in intermediate temperature $T_{1/2}$; (iii)
  we calculate the intermediate diffusion coefficient $D_{1/2} =
  D(T_{1/2})$; (iv) starting again from $T_1$, we
  solve the diffusion equation for the full time step $\Delta t$ with
  $D = D_{1/2}$ taken to be constant over time, resulting
  in the final temperature state $T_2$. This procedure turned
  out to be perfectly adequate: comparison with a test calculation
  using $10$ intermediate time steps instead of $1$ gave practically
  identical results.
  
  In the framework of the diffusion approximation, Eq.\,(\ref{E1}),
  saturation effects can be crudely accounted for by limiting the mean
  free path $\lambda$. However, there is some arbitrariness involved
  in this approach, and it is not clear which recipe gives the most
  realistic results. We have tried two extreme methods:
 
  In {\bf method~1}, we limit the  electron mean free path to a fraction 
  of the spatial resolution of the numerical grid, i.e.\ we calculate 
  $\overline{\lambda}$ as
  \begin{equation}
  \overline{\lambda}_1 = 
  \min \{f \times \Delta r, \;\; 2.625 \times 10^{5} \;
         T_{\rm e}^{2} / n_{\rm e} / \ln \Lambda\},
  \label{E7}
  \end{equation}
  where $\Delta r$ is the local spacing of the radial grid. Choosing
  the constant $f = 0.244$ ensures that the conductive heat flux can
  never exceed the saturation heat flux given by Eq.\,(5): In regions
  where the limiter is active, i.e.\ where $\overline{\lambda}_1 =
  0.244 \cdot \Delta r$, we obtain 
  $\vec{q} = \vec{q}_{\rm sat} \cdot \Delta r \, \vec{\nabla} T_{\rm e} \, 
  / T_{\rm e} \approx  \vec{q}_{\rm sat} \cdot \Delta T_{\rm e}/T_{\rm e}$, 
  where $\Delta T_{\rm e}$ 
  is the temperature difference between two adjacent grid points. Hence, 
  the limiting flux can only be reached at the sharp edge of a hot region 
  where $|\Delta T_{\rm e}| \approx T_{\rm e}$. A drawback of this method 
  is that $\overline{\lambda}_1$, and hence the results, depend 
  explicitly on the numerical resolution $\Delta r$.  Sequences computed with 
  this approach are denoted in Table\,\ref{tab.mod} with label `HC'.

  In {\bf method~2}, we limit the electron mean free path according to 
  the interpolation formula
  \begin{equation}
  \frac{1}{\overline{\lambda}_2} = \frac{1}{2.625 \times 10^{5} \;
         T_{\rm e}^{2} / n_{\rm e} / \ln \Lambda}+4.105\;
         \frac{\left| \vec{\nabla} T_{\rm e}\right|}{T_{\rm e}}.
  \label{E8}
  \end{equation}
  As with the first method, the conductive heat flux cannot exceed the
  saturation heat flux given by Eq.\,(5): In regions where $\lambda \ll
  \Delta r$, the first term on the right hand side of Eq.\,(\ref{E8})
  dominates, and $\overline{\lambda}_2 = \lambda$, i.e.\ the limiter is 
  inactive. When $\lambda \gg  \Delta r$, the second term on the right 
  hand side of Eq.\,(\ref{E8}) dominates, so $\overline{\lambda}_2 = 0.244 
  \cdot T_{\rm e} / \left |\vec{\nabla} T_{\rm e}\right|$, and $\vec{q} = 
  \vec{q}_{\rm sat}$.
  Sequences with this treatment of thermal conduction are indicated by the
  label `HC2'(see Table \ref{tab.mod}).

  The two methods represent the extreme cases. {Method~2} essentially gives
  the saturation flux Eq.\,(\ref{E5}) wherever the unlimited flux would
  exceed this value, while method~1 generally yields much smaller fluxes,
  reaching the saturation flux only under extreme conditions.

    As a final remark we note that we have not attempted to correct the
  calculation of the diffusion coefficient for the fact that the
  actual chemical composition of the nebular matter is not pure
  hydrogen.

\subsection{The CHIANTI code and the computation of the X-ray emission}
\label{X-ray.chianti}
  For obtaining the X-ray emission of our model PNe it is necessary to compute a
  synthetic optically thin X-ray spectrum for each radial shell of the model,
  since the different values for $T_{\rm e}$, $n_{\rm e}$ and $n_{\rm p}$ at each shell
  result in unique spectral characteristics. The spectra were calculated using the
  CHIANTI software package \citep{Deretal.97}, which has been used extensively
  by the astrophysical and solar communities. CHIANTI consists of an up-to-date
  set of atomic data for a large number of ions of astrophysical interest, and
  also includes a number of ancillary data and a suite of useful routines.

  Our synthetic spectra were computed with version 5.1 of CHIANTI \citep{LP.05},
  which includes the most recent atomic data. Only radial shells with
  \hbox{$T_{\rm e} > 10^5$ K} were considered, since the contribution of cooler
  layers is negligible in the X-ray range.   With the input of $T_{\rm e}$,
  $n_{\rm e}$ and $n_{\rm p}$ for a given
  volume element, individual spectra are then synthesized, including the contribution
  due to lines and various continua (free-free, free-bound and two-photon
  continuum).   We have used the ion fractions from \citet{Mazetal.98}, under
  the assumption of ionization equilibrium. The elemental abundances used are either
  those listed in Table\,\ref{tab.element} or solar values for the elements not
  considered in our hydrodynamics simulations.

  Each individual spectrum actually represents the emission
  (in erg\,s$^{-1}$\,cm$^{-3}$\,\AA $^{-1}$) from a unit volume. In order to
  determine the total X-ray emission, the averaged spectrum, and the brightness
  distribution in the plane of the sky, we must perform appropriate integrations
  over our spectra, taking the (spherical) geometry of the model into account.
  The multiplication of an individual spectrum with the volume of the corresponding
  radial shell gives the total emission from that shell, and the summation over
  all shells emitting in X-rays then yields the total spectrum.
  An integration over a certain wavelength range then results in a well-defined
  total X-ray luminosity, $L_{\rm X}$ (in erg\,s$^{-1}$).

  For determining the brightness distribution in the plane of sky
  (in erg\,cm$^{-2}$\,s$^{-1}$\,sr$^{-1}$), the integration is performed along the
  line of sight (perpendicular to the plane of sky) for a series of impact
  parameters (up to about 400).

  The computation of the complete X-ray spectrum and luminosity is rather time
  consuming and thus only performed for selected models along an evolutionary
  sequence.  Such a post-facto computation of the X-ray emission is somewhat
  inconsistent with the hydrodynamics since at least part of the energy lost from
  the bubble by X-rays, viz.\ the energy loss due to ions not considered in our
  simulations (cf.\ Table\,\ref{tab.element} and Sect. \ref{models}), is not 
  included in its energy budget.
  However, since the X-ray luminosity computed by means of the CHIANTI 
  code is only a very small fraction of the total radiation losses from the 
  hot bubble as already provided by the NEBEL code 
  (see Sect.\,\ref{bubble.time}, Fig.\,\ref{Qrad}), a somewhat incomplete 
  consideration of the X-ray loss in 
  the bubble's energy budget has absolutely no consequences for the dynamics 
  of the whole system.

\begin{figure}[t]                               
\includegraphics*[bb= 1cm 1cm 21cm 26.5cm, width=\linewidth]
                 {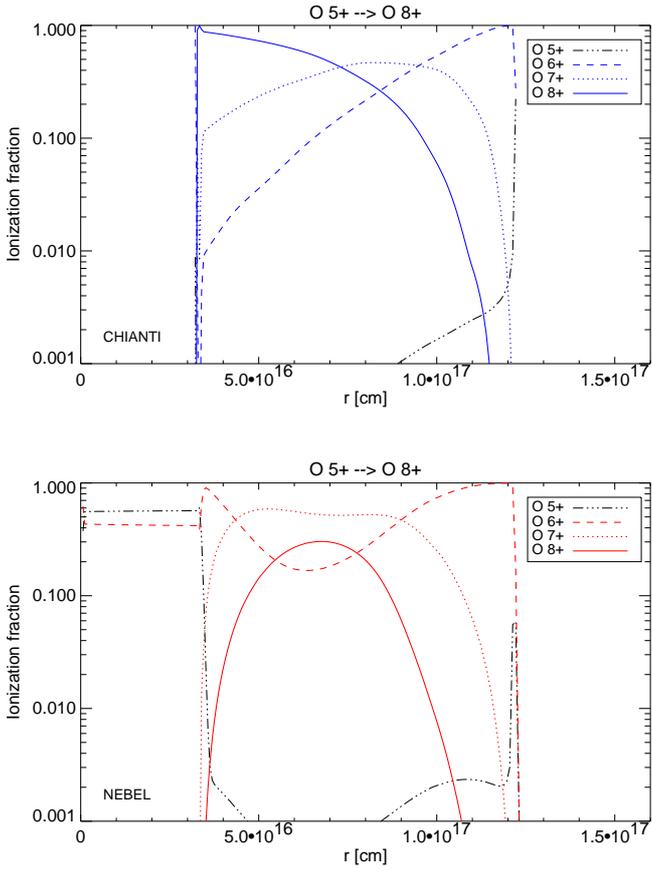}
\caption{The fractions of the last 4 ionization stages of oxygen within a typical
         PN bubble selected from sequence No. 6a-HC2.  The model parameters are
	 $L$\,=\,$5205$~\Lsun\ and $\teff$\,=$\,71\,667$~K at $t$\,=\,$5642$~yr.
	 Heat conduction is considered according to {method~2}.
         The central star is at the origin, the (reverse) wind shock at
	 $r$\,=\,$3.2\times10^{16}$~cm, and the conduction front at
	 $1.22\times10^{17}$~cm.
	\emph{Top}: predictions from CHIANTI. 	Note that the CHIANTI code is
	only applicable to the region embraced by the wind shock and the
	conduction front.
	\emph{Bottom}: predictions from our time-dependent computations with NEBEL.
	 Beyond the conduction front, i.e. in the `cool' nebular gas, the
	 fractions of the displayed ions drop rapidly to virtually zero.
        }
\label{bubble.ion}
\end{figure}

  We checked also whether the ionization structures predicted by our hydrodynamics code
  and by CHIANTI are consistent with each other.  Figure~\ref{bubble.ion} gives an
  example for the ionization structure of oxygen within a typical bubble and compares
  the predictions of the CHIANTI code (top panel) with our time-dependent
  NEBEL/CORONA code (bottom panel).
  Note that the ionization fractions computed by CHIANTI are \emph{equilibrium values} 
  and a function of the electron temperature only. In contrast, the ion densities
  computed by NEBEL/CORONA are the result of solving explicitly the time-dependent
  rate equations for the local temperatures and densities, and accounting for advection.
  Hence, the NEBEL results are expected to approach the equilibrium solution
  provided by the CHIANTI code only for sufficiently high electron density 
  and low flow velocity.

  The bottom panel of Fig.~\ref{bubble.ion} illustrates nicely how the ionization
  of oxygen changes while the gas passes through the wind shock: The main
  ionization stages of oxygen in the freely streaming wind are those
  of {O}$^{5+}$ and {O}$^{6+}$. After passing through the wind shock,
  the ionization switches to {O}$^{6+}$, {O}$^{7+}$ and {O}$^{8+}$ because of the
  very large  post-shock temperatures. Also close to the conduction front where 
  heated matter streams inwards \citep[relative to the front, see][Fig.\,4]{BBF.90},
  we see a small influence of advection (compare the distribution of {O}$^{5+}$
  around $r$\,=\,$1.2\times10^{17}$~cm in both panels of Fig.~\ref{bubble.ion}).

  The dense regions above $10^6$~K close to the conduction front, i.e.
  for $r\ga10^{17}$~cm in Fig.~\ref{bubble.ion}, contribute most to the bubble's
  X-ray emission, as we will see later.  There we have reasonable agreement
  between the CHIANTI and NEBEL predictions.  The remaining differences are likely
  due to differences in the atomic data used.  Considering the other uncertainties
  involved in this investigation, like, e. g., the wind model, the assumption of
  sphericity, and, last but not least, the observational data, we think that our
  approach of computing the X-ray emission post-facto from the nebular models by
  using the CHIANTI code is well justified.

\begin{figure*}[ht]                               
\sidecaption
\includegraphics*[bb=0.5cm 7.6cm 20.3cm 27.2cm, width=0.67\textwidth]
{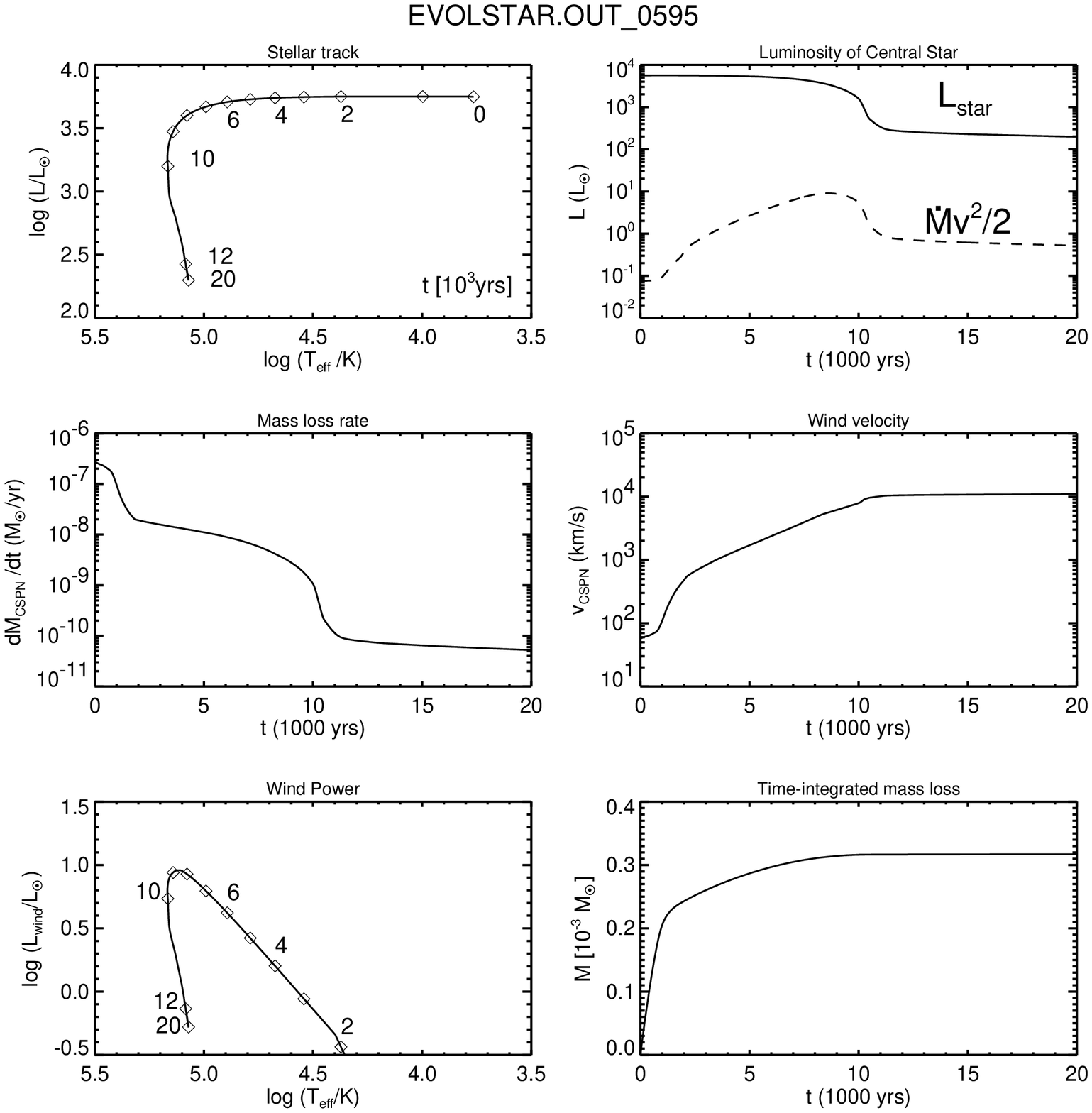}
\caption{\emph{Top}: evolutionary path of the 0.595~\Msun\ model in the
         Hertzsprung-Russell diagram with the post-AGB ages, $t$, indicated along the
	 track (\emph{left}), and the stellar bolometric (solid) and wind (dashed)
	 luminosities vs.\ age (right).
         \emph{Middle}: mass-loss rate (\emph{left}) and (terminal) wind velocity
         (\emph{right}).  We followed the recommendations of \citet{Pauletal.88}
         for the central-star wind ($\teff \ge 25\,000~{\rm K}$), while we assumed
         a Reimers wind \citep{Reim.75} during the transition to the PN domain
         \citepalias[cf.\ also][]{Peretal.04}.
         \emph{Bottom}: stellar wind luminosity (power) vs.\ stellar effective
         temperature (\emph{left}), again with the post-AGB ages indicated, and
         the total mass lost by the wind during the post-AGB evolution 
         (\emph{right}).
        }
\label{mod.prop}
\end{figure*}

\section{Heat conduction and bubble structure}
\label{bubb.struc}
  In this section we discuss in detail the results following from our numerical
  concept introduced in Sect.~\ref{heat.cond} and compare them with our previous
  simulations without heat conduction. For this purpose we used the sequences
  Nos.~6a, 6a-HC and 6a-HC2, all of which are based on the 0.595~\Msun\ post-AGB
  model with the same initial envelope (cf.\ Table~\ref{tab.mod}) but with a 
  different treatment of heat conduction. 

\subsection{The wind model}
\label{wind}
  Figure \ref{mod.prop} illustrates the evolutionary properties of the
  central star and its wind in terms of post-AGB time and stellar
  effective temperature.  The relevant quantity for powering any X-ray
  emission is the mechanical luminosity of the stellar wind,
  $\Lwind$\,=\,$\dotMw\,\Vwind^2/2$. It is important to emphasize that,
  according to the theory of radiation-driven winds for standard
  hydrogen-rich chemical composition in the formulation of
  \cite{Pauletal.88}, \emph{the mass-loss rate and the wind speed depend
  on the stellar parameters (mass, luminosity, effective temperature).}
  Based on these wind prescriptions, the mechanical energy transported
  by the wind increases during the evolution across the
  Hertzsprung-Russell diagram, simply because the slowly decreasing
  mass-loss rate is over-compensated by the increasing wind speed
  (Fig.\,\ref{mod.prop}, upper right). However, when the hydrogen shell
  becomes exhausted, the mass loss rate drops sharply in line with the
  stellar bolometric luminosity, causing also the mechanical wind power
  to drop considerably since the wind speed remains now virtually
  constant at its maximum value of about 10\,000 \kms.

  In any case, the mechanical power remains always rather small and, in this
  particular case, does not exceed 1\% of the stellar photon luminosity
  (Fig.\,\ref{mod.prop}, lower left).
  According to the wind model used in this work, the maximum of the
  mechanical power occurs close to maximum stellar temperature.
  Only very little mass is carried away by the wind during the whole transition
  to the white-dwarf domain, viz.\ \hbox{$\approx\!3\times 10^{-4}$~\Msun}
  (lower right panel), which may be compared with the typical PN mass of
  a tenth of a solar mass. 

  We emphasize that \emph{most} of
  this mass is already lost with low speed during the first 1000 years of
  transition to the PN stage! Only during this phase we have wind speeds as low as
  a few 100 \kms, i.e.\ low enough to provide post-shock temperatures of the order 
  of $10^6$ K. 
  These mass-loss parameters are typical for those of the `early wind' and 
  are here based on the \citet{Reim.75} prescriptions (see caption of 
  Fig.\,\ref{mod.prop}).
  The `early-wind' phase is included consistently in our simulations with the 
  appropriate treatment of radiative cooling.
  The stellar mass lost during the following PN stage is only about 
  $8\times10^{-5}$  \Msun, but this
  material has a very high kinetic energy because of its large speed exceeding
  1000~\mbox{\kms}, leading to post-shock temperatures in excess of $10^7$ K.

\begin{figure}[t]
\includegraphics[bb= 1.8cm 1.0cm 20.4cm 12.5cm, width=\linewidth]
                {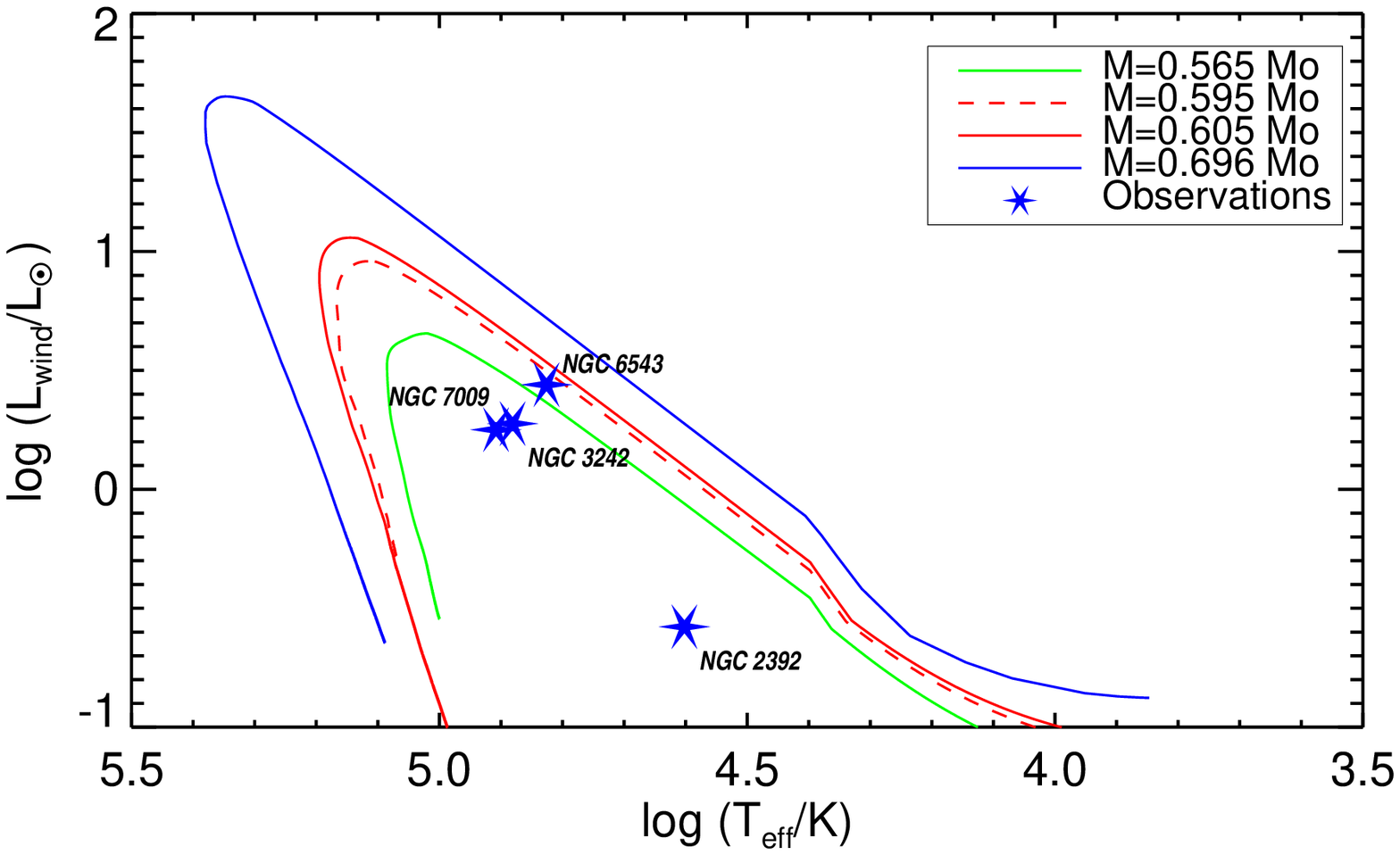}
\caption{Stellar wind luminosities vs. effective temperatures for the four mass
         sequences considered in this work. The observed wind luminosities of
         PNe with diffuse X-ray emission, taken from the compilation in Table
         \ref{tab.xray}, are shown as `star' symbols.
        }  
\label{wind.models}
\end{figure}

  We close this section on our wind model with a discussion about its 
  relevance for real objects. For this purpose we compared the observed 
  wind luminosities of PNe with  diffuse X-ray emission used in 
  Sect.\,\ref{comp.obs} and listed in Table 
  \ref{tab.xray} with the predictions of our post-AGB models 
  (see Fig. \ref{wind.models}). There is only one apparent discrepancy between
  theory and observation: The wind power of \object{NGC~2392} is about a 
  factor of 5 below our predictions, a consequence of the exceptionally low 
  wind speed measured for this particular object. The remaining objects from 
  Table \ref{tab.xray} (\object{NGC~3242}, \object{NGC~6543}, 
  \object{NGC~7009}) have wind luminosities 
  which are, on the average, only a factor 2 below the theoretical predictions.
     
  Given the large uncertainties of the mass-loss rate determinations, we do not
  consider a factor of two difference between observed and computed wind 
  powers to be alarming. Moreover, the distances of \object{NGC~3242}, 
  \object{NGC~6543}, and \object{NGC~7009} used here provide stellar 
  luminosities that are somewhat \emph{below} those for a typical central 
  star of 0.6 \Msun\ which is about 5000 \Lsun\ (cf. Table \ref{tab.xray}). 
  A corresponding increase of the distances would bring theory and observation
  in much closer agreement ($L_{\rm wind} \propto \rm distance^{1.5}$). 

\begin{figure}[h]                 
\includegraphics*[bb= 1.5cm 0.5cm 19.6cm 12.9cm, width=0.95\linewidth]
                 {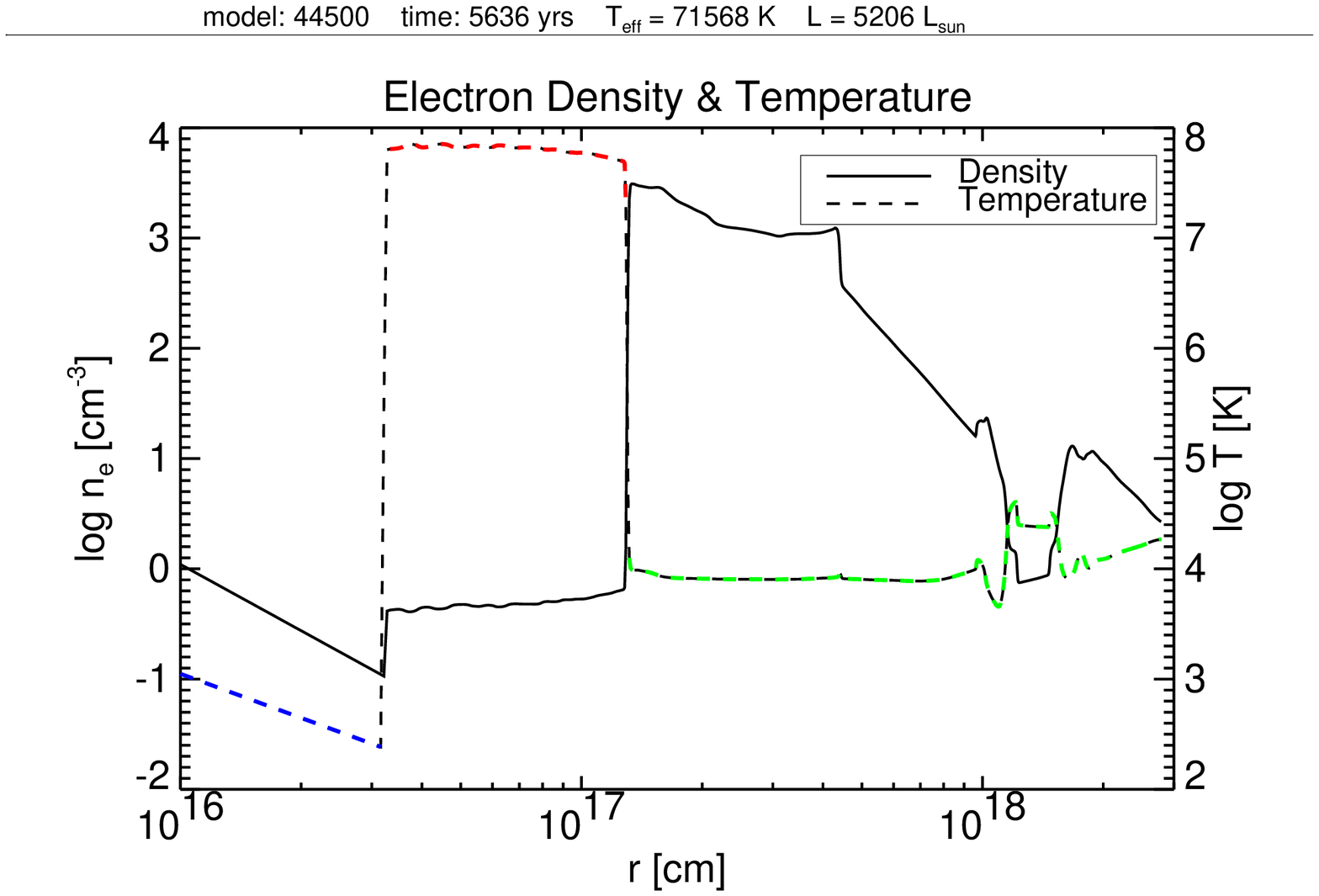}
\includegraphics*[bb= 1.5cm 0.5cm 19.6cm 12.9cm, width=0.95\linewidth]
                 {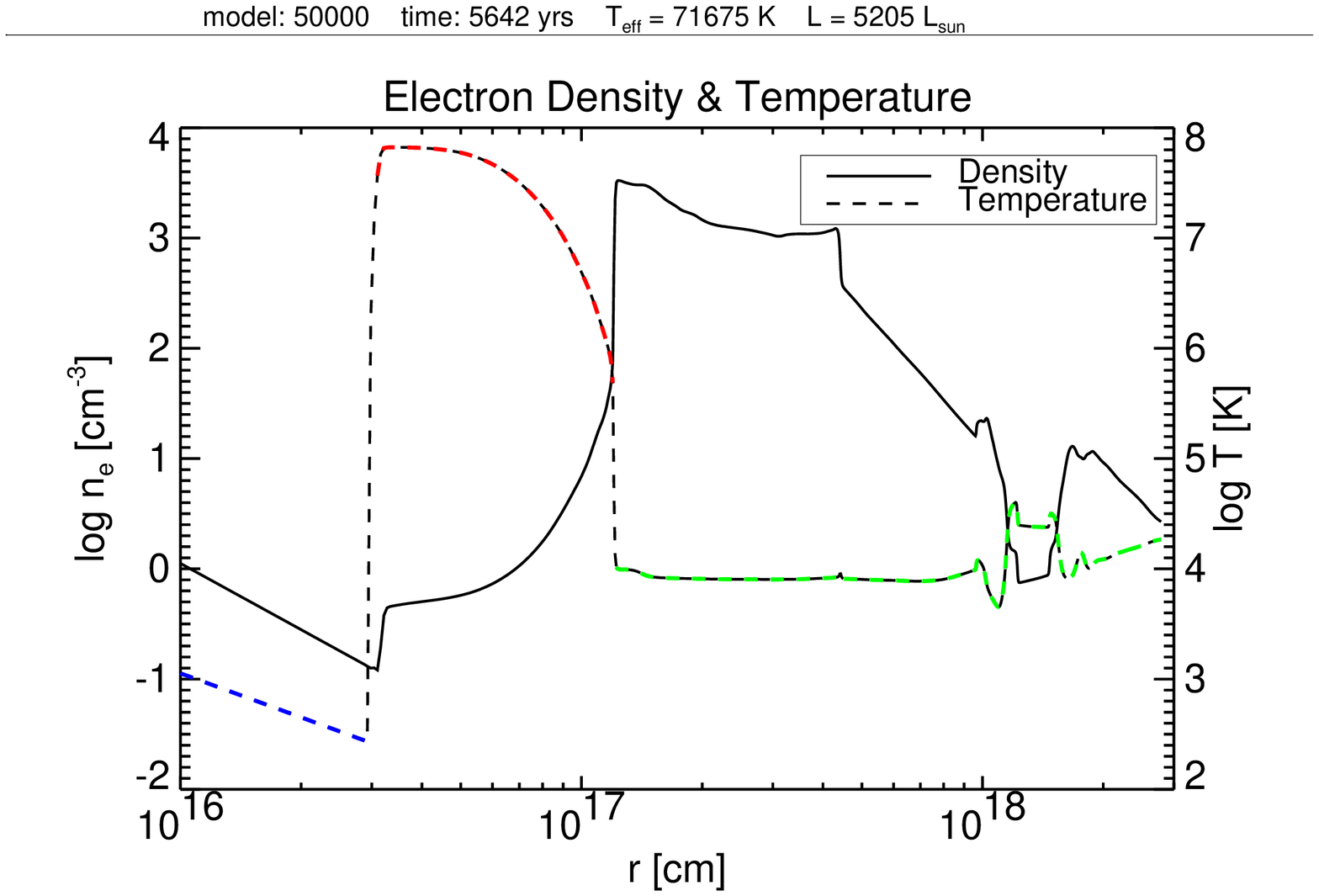}
\includegraphics*[bb= 1.5cm 1.0cm 19.6cm 12.9cm, width=0.95\linewidth]
                 {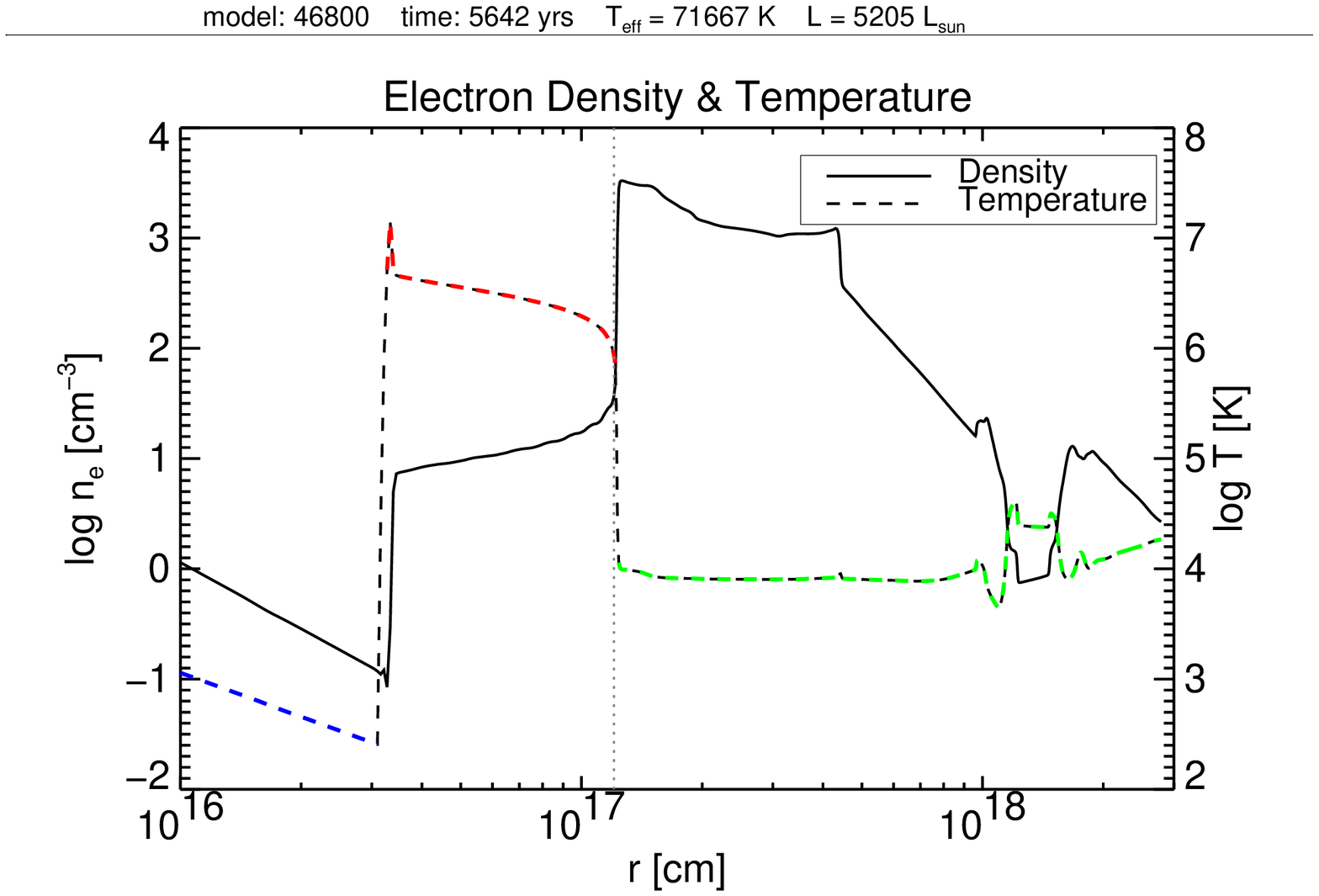}
\caption{Radial profiles of electron density (solid, left ordinate) and temperature
         (dashed, right ordinate) of
         three models taken from sequences No.~6a (\emph{top}), No.~6a-HC 
         (\emph{middle}) and No.~\mbox{6a-HC2} (\emph{bottom}), respectively, 
         at about the same positions along the
	  stellar path shown in Fig.\,\ref{mod.prop}.  The approximate stellar
	  parameters are ${L\simeq5200~\lsun}$, ${\teff \simeq 71\,600~\mbox{K}}$ at
	  ${t\simeq5640~\mbox{yr}}$.
         The central star is at the origin, and the (reverse) wind shock at
	 \mbox{$r\simeq 3\times10^{16}~\mbox{cm}$}.  The shocked wind gas,
	 i.e. the bubble, is (in all three cases) between the wind shock and the
	 contact surface/conduction front ($r\simeq1.3\times10^{17}~\mbox{cm}$).
         The PN proper is  bounded by the contact surface/conduction front and an outer
	 shock at \mbox{$r\simeq4.4\times10^{17}~\mbox{cm}$}. The PN is surrounded by 
         the ionized AGB wind whose radial density profile reflects the mass-loss
         history of the late AGB evolution \citep{Schetal.97, Stetal.98}.
        }
\label{comp.HC}
\end{figure}

\subsection{The influence of thermal conduction}
  The influence of thermal conduction is illustrated in Fig.~\ref{comp.HC} where
  the density and temperature structures of three models at about the same position
  along the evolutionary path shown in Fig.\,\ref{mod.prop} are compared.
  As expected, thermal conduction across the bubble/PN interface has a profound
  impact on the density and thermal structure of the bubble.
  As a reference, the top panel of Fig.~\ref{comp.HC} shows the typical temperature /
  density structure of the shocked wind if thermal conduction is ignored.
  The gas is very hot (${T_{\rm e} \simeq 8\times10^7~{\rm K}}$)
  and tenuous (${n_{\rm e} \approx}$ 0.5~\ccm), and these properties do not change 
  much with radius. Such a bubble structure is in
  sharp contrast to the cases where heat conduction is explicitly considered, as is
  evident from the middle and bottom panels of Fig.~\ref{comp.HC}.

  Already the treatment of heat conduction according to {method~1} leads to a
  completely different bubble structure:  a significant temperature gradient is
  established, whereby the region close to the conduction front
  reaches temperatures as low as $10^6~\mbox{K}$ (Fig.~\ref{comp.HC}, middle panel).
  The bubble remains virtually isobaric (because of the high sound speed), and
  an increased matter density must compensate the temperature decrease.
  This additional mass is provided by a (subsonic) flow of nebular gas heated
  (`evaporated') at the conduction front \citep[cf.][]{Wetal.77}.
  Note that the physical conditions immediately behind the (reversed) wind shock
  are virtually not influenced by heat conduction, i.e. the shock remains
  fully adiabatic.

  {Method~2} provides more efficient heat conduction, and hence the thermal
  structure of the bubble is more homogeneous with smaller temperature and density
  contrasts between wind shock and conduction front (Fig.~\ref{comp.HC}, bottom panel).
  The bubble gas is not hotter than about $5\times10^6~{\rm K}$ (except in a thin region
  immediately behind the wind shock), and the minimum (electron) density is about
  10~\ccm.  Because thermal conduction carries energy away from the wind shock, the
  latter is now not adiabatic anymore: the density jump across
  the shock is nearly two orders-of-magnitude, in sharp contrast to the heat conduction
  treatment based on {method~1} and the case without heat conduction in which
  the densities increase only by a factor of four (see Fig.~\ref{comp.HC}, top and
  middle panel).
  We note also that the
  bubble structures seen in Fig.~\ref{bubb.struc} are typical ones which
  do not change much along the main evolution across the HR diagram.
  The concept of an ideal contact discontinuity with no mass transfer from one side
  to the other does not hold anymore:  The outer edge of the bubble (= inner edge of the
  nebula) is now defined by the heat conduction front.

  Figure~\ref{comp.HC} indicates also that \emph{the dynamics of the whole system remains
  practically unaffected by the physical treatment of the bubble gas}:  in all three
  cases considered, the bubble sizes and the nebular structures are virtually identical
  (see also Fig. \ref{comp.surface}).

  Heat conduction does not change the total energy budget of the bubble but only
  transfers thermal energy across the bubble from the wind shock towards
  the contact surface/conduction front, where it is used to heat and
  `evaporate' nebular gas. The latter flows inwards relative to the conduction front
  and remains thereby inside the bubble \citep[see e.g.][]{Wetal.77}.
  Line cooling can change the bubble's energy content substantially 
  (see Sect.\,\ref{bubble.time}).
  However, radiative cooling is confined to a thin surface layer of the hot bubble 
  ($10^4\ldots 10^5$~K) whose temperature and density structure is not much affected
  by thermal conduction. Radiative losses are therefore only slightly enhanced in
  models with heat conduction (cf.\ Fig.\,\ref{Qrad}).
  Although the amount of mass `evaporated' from the cold nebular gas exceeds by far
  the mass injected into the bubble by the stellar wind, it can be totally neglected for
  the  mass budget of the PN proper (see Sect.\,\ref{bubble.time}).

\begin{figure}[t]                 
\includegraphics*[bb= 1cm 1cm 20cm 12.85cm, width=\linewidth]{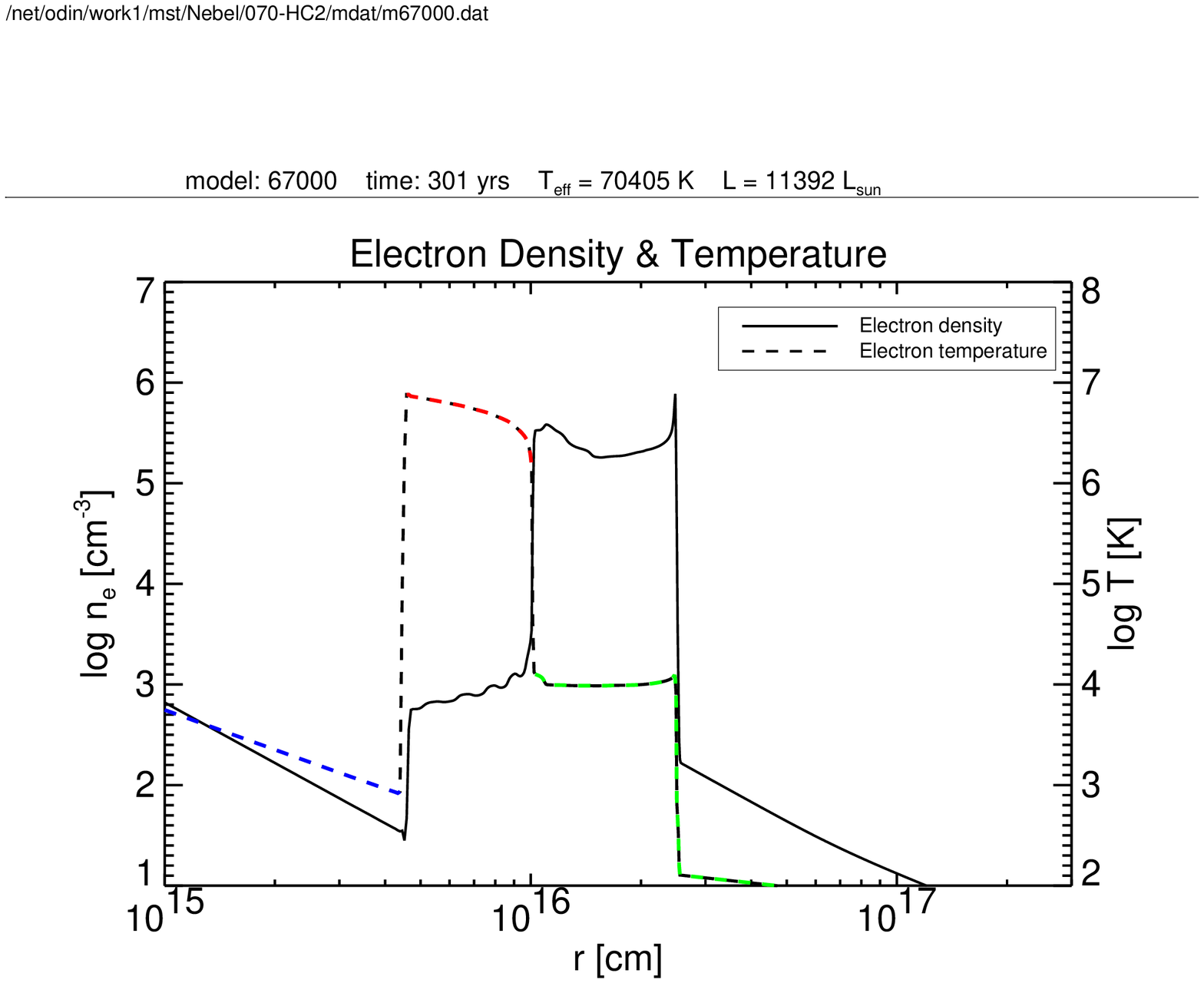}
\includegraphics*[bb= 1cm 1cm 20cm 12.85cm, width=\linewidth]{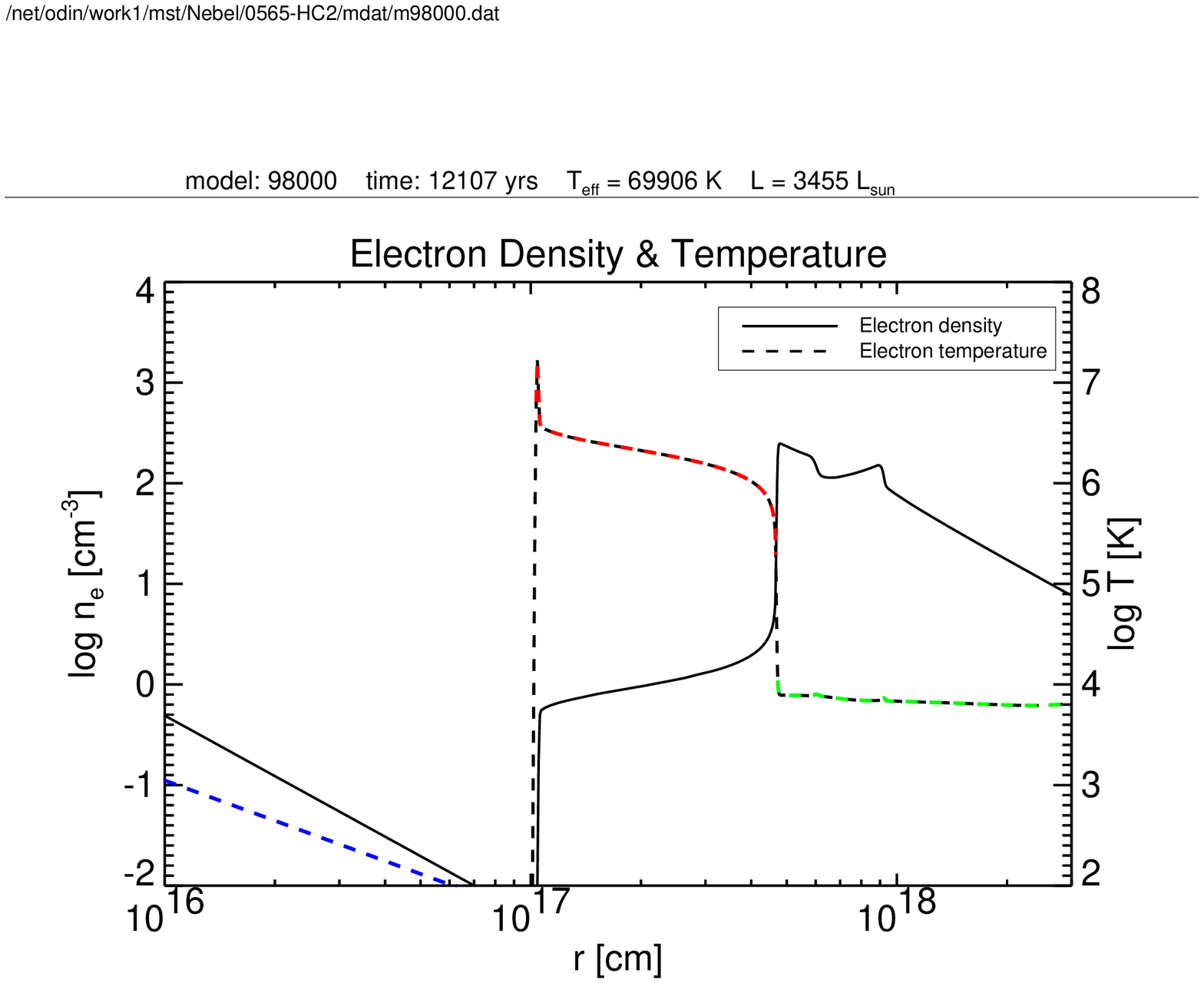}
\caption{Radial profiles of electron density (solid, left ordinate) and
         temperature (dashed, right ordinate) of two models with different
	 central-star masses.
	 \emph{Top}: 0.696~\Msun\ (sequence No. 10-HC2) at \hbox{$t$\,=\,$301$}~yr and
	 with \hbox{$L$\,=\,$11\,392$}~\Lsun, $\teff$\,=\,$70\,405$ K. The large drop of
	 electron density and temperature at $r\simeq2.5\times 10^{16}$~cm indicates
	 the position of the ionization front. The reverse wind shock is at
	 $r$\,=\,$4.5\times 10^{15}$ cm, the conduction front at 
         $r$\,=\,$1\times10^{16}$ cm.
	 \emph{Bottom}: 0.565~\Msun\ (sequence No. 22-HC2) at $t$\,=\,$12\,107$ yr 
         with $L$\,=\,$3455$~\Lsun, $\teff$\,=\,$69\,906$~K. The wind shock is at
	 $1\times10^{17}$ cm, the conduction front at $r$\,=\,$4.5\times10^{17}$ cm.
	 Note the different ranges the electron density.
        }
\label{comp2.HC}
\end{figure}

  Figure \ref{comp2.HC} illustrates the typical bubble structures for two models
  with different central stars and hence different wind powers and time scales of
  evolution.  The model with the massive central star (0.696~\Msun) has a very
  compact and dense bubble due to its fast evolution and powerful stellar wind.
  The other extreme occurs for the model with the least massive central star,
  (0.565~\Msun), which evolves most slowly and has only a rather modest stellar wind:
  the bubble is extended and relatively tenuous.  Also the mean bubble
  temperatures reflect the different wind properties: the 0.696~\Msun\ model has
  the largest bubble temperatures of our sequences because the mechanical energy
  input by the the stellar wind is largest and the bubble size is smallest, as explained
  in more detail in Sect.\,\ref{bubble.time}. The case of 0.595~\Msun\
  shown in the bottom panel of Fig.~\ref{comp.HC} is intermediate.

\subsection{The evolution of the bubble masses}
\label{bubble.time}

\begin{figure}[t]                 
\includegraphics*[bb= 1.0cm 14.5cm 20.5cm 25.7cm, width=\columnwidth]{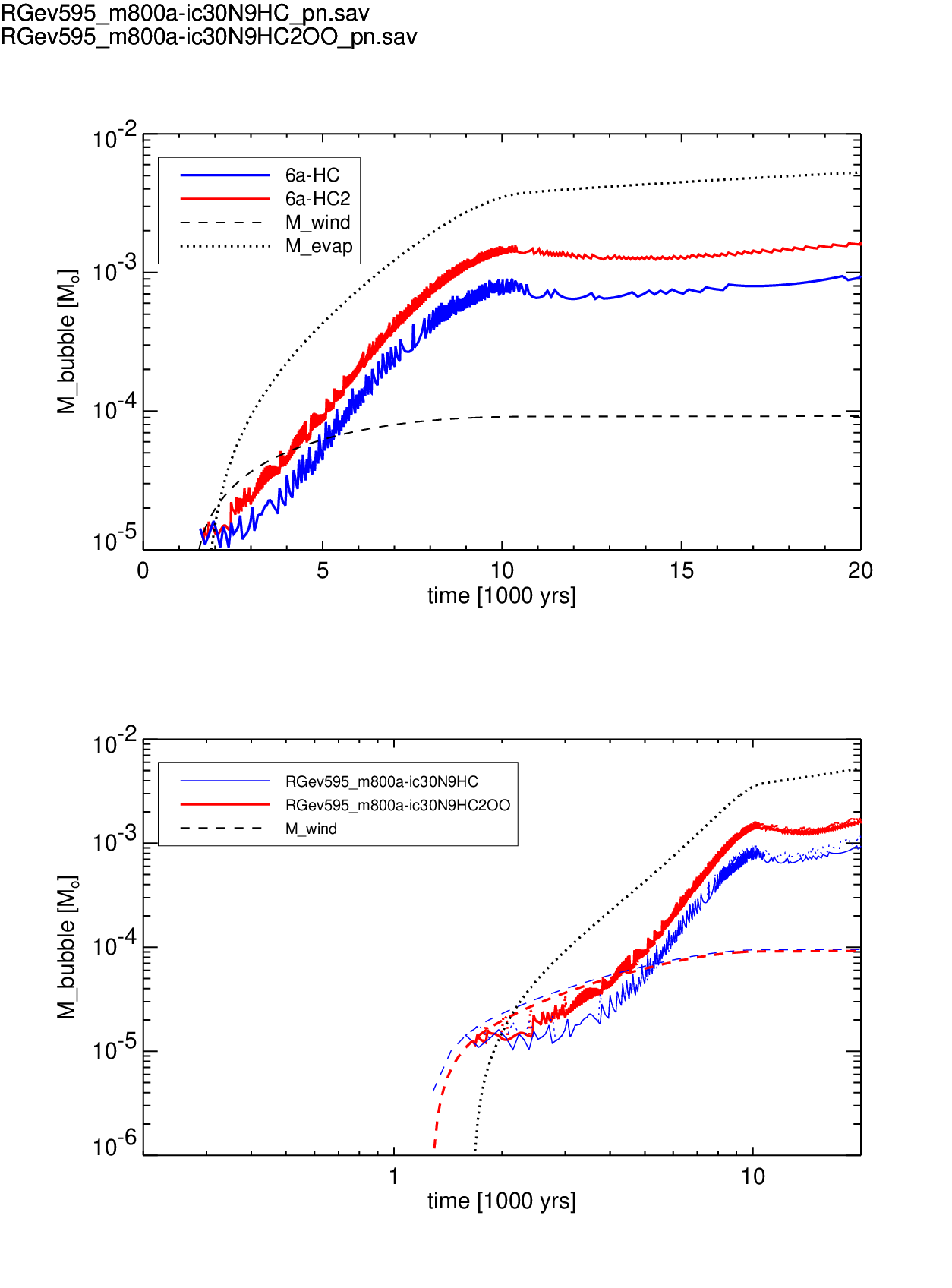}
\caption{Evolution of the bubble mass with time for the sequences Nos.~\mbox{6a-HC}
         and 6a-HC2.  Gas hotter than $10^5$~K is considered to belong to the bubble.
         The scatter is numerical noise.
	 The dashed line indicates the mass blown into the bubble by the central-star 
         wind, counted from $t = 1200$ yr. The dotted line gives a theoretical upper 
         limit of the `evaporated' mass based on Eq.\,(\ref{meva3}).
        }
\label{bubble.mass}
\end{figure}

  The evolution of the bubble's mass with time is shown in Fig.~\ref{bubble.mass} for
  the two heat conduction cases considered here.  One sees that the `evaporated'
  nebular  matter soon exceeds the matter blown into the bubble by the stellar wind.
  After 10\,000 years of evolution across the Hertzsprung-Russell diagram, the
  mass contained in the bubble reaches a maximum which depends on the treatment of
  thermal conduction: \hbox{$\simeq\!8\times10^{-4}$}~\Msun\ (HC) and
  \hbox{$\simeq\!15\times10^{-4}$}~\Msun\ (HC2), respectively, which is still
  negligible compared to a typical PN shell mass of about 0.1~\Msun.
  Obviously, the reduced thermal conduction efficiency of method~1 results in
  a lower evaporation rate of cold gas relative to that of method~2.
  In either case, the whole bubble mass is virtually confined in a narrow outer shell.

  Note that most of the mass blown-off by the central-star wind during the post-AGB
  evolution is not contained in the hot bubble but in the inner part of the nebula.  
  The (reverse) wind shock forms only when
  the wind speed exceeds a certain value, which happens at about 1200 yr after
  departure from the tip of the AGB in the case of a 0.595 \Msun\ central-star model
  (for the zero point see Fig.~\ref{mod.prop}).
  Only about \hbox{$1\times10^{-4}$} \Msun\ are blown into the bubble during the
  remaining part of evolution (Fig.~\ref{bubble.mass}).

  The general time dependence of the bubble mass as shown in Fig.~\ref{bubble.mass} can
  be interpreted in the following way.  According to \citet{BBF.90}, one can distinguish
  three different phases for thermal conduction fronts: (i) the `evaporation' phase
  in which the front advances relative to the initial interface and heats cool gas;
  (ii) the quasi-static phase in which the front stalls because heating and `evaporation'
  balance radiative cooling and `condensation'; (iii) the `condensation' phase in which 
  the front recedes relative to the previous position because cooling/`condensation' 
  dominates.

  As long as the wind power and hence the energy input into the bubble
  increases steadily, the conduction front is obviously in the `evaporation' phase, and
  cooler nebular gas is heated and added to the bubble.  This phase ends
  when the stellar luminosity, and hence also the wind power, drops because hydrogen
  burning ceases.  This occurs in our simulations after about 10\,000~years
  of post-AGB evolution for the 0.595~\Msun\ model shown in Fig.~\ref{bubble.mass}
  (see also bottom left panel of Fig.~\ref{mod.prop}).   For the following $\approx$
  3500~years radiation cooling dominates, and the front becomes `condensation'
  dominated, leading to a small reduction of the bubble mass (Fig.~\ref{bubble.mass})
  and the rate of expansion.  After the rapid fading of
  the central star has stopped at about 12\,000 years, the wind
  is blowing with nearly constant albeit lower strength (Fig.~\ref{mod.prop}).
  Due to the continued expansion, densities decrease and radiative losses
  becomes less important; the conduction front turns slowly back into the 
  `evaporating' stage, and the bubble mass starts to increase again, but at a 
  reduced rate (Fig.~\ref{bubble.mass}).

\begin{figure}[t]                 
\includegraphics*[bb= 1.0cm 14.5cm 20.5cm 25.7cm, width=\columnwidth]{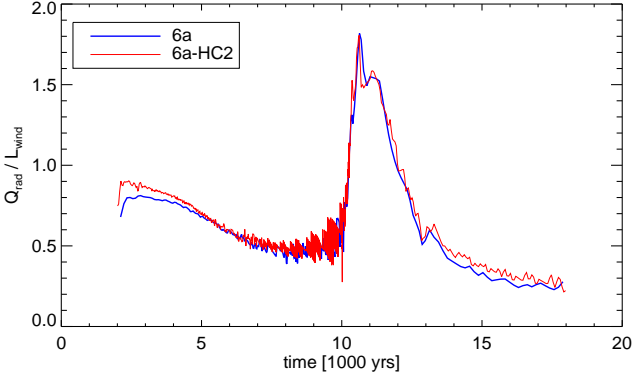}

\caption{The ratio $Q_{\rm rad} / \Lwind$ as a function of time for
models 6a (thick) and 6a-HC2 (thin). Radiative losses are 
confined to a thin layer at the outer surface of the hot bubble;
they are slightly larger in the presence of thermal conduction.
        }
\label{Qrad}
\end{figure}

  For spherical cases \citet{Wetal.77} estimated analytically an evaporation rate
  of
\begin{eqnarray}
  \dotMb & = &
  C_1 \,\langle T_{\rm hb} \rangle^{5/2}\, R_{\rm hb}^2/(R_{\rm hb} - R_{\rm s})
  \nonumber \\
  & \simeq & C_1 \,\langle T_{\rm hb} \rangle^{5/2}\, R_{\rm hb} \quad \mbox{since}
  \quad R_{\rm s} \ll R_{\rm hb}\,,
  \label{meva}
\end{eqnarray}
  where $C_1$$\,\approx\,$$4.13\times10^{-14}$ and
  $\langle T_{\rm hb} \rangle$, $R_{\rm hb}$, and  $R_{\rm s}$ are the
  mean temperature of the hot bubble, its radial size, and the position of the wind 
  shock, respectively. {The central star defines the origin, and the wind shock at
  $R_{\rm s}$ constitutes the inner boundary of the bubble.} 
  It must be noted, however, that Eq.\,(\ref{meva}) assumes that radiative 
  cooling of the bubble is negligible, and thus provides only an
  upper limit of the evaporation rate \citep[cf.][Eq. 61 therein]{Wetal.77}.

  Equation (\ref{meva}) allows us to compute the (equilibrium) evaporation rate 
  of a bubble with given temperature and radius. However, what we really want
  to know here is how the mean temperature of the hot bubble, and hence its
  evaporation rate, depend on the power of the stellar wind, 
  $\Lwind=\dotMw \Vwind^2/2$. 
  To answer this question, we consider the energy balance of the hot bubble,
  first without heat conduction.  In this case, the kinetic energy of the stellar 
  wind is used to increase the thermal and kinetic engery of the bubble, to cover
  the work done by the expansion of the bubble, and to compensate possible radiative 
  losses from the bubble:
  \begin{equation}
    \Lwind = \dot{E}_{\rm th}\, +  \dot{E}_{\rm kin}\, + 
                      W_{\rm exp} + Q_{\rm rad}\, ,
    \label{thb0}
  \end{equation}
  where $ Q_{\rm rad}$ represents the net radiative cooling rate 
  integrated over the volume of the bubble. Denoting by $f_{\rm th}$ the 
  fraction of the wind power that is converted into thermal energy, we can 
  write
  \begin{equation}
    f_{\rm th}\, \Lwind = \dot{E}_{\rm th} = 
    \dotMw\, \frac{3}{2}\,\frac{k}{\mu}\, T_0\, ,
    \label{thb1}
  \end{equation}
  where $\dotMw$ is the stellar mass loss rate and $T_0$ the temperature of the
  shock-heated bubble. For a strong adiabatic shock, $f_{\rm th}$$=$$9/16$ and
  $T_0$ is given by the relation
\begin{equation}
  T_0 = \frac{3}{16}\,\frac{\mu}{k}\, \Vwind^2 \, ,\quad \mbox{or} \quad 
  T_0 \approx 1.4\times 10^7\, 
  \left(\frac{\Vwind}{1000\;\mathrm{km/s}}\right)^2\, [\mathrm{K}]\, .
  \label{thb2}
\end{equation}
Here $\Vwind$ is the velocity of the stellar wind, and $\mu$ is the average mass of 
the gas particles. The numerical factor in Eq.\,(\ref{thb2}) was obtained assuming 
$\mu$$\,=\,$$0.6\,m_{\rm H}$ for fully ionized plasma of solar composition. According
to Eq.\,(\ref{thb2}), the temperature of the hot bubble depends only on the wind
velocity. We note that the bubble temperatures of the two models shown in 
Fig.\,\ref{comp2.HC} cannot be understood with this relation: the stellar wind 
of the more massive central star (top) is about two times slower than that of the
less massive one (bottom), but its bubble temperature is about 3 times higher!
Obviously, thermal conduction makes a difference.

In the presence of heat conduction, the mass of the of bubble increases both due
to the mass loss through the stellar wind with rate $\dotMw$, and due to 
`evaporation' through thermal conduction with rate $\dotMb$. Under these 
circumstances, the wind must supply the power to heat and expand both the
stellar wind entering the bubble through the inner shock and the mass being
added at the surface of the bubble by `evaporation'. The mean temperature of the 
bubble, $\langle T_{\rm hb} \rangle$, is then given by the relation
\begin{eqnarray}
 f_{\rm th}\, \Lwind = \dot{E}_{\rm th} 
    &=& \frac{3}{2}\,\frac{k}{\mu}\, \left(\dotMw + 
        \dotMb \right) \langle T_{\rm hb} \rangle \nonumber \\
    &=&  \frac{3}{2}\,\frac{k}{\mu}\, \left(\dotMw \,\langle T_{\rm hb} \rangle + 
         C_1 \,\langle T_{\rm hb} \rangle^{7/2}\, R_{\rm hb} \right) \,  ,
  \label{thb3}
\end{eqnarray}
where we have used Eq.\,(\ref{meva}). Equation (\ref{thb3}) can be used to estimate
the mean temperature of the hot bubble for given wind power, wind mass loss rate, 
and bubble size. In turn, the evaporation rate follows from Eq.\,(\ref{meva}). 

\begin{figure}[t]                 
\includegraphics*[bb= 1.0cm 14.5cm 20.5cm 25.7cm, width=\columnwidth]{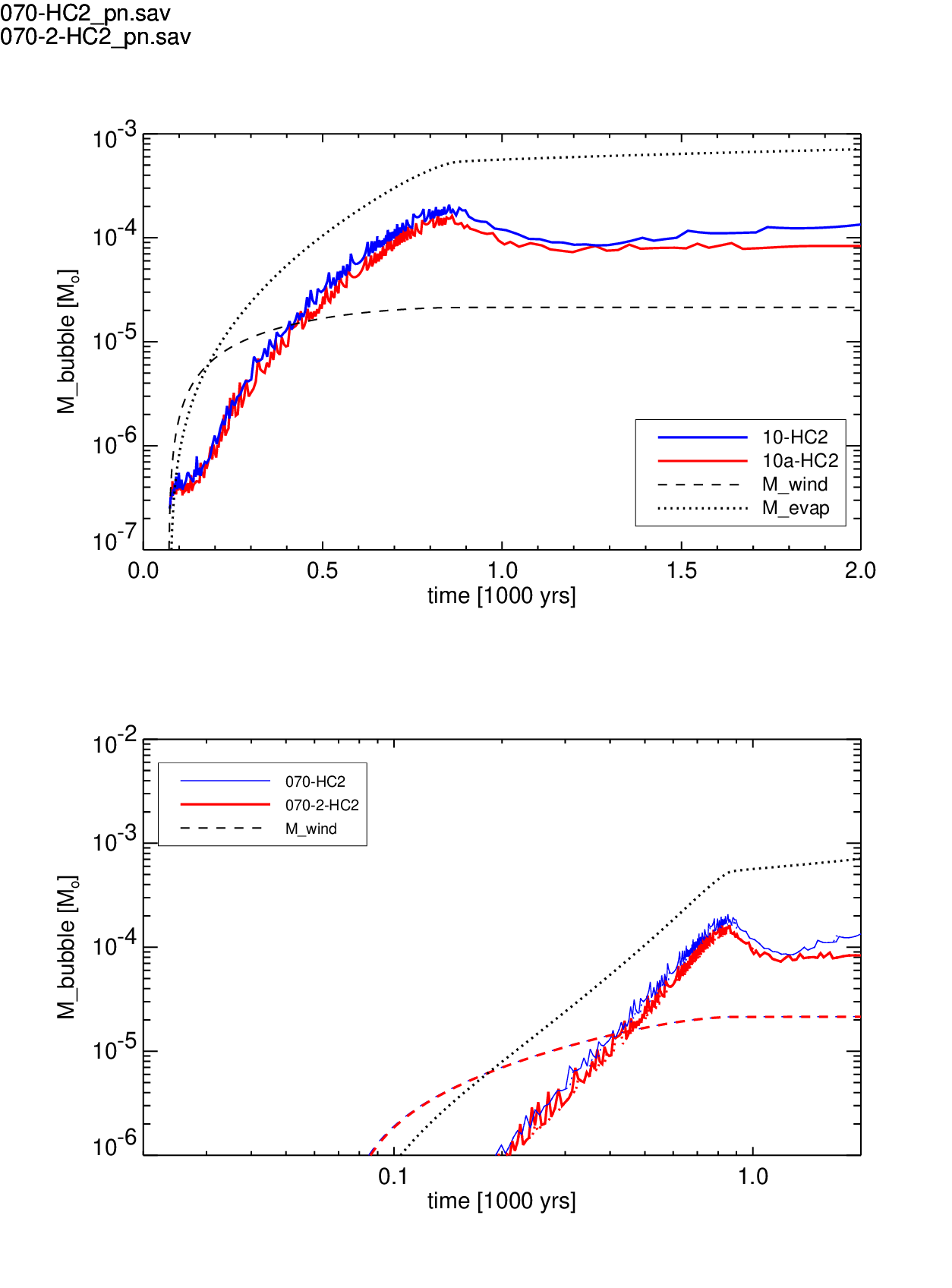}
\includegraphics*[bb= 1.0cm 14.5cm 20.5cm 27.3cm, width=\columnwidth]{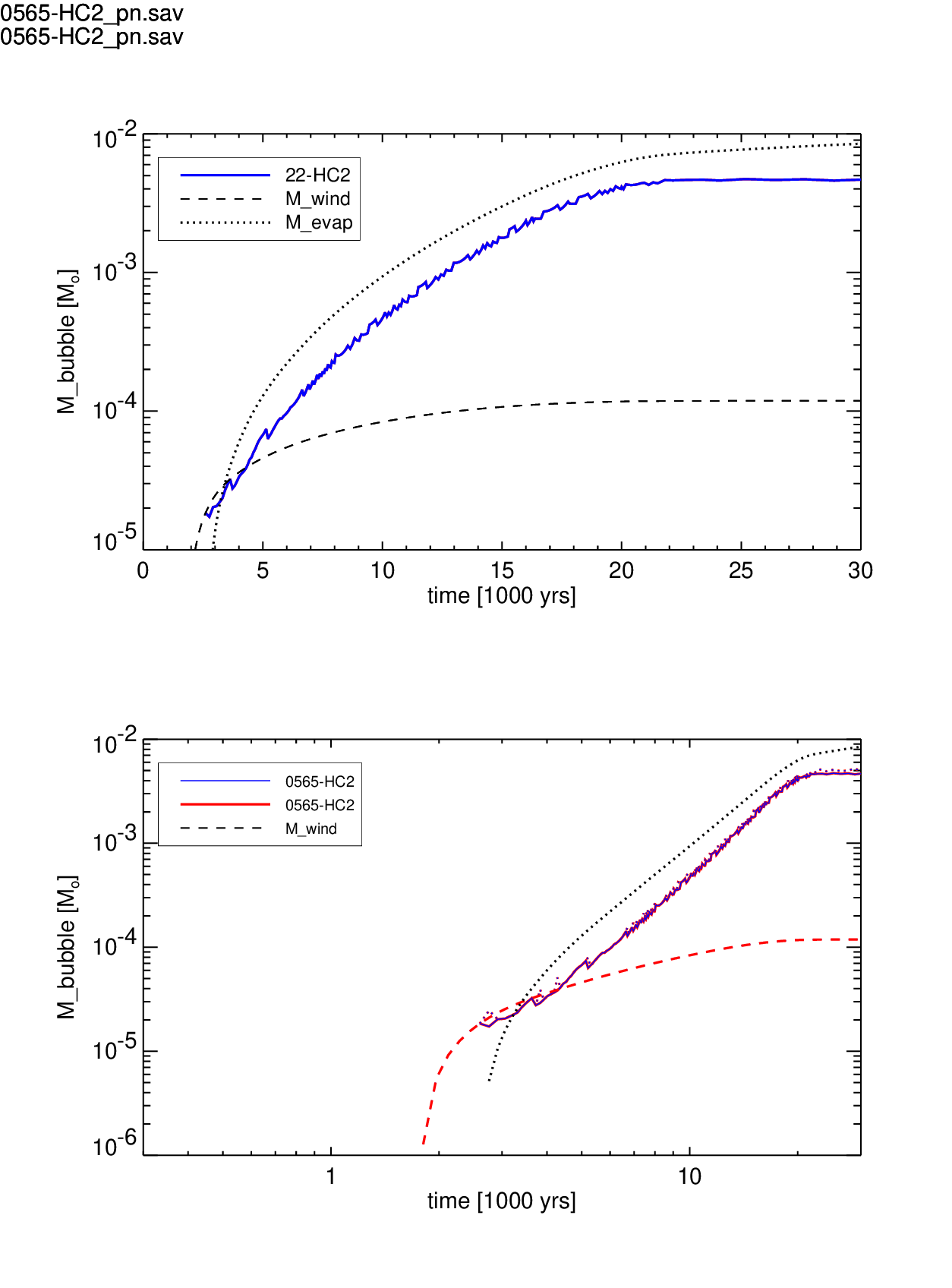}

\caption{The bubble masses vs. time for two sequences with a 0.696 \Msun\ central star
         (Nos.~10-HC2 and 10a-HC2) (\emph{top}) and the 0.565 \Msun\ sequence (No.
	 22-HC2) (\emph{bottom}). Again, only gas hotter than $10^5$~K is
	 considered to belong to the bubbles.
	 The dashed line indicates the masses blown into the bubble by the
	 central-star wind, counted from $t$\,=\,$75$ yr (0.696 \Msun) and
	 $t$\,=\,$2600$ yr (0.565 \Msun), respectively.
         The dotted line gives the theoretical upper 
         limit of the `evaporated' mass based on Eq.\,(\ref{meva3}).
        }
\label{bubble.mass.07}
\end{figure}

If $\dotMw \ll \dotMb$, which is a valid approximation during the main
part of the evolution (once the bubble is well established until the central star 
begins to fade, see Figs.\,\ref{bubble.mass} and \ref{bubble.mass.07}), we obtain
\begin{equation}
  \langle T_{\rm hb} \rangle^{7/2} = \frac{ f_{\rm th}\,\Lwind}
  {\left(\frac{3}{2} \frac{k}{\mu} C_1\, R_{\rm hb} \right)} 
  \quad \mbox{or} \quad 
  \langle T_{\rm hb} \rangle \approx 24\, 
  \left(\frac{\Lwind}{R_{\rm hb}}\right)^{2/7}\, .
  \label{thb4}
\end{equation}
According to Eq.\,(\ref{thb4}), the temperature of the hot bubble
depends on the wind power and the bubble radius. This relation
explains the behavior of the two models shown in Fig.\,\ref{comp2.HC}:
the 3 times higher wind power of the more massive central star, and
its 10 times smaller bubble size imply a mean bubble temperature which
is about 2.6 times higher that that of the less massive model.

Combining Eqs.\,(\ref{meva}) and (\ref{thb4}), we can write the evaporation rate
as
\begin{equation}
 \dotMb = \frac{f_{\rm th}^{5/7}\,\Lwind^{5/7}\,C_1^{2/7} 
 \,R_{\rm hb}^{2/7}}{\left(\frac{3}{2} \frac{k}{\mu}\right)^{5/7}}
 \label{meva2}
\end{equation}
or
\begin{equation}
 \dotMb=1.3\times 10^{-7}\,
 \left(\frac{\Lwind}{\lsun}\right)^{5/7}\ 
 \left(\frac{R_{\rm hb}}{10^{17}\;\mathrm{cm}}\right)^{2/7}\,[\mathrm{\mdot}] \, .
 \label{meva3}
\end{equation}
The dotted curves shown in Figs.\,\ref{bubble.mass} and \ref{bubble.mass.07} 
were obtained by integrating Eq.\,(\ref{meva3}) over time, 
given $\Lwind(t)$ and $R_{\rm hb}(t)$ for the respective model sequence.
Note again that Eq.\,(\ref{meva3}) gives an upper limit of the evaporation rate,
since both $f_{\rm th}$ and $C_1$ will be smaller than assumed here if 
radiation losses are important.

In our model simulations, `evaporation' due to thermal conduction is significantly 
less efficient than predicted by Eq.\,(\ref{meva3}). As shown in 
Fig.\,\ref{bubble.mass}, the mass of the hot bubble measured in the numerical model 
is a factor 3 to 5 lower
than suggested by the analytical estimate. In-depth investigations have revealed 
that this discrepancy is caused by substantial radiative energy losses
at the conduction front / contact discontinuity. As illustrated in Fig.\,\ref{Qrad},
these losses are most severe in the early phases, when a large fraction of the wind 
power is radiated away. This explains also why the bubble mass grows very slowly at 
the beginning: a considerable part of the mass added to the bubble by the stellar 
wind cools efficiently and `condensates' outside the bubble. Later, `evaporation' 
becomes more efficient and the mass of the bubble grows much faster than implied 
by the wind moss loss rate. But even at $t$ = 8000 yr, 50\% of the wind power 
is still lost as radiation. It is interesting to note that the radiative cooling at
the surface of the hot bubble is hardly altered by thermal conduction (see 
Fig.\,\ref{Qrad}).

  For completeness we show in Fig.~\ref{bubble.mass.07} the evolution of
  bubble masses for two sequences with a more massive (0.696 \Msun) and for one
  sequence with a less massive central star.
  The two sequences Nos. 10-HC2 and 10a-HC2 with the massive central-star model 
  (top panel) differ only by their AGB mass-loss rates of $1\times10^{-4}$ and 
  $2\times10^{-4}$ \Mdot, respectively (see Table~\ref{tab.mod}).

  For the 0.696 \Msun\ cases, the time evolution is similar as already discussed 
  above for 0.595 \Msun. A phase of rapidly increasing bubble mass is followed 
  by a substantial decrease when the heat supply by the fading central-star wind 
  drops. The `condensation' of bubble mass is more severe in the 
  0.696 \Msun\ cases because the bubbles are small and rather dense 
  \hbox{($n_{\rm e}\approx 200\ldots500$ cm$^{-3}$)}, affording a more efficient 
  radiative cooling. The discrepancy between the evaporation rate obtained
  from the simulation and the prediction of Eq.\,(\ref{meva3}), respectively,
  is even more  severe than in the case of the central star with 0.595 \Msun.

  Since the massive central-star models  evolve on very short time scales,
  their bubbles are the smallest in size and mass of all sequences computed.
  The slightly smaller bubble masses found for sequence No. 10a-HC2 
  (as compared to those of sequence No. 10-HC2) are due to their somewhat smaller
  sizes, a consequence of the denser bubble environment caused by
  the larger AGB mass-loss rate, $\dot{M}_{\rm agb} = 2\times10^{-4}$ \Mdot\
  instead of $\dot{M}_{\rm agb} = 1\times10^{-4}$ \Mdot.

  In the case of the low-mass central star (0.565 \Msun),
  the wind power drops slowly after 20\,000 years of post-AGB evolution. As a
  result, the mass of the hot bubble ceases to increase. But in contrast to the
  simulations with more massive central stars, there is no real 'condensation'
  phase. Rather, the mass of the hot bubble remains constant with time, obviously
  because radiative cooling is inefficient due to the low plasma densities. Hence,
  we find the best agreement between analytical and numerical evaporation rate
  for this sequence.

  We finally note that 'evaporation' due to heat conduction (HC2)
  increases the bubble mass by as much as a factor of 10, relative to the 
  wind-blown mass, for the 0.696 \Msun\ cases, and by a factor of about 40
  for the model with $M$$=$$0.565$ \Msun\ near the end of the evolution
  ($t$$\,>\,$$20\,000$~yr), when 'evaporation' and 'condensation' balance and
  the bubble mass is essentially constant (see Fig.~\ref{bubble.mass.07}).
  The case of 0.595~\Msun\ is intermediate.
\section{The X-ray emission}
\label{x-rays}
\subsection{Spectra}
\label{spec}

\begin{figure}[t]                 
\includegraphics*[bb= 0.9cm 1.0cm 20.5cm 12.8cm, width=\linewidth]{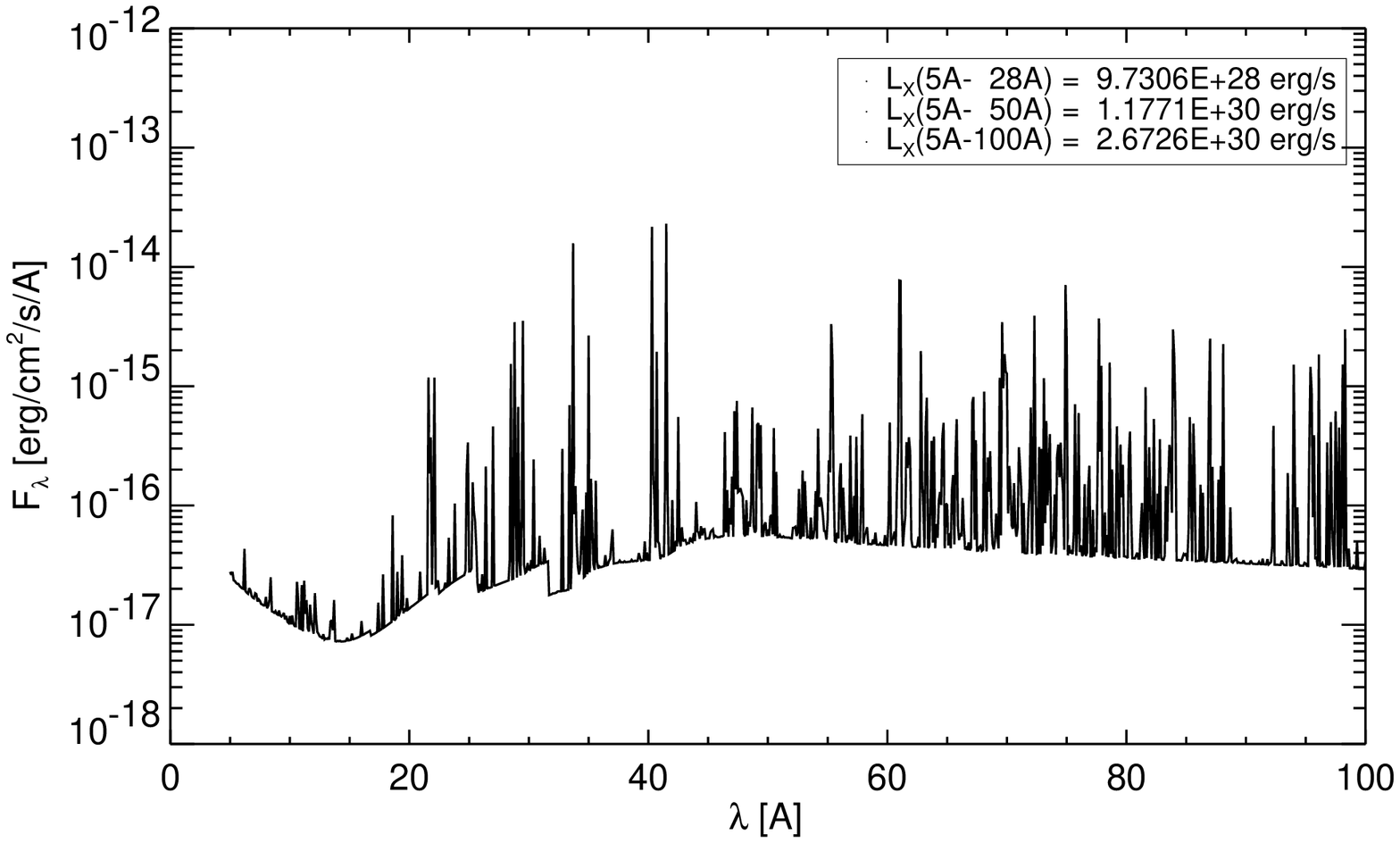}
\includegraphics*[bb= 0.9cm 1.0cm 20.5cm 12.8cm, width=\linewidth]{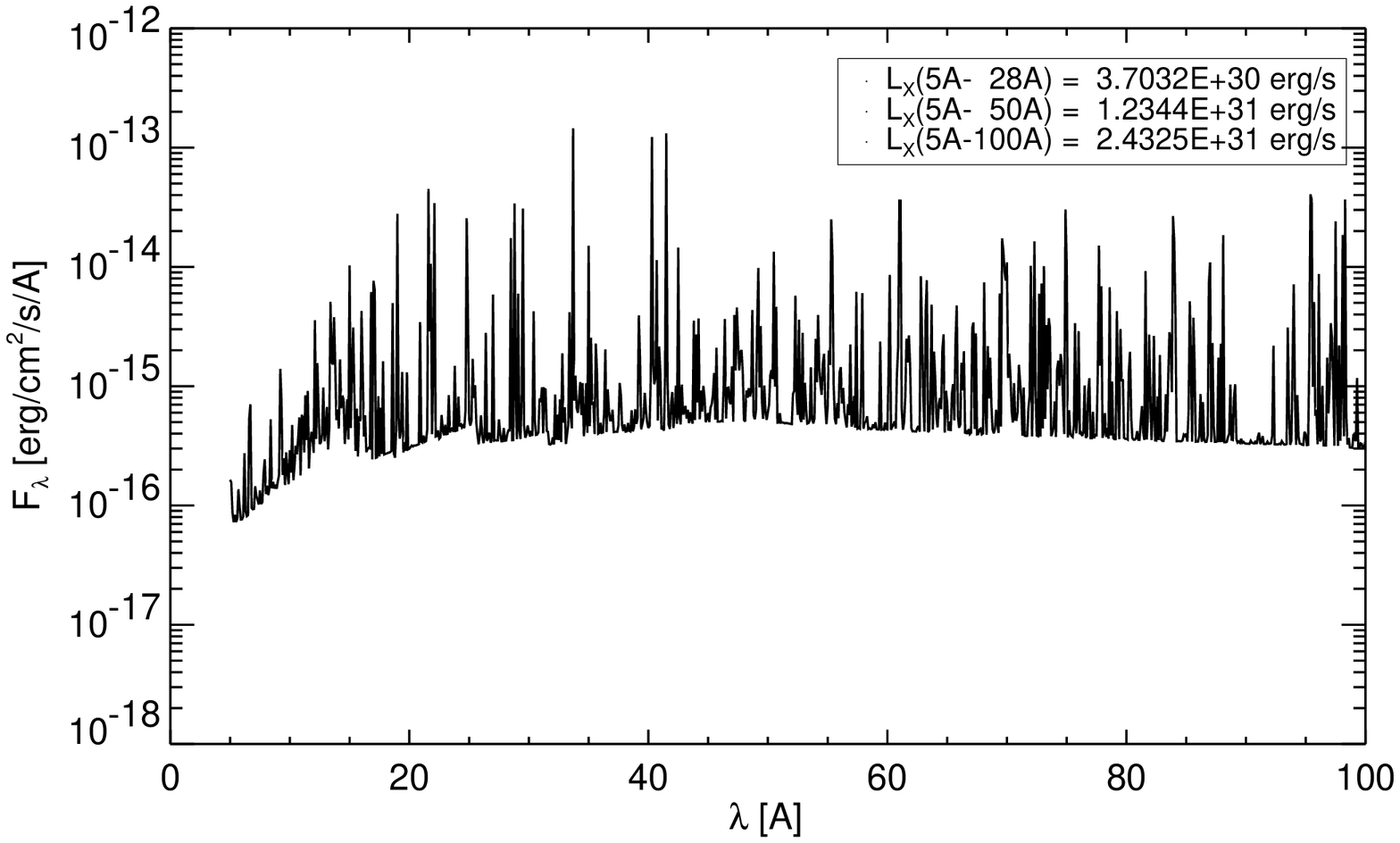}
\includegraphics*[bb= 0.9cm 1.0cm 20.5cm 12.8cm, width=\linewidth]{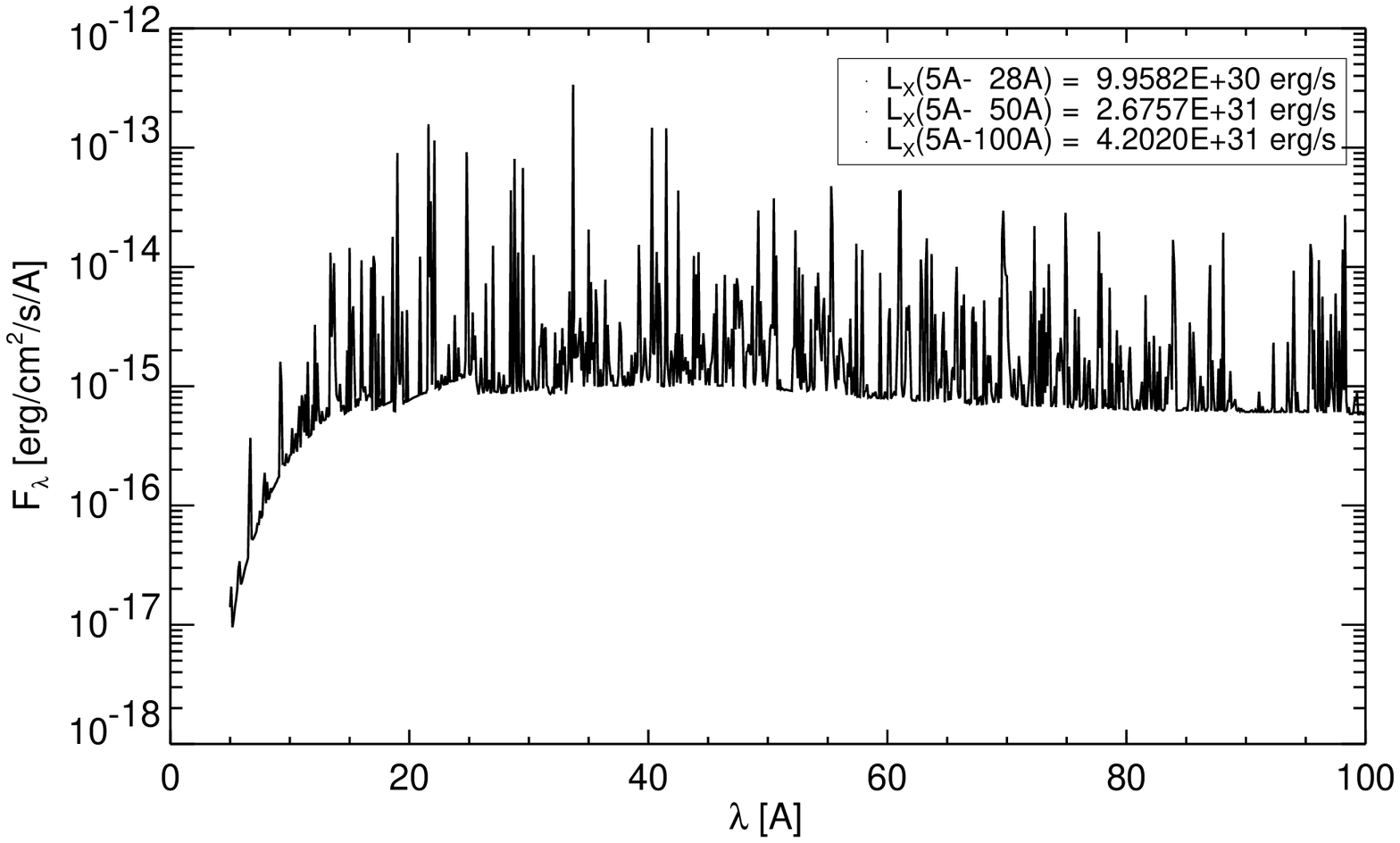}
\caption{The fluxes emitted in the wavelength band of 5--100~\AA\ (2.5--0.125~keV)
         computed from the bubbles of the models shown in Fig.~\ref{comp.HC},
	 without heat conduction (\emph{top}), and with heat conduction included:
	 HC (\emph{middle}) and HC2 (\emph{bottom}). 
         The fluxes given in the ordinate refer to an assumed distance of 1~kpc,
         and the contribution of regions with $T_{\rm e} < 10^5$ K is ignored.
         The insets give the luminosities in three different wavelength bands.
        }
\label{comp.X-ray}
\end{figure}

\begin{figure}[t]                 
\includegraphics*[bb= 0.9cm 1.0cm 20.5cm 12.8cm, width=\linewidth]{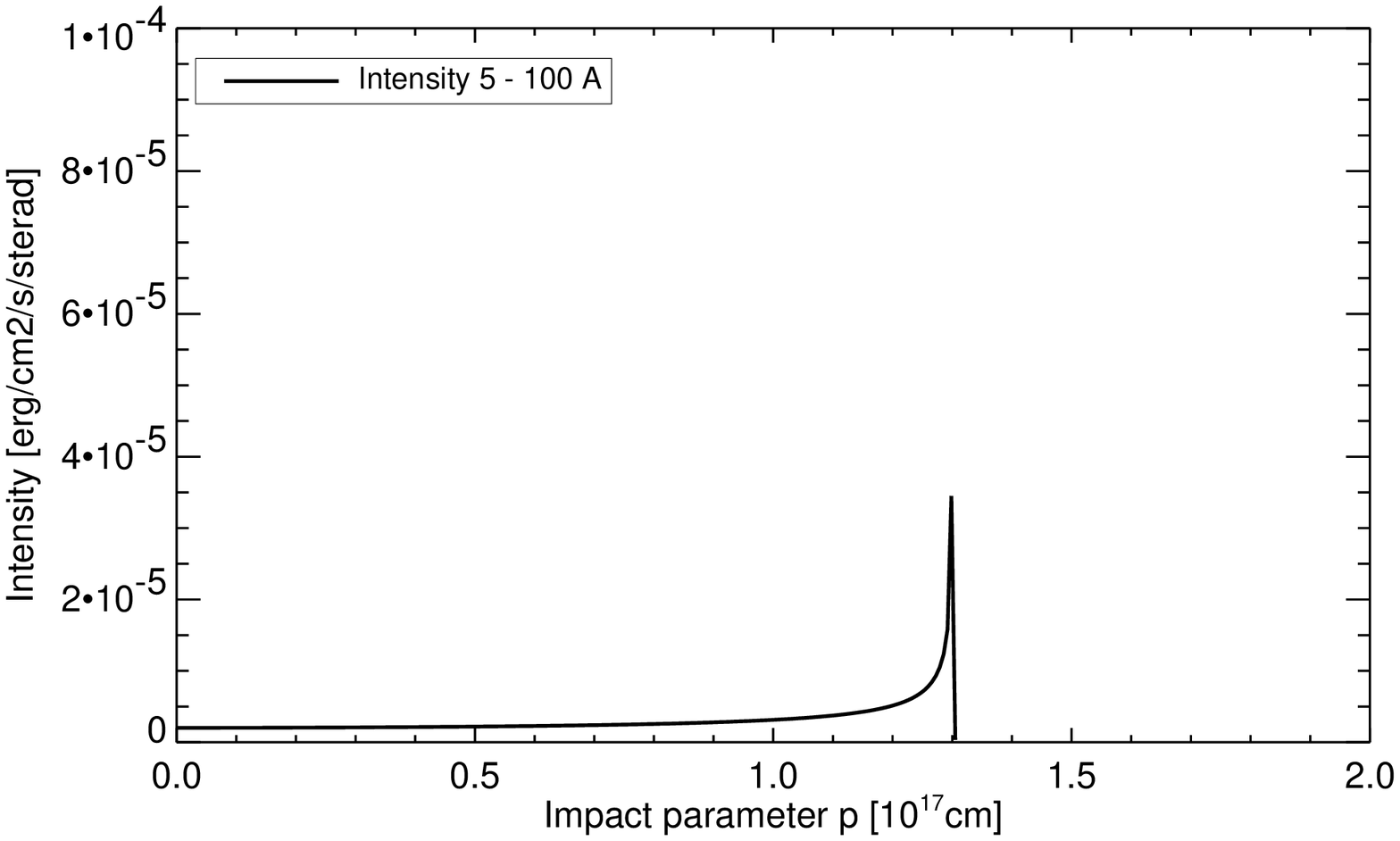}
\includegraphics*[bb= 0.9cm 1.0cm 20.5cm 12.8cm, width=\linewidth]{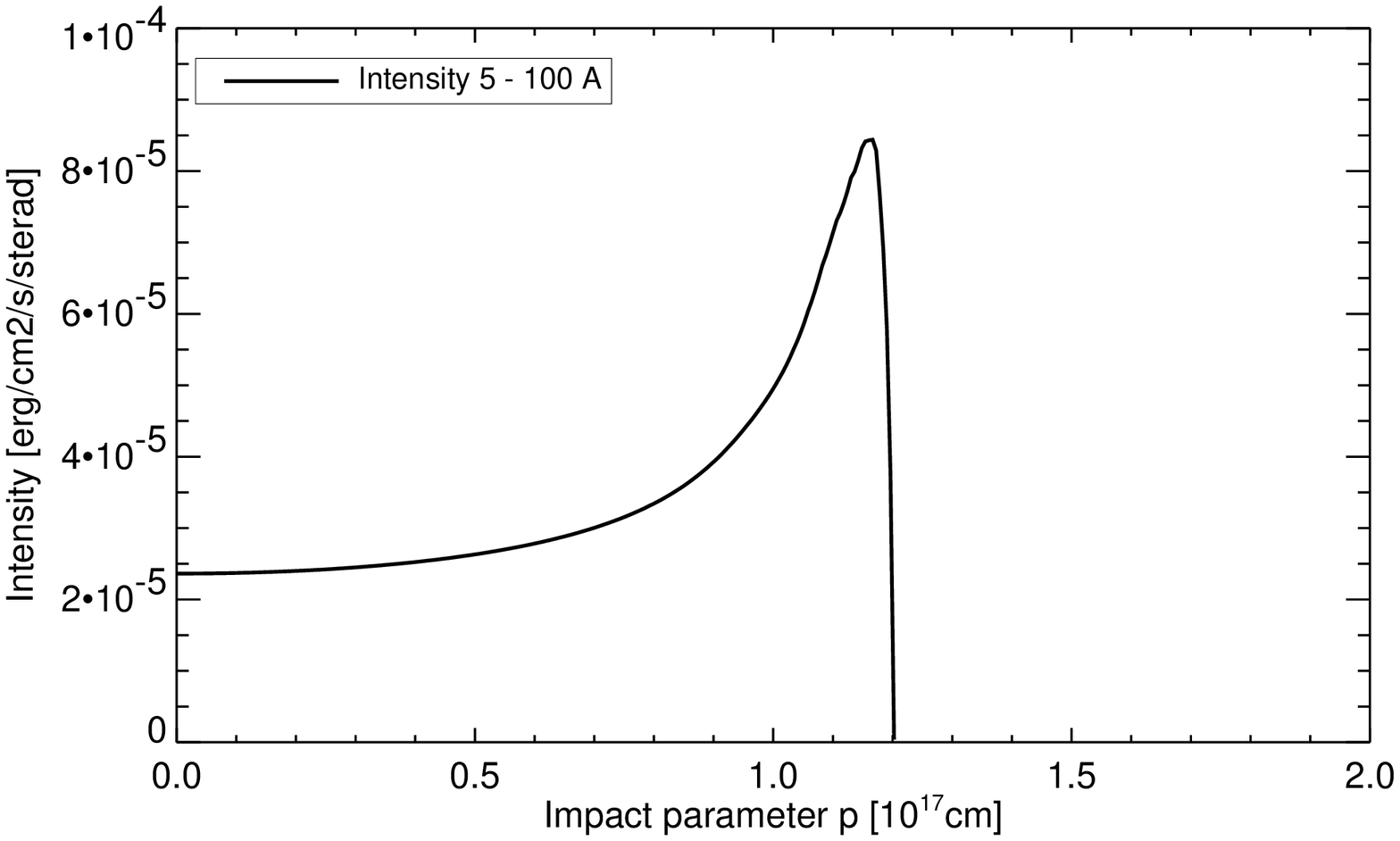}
\includegraphics*[bb= 0.9cm 1.0cm 20.5cm 12.8cm, width=\linewidth]{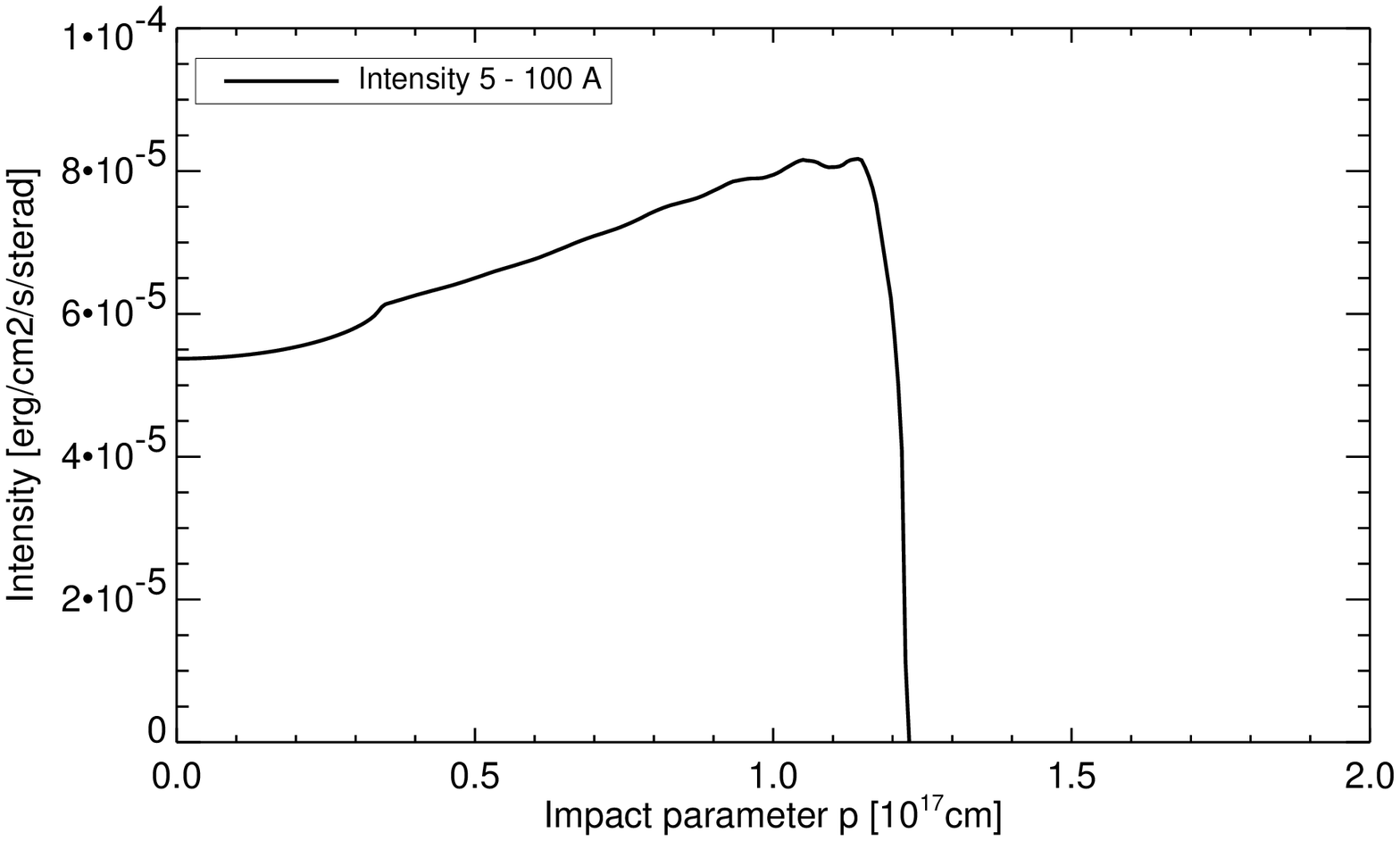}
\caption{Radial intensity profiles, integrated over the wavelength band of
         5--100~\AA\ (2.5--0.125~keV), of the models from Fig.~\ref{comp.X-ray} without
         (\emph{top}) and with heat conduction considered: HC (\emph{middle})
	 and HC2 (\emph{bottom}).  At impact parameter 
         \hbox{$p$$=$$3.3\times10^{16}$~cm} the inner edge of the bubble, 
         i.e.\ the position of the (reverse) wind shock, is visible in the bottom 
         panel.
	}
\label{comp.surface}
\end{figure}

  We computed detailed X-ray spectra, as described in Sect.~\ref{X-ray.chianti},
  for the wavelength interval 5--100~\AA, corresponding to
  2.5--0.125~keV.  Examples are shown in Fig.~\ref{comp.X-ray} for the three
  models from Fig.~\ref{comp.HC}. The X-ray emission consists of a continuum 
  and numerous strong lines of highly ionized species.
  The models with thermal conduction included (middle and bottom panel of
  Fig.~\ref{comp.X-ray})
  have much larger fluxes in the 5--100~\AA\ wavelength band
  than the model without conduction.  The total X-ray luminosities
  (in the same spectral band) are  $2.4\times10^{31}~{\rm erg\,s}^{-1}$ (HC) and
  $4.2\times10^{31}~{\rm erg\,s}^{-1}$ (HC2), corresponding to
  $1.2\times10^{-6}\,{L}_\mathrm{star}$ (or $6.3\times10^{-3}$~\Lsun) and
  $2.1\times10^{-6}\,L_\mathrm{star}$ (or $1.1\times10^{-2}$~\Lsun), respectively.
  Without thermal conduction, the X-ray luminosity is only $ 2.7\times10^{30}$~\ergs\
  in the same wavelength band.  In any case, these
  luminosities are only minute fractions of the mechanical energy carried away
  by the central-star wind, which is about 3.4~\Lsun\ for the models used in
  Fig.~\ref{comp.X-ray} (cf.\ Fig.\,\ref{mod.prop}).

  We note that our models with thermal conduction included provide X-ray
  luminosities which are smaller than those computed by \citet{ZP.96}
  (by roughly a factor 10). However, radiative line cooling is not included
  in their analytical models, leading to a severe overestimation of the 
  evaporation rate (see Sect.\,\ref{bubble.time}). Moreover, the
  simplified assumptions that only wind interaction is responsible for the 
  compression of the AGB wind, ignoring the effects of photoionization, result
  in an unrealistic expansion behavior of the bubble.

  We conclude that an approach like the one conducted by \citet{ZP.96} is 
  likely too simple for a reliable computation of the X-ray emission from 
  planetary nebulae.

  We emphasize in this context that the X-ray flux depends also on the chemical
  composition of the bubble gas. A detailed investigation of how the X-ray flux
  depends on the chemical composition, especially for cases with hydrogen-poor
  stellar winds, will be the subject of future work.

\subsection{Surface brightness distributions}
\label{surface}

  The X-ray intensity, integrated over the wavelength band of 5--100~\AA, is
  limb-brightened and reflects the fact that the regions with physical properties
  favorable for emitting X-rays are mainly the outer layers of the hot bubble
  (Fig.~\ref{comp.surface}). The absolute value of the intensity and its radial 
  profile depend therefore sensitively on the way thermal conduction is treated.

  In the case without any thermal conduction, the X-rays
  are emitted only from the very thin interface between the hot shocked
  stellar wind and the PN gas,
  resulting in a ring-like, comparatively weak emission.
  (Fig.~\ref{comp.surface}, top panel).  In the case of low-efficiency heat
  conduction ({method~1}) the X-ray emission is still rather strongly peaked
  towards the bubble's surface, with a center-to-limb variation of a factor 4 (middle
  panel). {method~2} results in a more homogeneously distributed X-ray emission
  with a modest center-to-limb variation of only a factor 1.5. A larger part of
  the bubble is now contributing to the X-ray emission (bottom panel).

  One notices that, although all the three models shown in Fig.~\ref{comp.surface}
  originated from the same initial configuration and are virtually of equal age,
  the sizes of their bubbles differ by a small amount:
  with thermal conduction included the bubble is a little bit smaller,
  \hbox{$r_\mathrm{hb}=1.2\times10^{17}~{\rm cm}$} instead of
  \hbox{$r_\mathrm{hb}=1.3\times10^{17}~{\rm cm}$} in the case without
  conduction\footnote{
  In the cases with heat conduction, the size of the bubble, $r_{\rm hb}$, 
  is defined by the outer edge of the conduction front.}.
  This appears to be related to the fact that radiative energy losses from the
  bubble are somewhat larger for models including thermal conduction 
  (see Fig.\,\ref{Qrad}). This additional cooling causes a slightly slower 
  expansion of the bubble, but is too small to significantly change the 
  dynamics of the whole nebula: The PN evolution
  is virtually not influenced by heat conduction (cf.\ Fig.~\ref{comp.HC}).

\subsection{Time evolution}
\label{time.evol}

  An important criterion for the usefulness of thermal conduction in explaining the
  X-ray emission from planetary nebulae is not only the predicted X-ray luminosity
  itself, but also how it evolves with time, i.e. as a function of the stellar 
  parameters.
  In general, the X-ray luminosity is determined, for a given temperature, by the
  total volume emission measure of the X-ray emitting region,
  $\eta_{\rm X} \simeq\rho_{\rm X} M_{\rm X}$, with $\rho_{\rm X}$ being a 
  characteristic mean density of the emitting volume and $M_{\rm X}$ its mass.  
  For simplicity
  we set ${\rho_{\rm X}\propto M_{\rm X}/R_{\rm hb}^3}$.  If $M_{\rm X} \propto t^{\,a}$
  and $R_{\rm hb}\propto t^{\,b}$, we have 
  $L_{\rm X}\propto\rho_{\rm X} M_{\rm X}\propto t^{2a - 3b}$.

  Considering for the moment no thermal conduction,
  and assuming a central-star wind with constant mass-loss rate and velocity, we have
  ${a=1}$ and ${b=1}$, where the latter expression
  holds if the bubble expands into an environment with ${\rho\propto r^{-2}}$ 
  \citep[][Eq. 3.1 therein]{KMcK.92}.
  In this case, ${L_{\rm X}\propto t^{-1}}$ \citep[see also][Eq. 33]{VK.85}.
  This situation changes somewhat if the X-rays are only observed within a limited
  energy range and if the wind is evolving in time, i.e. if the wind speed increases
  (cf. Fig.~\ref{mod.prop}).  Since the post-shock temperature scales with wind speed
  squared, the X-ray emission increases first, reaches a maximum at an appropriate
  wind velocity, and drops then rapidly \citep[see][Figs. 11--13 therein]{VK.85}.
  Note that in the case of increasing stellar wind power the bubble expansion is
  accelerated (${b>1}$), leading to a faster X-ray luminosity drop than in the case
  of a constant wind.

\begin{figure}[t]                 
\includegraphics*[bb= 2.0cm 0.5cm 20.2cm 12.05cm, width=\linewidth]{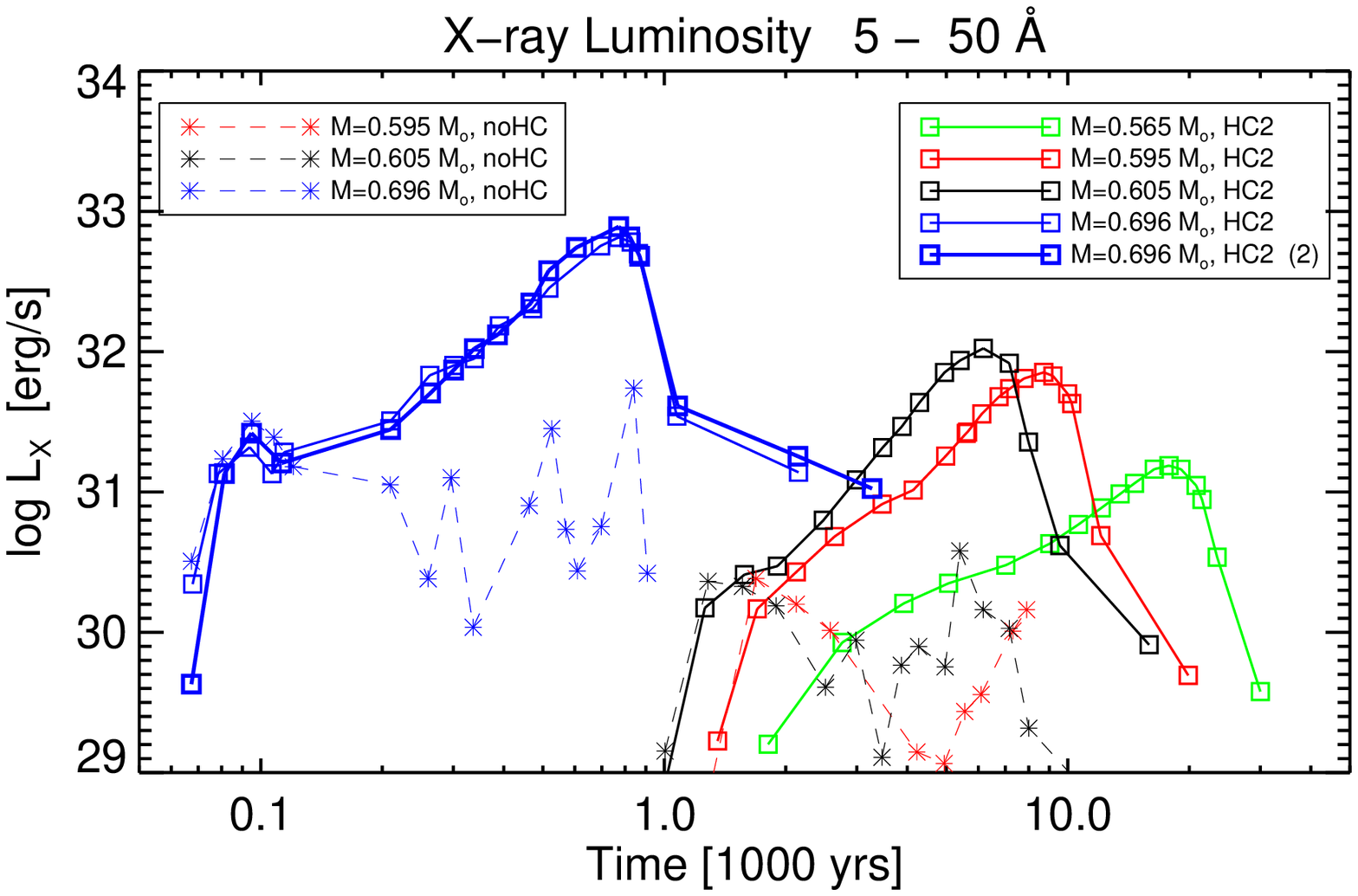}
\includegraphics*[bb= 2.0cm 1.1cm 20.2cm 12.10cm, width=\linewidth]{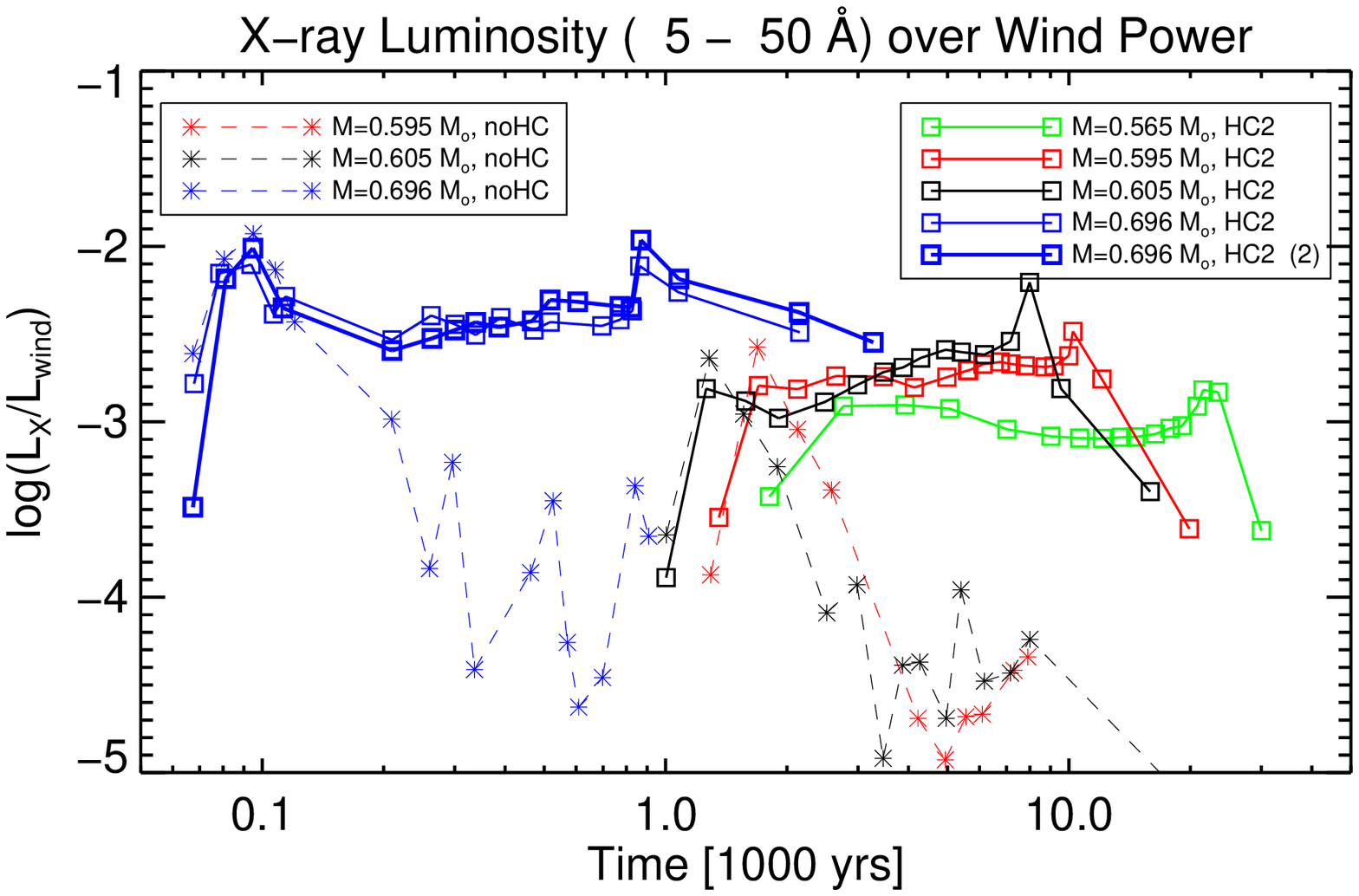}
\caption{Development of the X-ray luminosity
         in the wavelength range  5 to 50~\AA\ (2.5--0.25 keV)
         as a function of time for the hydrodynamical sequences
	 indicated in the inset. Five sequences have heat conduction included
	 (22-HC2, 6a-HC2, 6-HC2, 10-HC2, and 10a-HC2 of Table~\ref{tab.mod}), 
         and three are without heat conduction for comparison (6a, 6, and 10, 
         dashed). The strong fluctuation of X-ray luminosities of these models 
         reflect the large numerical uncertainties related to the thinness of 
         the X-ray emitting region. The symbols indicate the times at which
	 the X-ray emission was computed by means of
         the CHIANTI code (cf.\ Sect.~\ref{X-ray.chianti}).
         The two panels show $L_{\rm X}$ vs. time (\emph{top}) and
         $L_{\rm X}/\Lwind$ vs.\ time (\emph{bottom}), respectively.
	}
\label{X-ray.evol}
\end{figure}

  We have already seen in Sect.~\ref{bubb.struc} that thermal conduction across
  the bubble/PN interface causes the bubble mass to increase rapidly with
  time thanks to `evaporation' from the dense nebular matter.
  For the case shown in Fig. \ref{bubble.mass}, ${a\simeq4.5}$ and ${b\simeq2}$,
  leading to $L_{\rm X}\propto t^3$.
  Any variation of $T_{\rm X}$, the characteristic temperature of the X-ray 
  emitting volume, during the course of evolution is thereby neglected.

  This variation of the emission measure is the consequence of the adopted wind model
  in which wind power and speed increase during the evolution as shown
  in Fig.~\ref{mod.prop}.  Obviously the bubble mass increases faster by heat
  conduction than the density decreases by expansion.
  In contrast, the X-ray luminosity would increase only linearly with time
  if we assume $\dotMb\propto R_{\rm hb}$ (Eq.\,\ref{meva}) and a
  linear expansion law, $R_{\rm hb} \propto t$, instead.

  According to the approximate expression for the emission measure given above, it
  is also possible that $L_{\rm X}$ decreases with time, provided ${2a-3b}$ becomes
  negative, or ${a<1.5\,b}$, at some point during the evolution.  This happens for
  instance in our simulations when the wind power decreases at the end of the
  evolution ($a$ becomes even negative, cf. Figs.~\ref{bubble.mass} and
  \ref{bubble.mass.07}).  Another possibility is a weak wind generating only
  a small evaporation rate.

  The temporal evolution of the X-ray emission for a selection of our sequences
  listed in Table\,\ref{tab.mod} is presented in Fig.\,\ref{X-ray.evol}.
  We consider only models computed with heat conduction according to {method~2}
  (HC2). All sequences
  are plotted from the early transition phase until the white-dwarf domain is reached.
  Note that the evolutionary time scales decrease rapidly with stellar mass; it takes
  only ${\simeq\! 1000}$ yr for the 0.696 \Msun\ central star to exhaust its
  hydrogen-burning shell.

  At first glance it appears to be astonishing that, according to our simulations,
  X-ray luminosities increase systematically with central-star mass, despite the fact
  that the bubble masses decrease (cf. Sect~\ref{bubble.time}).  This is, however, a
  consequence of the dependence of post-AGB evolution on central-star mass: Both the
  stellar wind power and evolutionary speed increase substantially with remnant mass.
  As a result, the hot bubbles around more massive objects remain smaller and more
  dense and have larger emission measures despite of their smaller masses as compared
  to those around less massive and more slowly evolving central stars.

  The trend of the X-ray emission with age is expected from the
  previous discussions.  At early times, at the beginning of the sequences shown 
  in Fig.\,\ref{X-ray.evol}, the X-ray luminosities increase
  sharply with time, reflecting the increase of wind speed and the resulting increase
  of post-shock temperatures to values suitable for soft X-ray emission (top panel).
  Note that both types of models, with and without heat conduction, give virtually
  the same X-ray emission during this phase, because the fraction of `evaporated'
  matter is still insignificant.
  This is typical for the `early-wind' phase of evolution when heat conduction 
  is still unimportant.

  For the energy range considered here (0.25--2.5 keV), there appears to be an 
  optimum velocity, \hbox{$\Vwind\simeq 400$} \kms\
  \hbox{($T_{\rm post-shock}\simeq 2.3\times10^{6}$ K, see Eq.\,\ref{thb2})} 
  for which the X-ray 
  emission of \emph{models without thermal conduction} reaches a maximum.
  This happens at  $t\simeq95$ yr for the 0.696 \Msun\ sequence and
  \hbox{$t\simeq 1800$ yr} for 0.595 \Msun, respectively, when the central stars are
  still very cool \hbox{($\teff < 25\,000$ K)} and the planetary nebula formation is
  just beginning.
  The luminosities achieved (${\simeq\!10^{30}}$--$10^{31}$ \ergs) are in the observed
  range, but the stellar parameters like temperature and wind speed are inconsistent
  with the observed values!
  Beyond this maximum, the X-ray emission decreases with time due to expansion
  and because the bubble becomes too hot in the absence of heat conduction.
  The memory of the `early-wind' phase is lost.

  The situation changes completely if heat conduction is included in the simulations.
  The X-ray luminosity increases with time to levels
  up to 2 dex \emph{above} the values achieved without conduction. The trend
  with age is in line with the discussion above, i.e. an increase with time
  (or buble size) until the central star is rapidly fading (top panel of 
  Fig.\,\ref{X-ray.evol}).
  This occurs after about 9000~years of evolution for 0.595~\Msun,
  after about 6000~years for 0.605~\Msun, and after only 800~years for 0.696~\Msun.

  The final evolution, once the fading of the PN nucleus has stopped, is interesting.
  We see from Fig. \ref{bubble.mass} that the heat conduction turns back into the
  `evaporation' stage (at $t\simeq 15\,000$ yr for the $M=0.595$ \Msun\ example),
  albeit with much reduced efficiency.  The `evaporated' nebular mass
  is, however, not sufficient to compensate for the emission measure decrease due to
  expansion, and thus the X-ray luminosity decreases with time as well.

  Altogether we see that the numerical simulations confirm the qualitative estimates
  made above:  During the horizontal part of the evolution across the HR diagram
  the high evaporation rate triggered by a powerful and accelerating stellar wind
  dominates and leads to an ever increasing X-ray power.  At the end of the evolution
  when the wind dies, the continued expansion reduces the emission measure for the
  X-rays.

  Figure~\ref{X-ray.evol} demonstrates also the dependence of
  the X-ray luminosity on details of the final AGB mass-loss rate.
  First of all, the bubble of sequence No. 10-HC2
  \hbox{($\dot{M}_{\rm agb}= 1\times10^{-4}$ \Mdot)} expands a little bit faster than
  that of sequence No. 10a-HC2 \hbox{($\dot{M}_{\rm agb}= 2\times10^{-4}$ \Mdot)}
  because the latter has a denser envelope.  Consequently, they differ also somewhat
  in their X-ray luminosity.  At the beginning of the evolution, the bubble of
  sequence No. 10a-HC2 with its denser shell 
  develops a slightly lower X-ray luminosity because its emission measure
  $\rho_{\rm X}M_{\rm X}$ is lower:  the smaller $M_{\rm X}$ is not adequately
  compensated for by a larger $\rho_{\rm X}$.  Later on, the radius difference
  between both sequences becomes larger, and the X-ray emission from sequence
  No. 10a-HC2 exceeds that of sequence No. 10-HC2 because of the large radius
  dependence of $\eta_{\rm X}$.

  Our detailed hydrodynamical treatment does not confirm the
  scaling law derived by \citet{ZP.96} according to which
  \hbox{$L_{\rm X}\propto\dot{M}_{\rm agb}^{0.75}$} for constant AGB wind speed
  and given age.  This law predicts an X-ray luminosity for sequence
  No. 10a-HC2 (\hbox{$\dot{M}_{\rm agb}= 2\times10^{-4}$ \Mdot}) which is
  larger than  the X-ray luminosity of sequence No. 10-HC2
  (\hbox{$\dot{M}_{\rm agb}= 1\times10^{-4}$ \Mdot}) by 0.23 dex for the whole
  evolution (cf. also the previous discussion of this matter in 
  Sect.\,\ref{spec}). Our dependence is smaller and more complicated.

  The bottom panel of Fig. \ref{X-ray.evol} illustrates in detail how the X-ray
  luminosity depends on the central-star's wind power.   In general, a fraction of
  between 0.01 and 0.001 of the wind power is radiated away by X-rays between 5 and
  50 \AA\ during the high-luminosity/high wind-power phase of evolution.
  These fractions become, of course, smaller (larger) if the wavelength band
  is reduced (increased).  The variations of $L_{\rm X}/\Lwind$ along 
  a sequence reflect the response
  of thermal conduction to the stellar wind evolution:  After the early
  maximum of X-ray emission the ratio $L_{\rm X}/\Lwind$ decreases first
  because it needs some time for the conduction to `evaporate' sufficient nebular
  matter to compensate for the expansion based reduction of the emission
  measure ($\simeq$ $\rho_{\rm X}M_{\rm X}$).  Then the X-ray luminosity turns out to
  be fairly proportional to the wind power during the following evolution across
  the HR diagram.   The sudden increase of
  $L_{\rm X}/\Lwind$ immediately before the stellar (wind) luminosity
  declines is due to the fact that `condensation' occurs on a longer time scale
  than the stellar (wind) luminosity drop.

  Without heat conduction included, $L_{\rm X}/\Lwind$ becomes very small 
  during the course of evolution, viz. $\simeq\!10^{-4}$--$10^{-5}$. This fact 
  demonstrates that thermal conduction by electrons across the contact surface is 
  an efficient means to convert wind power into X-ray radiation for the whole 
  lifetime of a PN!

  There is an additional factor that influences the X-ray luminosity:
  the spectral band used in the computations (and observations as well). Here it
  is the low energy limit that matters, since energies higher than $E$$\approx$ 
  0.65~keV ($\lambda$$\,<\,$20~\AA) contribute very little to X-ray luminosity. 
  Defining the X-ray band by $0.25$~keV $<$ $E$ $<$ 2.5 keV ($5$~\AA\ $<$ $\lambda$
  $<$ $50$~\AA), more massive and luminous central stars show an apparently larger 
  efficiency of converting wind power into X-ray power, as is clearly seen in 
  Fig.~\ref{X-ray.evol}. Defining instead the X-ray band by 
  $0.05$~keV $<$ $E$ $<$ 2.5 keV ($5$~\AA\ $<$ $\lambda$ $<$ $250$~\AA), the
  ratio $L_\mathrm{X}/\Lwind$ becomes roughly $0.01$ for the whole range of central 
  star masses.

\section{Comparison with observations}
\label{comp.obs}
   In this section we will compare the results of our thermal conduction simulations
   with existing observations from the X-ray satellites XMM-Newton and Chandra.
   Such a comparison is, however, extremely hampered by the fact that especially the
   soft X-ray emission is heavily absorbed by the interstellar medium.  Another
   shortcoming of existing X-ray observations is their rather low spectral resolution.

\begin{figure}[t]                 
\includegraphics[bb= 0.3cm 1.0cm 20.3cm 13.0cm, width=\linewidth]{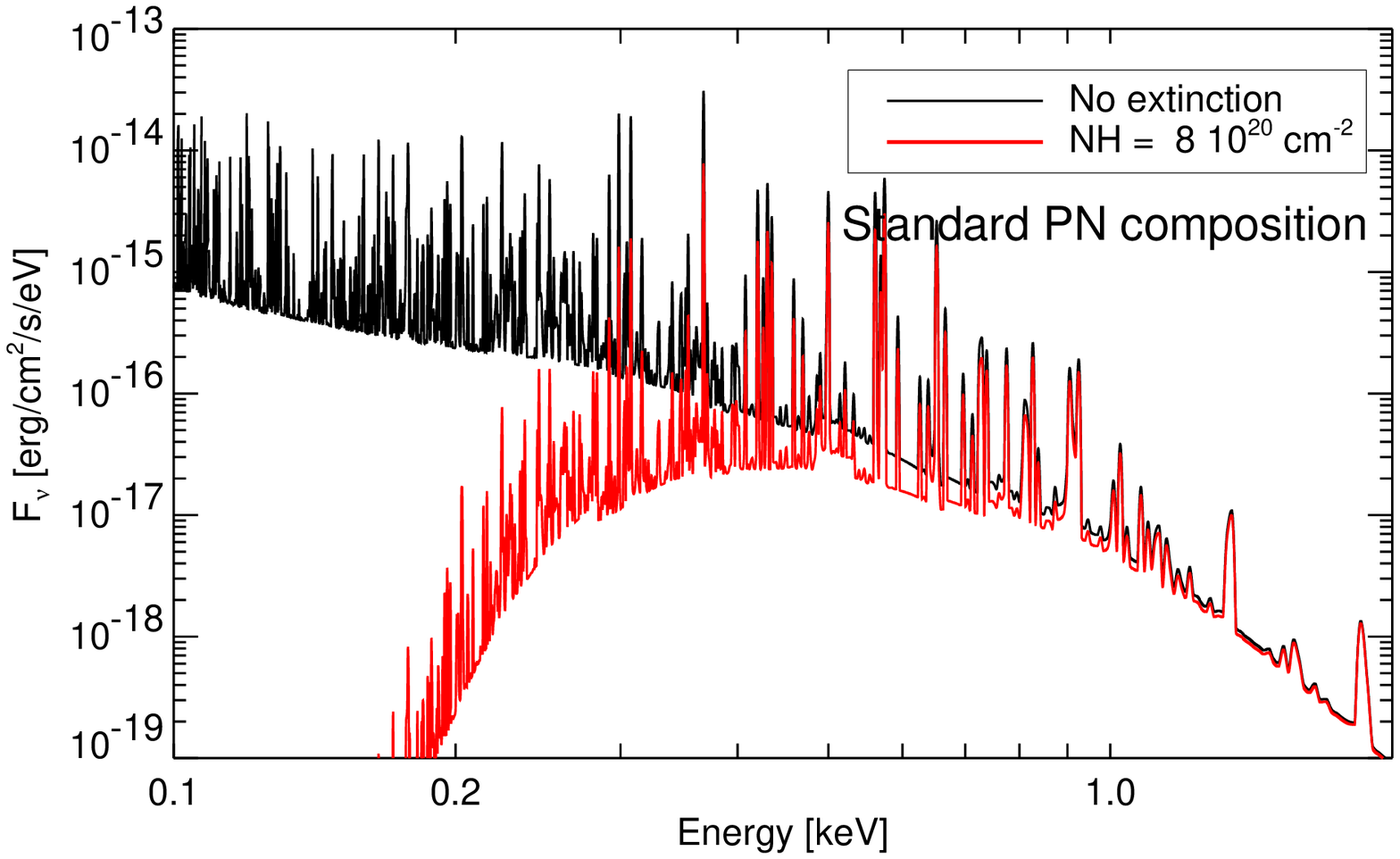}
\includegraphics[bb= 0.3cm 1.0cm 20.3cm 13.0cm, width=\linewidth]{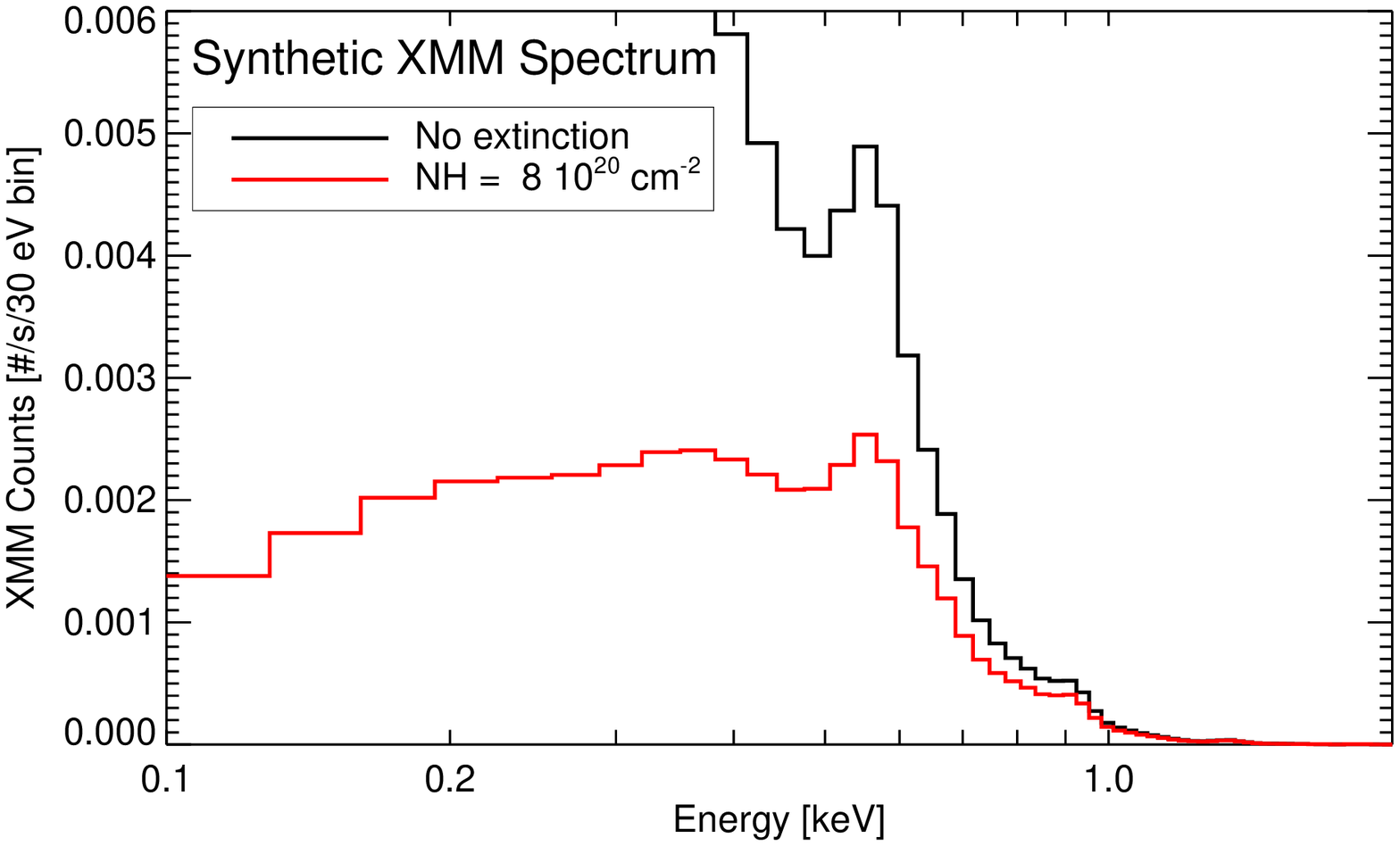}
\caption{\emph{Top}: Spectral flux densities computed from the model shown in the 
         bottom panel of Fig.~\ref{comp.HC} (sequence No.\ 6a-HC2) at a distance 
         of 1~kpc, non-attenuated (black) and attenuated by an intervening hydrogen 
         column density of $N_{\rm H} = 8\times10^{20}~\mbox{cm$^{-2}$}$ (grey).
	 \emph{Bottom}: Simulated count rates per spectral bin of 0.03~keV width for
	 the EPIC camera of the XMM-Newton satellite, with (gray) and without 
         (black) absorption.
}
\label{extinction}
\end{figure}

  As an example, Fig.~\ref{extinction} gives the computed X-ray flux from a
  nebular model as seen from a distance of 1~kpc, with and without interstellar
  absorption.
  The model used here is the one shown in the bottom panel of Fig.~\ref{comp.HC}.
  It is of intermediate age, 5640~yr, and with stellar parameters of
  \hbox{$M =0.595$ \Msun}, \hbox{$L\simeq 5200~{\rm L}_{\sun}$}, and
  $\teff\simeq 71\,500$~K it is a typical representative of most objects with
  known X-ray emission.  The X-ray spectrum is not expected to change much while
  the star evolves across the HR diagram since the characteristic temperature of
  the X-ray emitting region remains rather constant during the high-luminosity
  part of evolution (cf. Fig.~\ref{tx} in Sect.~\ref{xray.lum}).

  The upper panel of Fig.~\ref{extinction} demonstrates the effect of (interstellar)
  extinction on the fully resolved X-ray spectrum, assuming a column density of 
  $8\times 10^{20}$ hydrogen atoms per cm$^2$. The extinction cross section per 
  hydrogen atom as a function of X-ray energy was calculated according to
  \citet[][Table~2]{MMcC.83}.
  In the lower panel we show a simulation of the expected
  count rates as they would be measured by the XMM-Newton satellite with its limited
  spectral resolution, using an appropriate response 
  matrix\footnote{http://xmm.vilspa.esa.es/external/xmm\_sw\_cal/calib/epic\_files.shtml}
  describing the complex response of the energy channels of the EPIC camera to 
  irradiation by X-ray photons of given energy. For comparison with
  published X-ray spectra of PNe, we have re-binned the XMM energy channels
  such that the full energy resolution of 5~eV is reduced to 30~eV. At this 
  resolution, the only prominent feature seen at $\approx\,$0.55~keV  (22~\AA) belongs 
  to the strong \ion{O}{vii} complex. Other line features cannot be seen.

\begin{figure}[t]                 
\includegraphics[bb= 0.3cm 1.0cm 20.3cm 13.0cm, width=\linewidth]{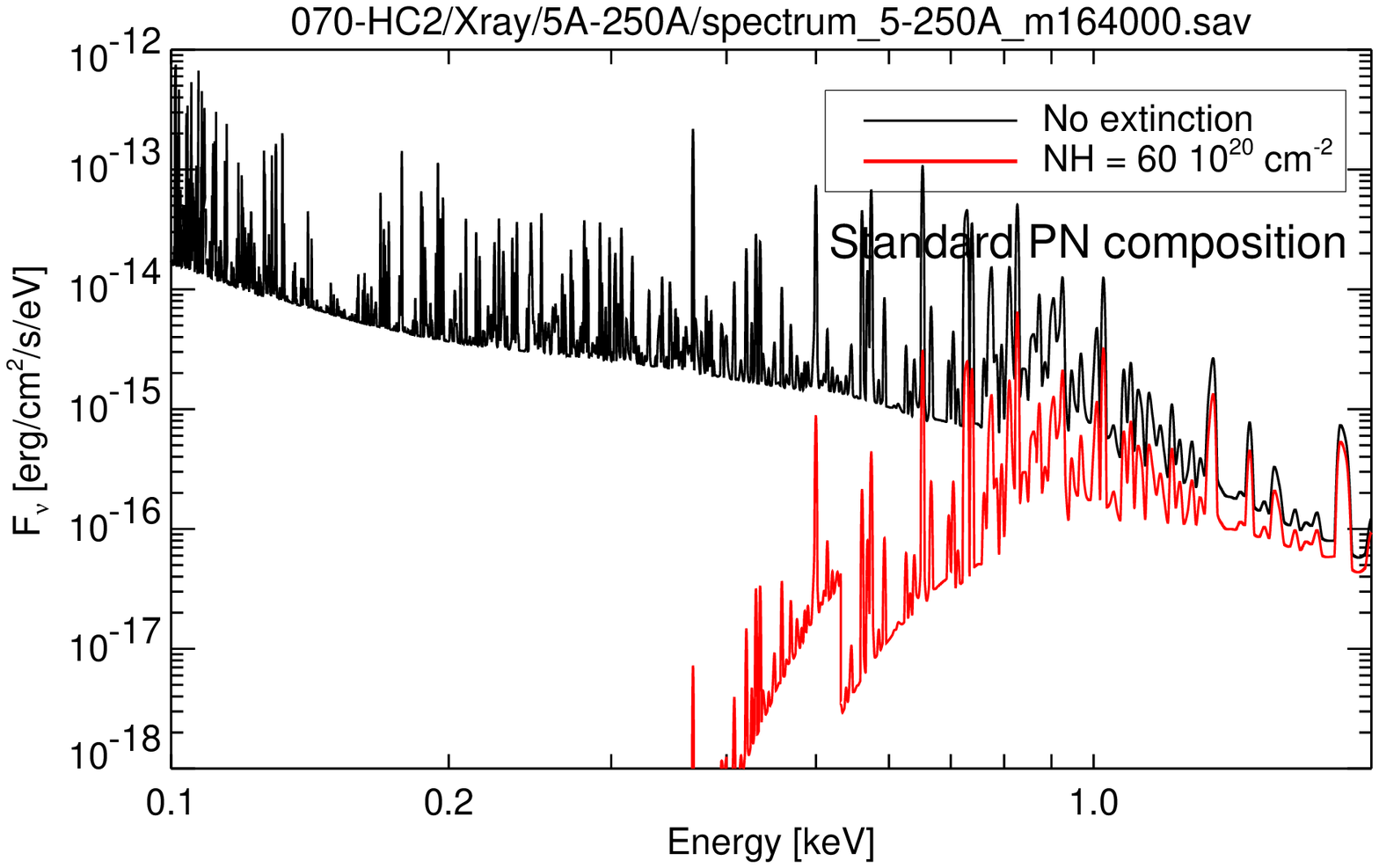}
\includegraphics[bb= 0.3cm 1.0cm 20.3cm 13.0cm, width=\linewidth]{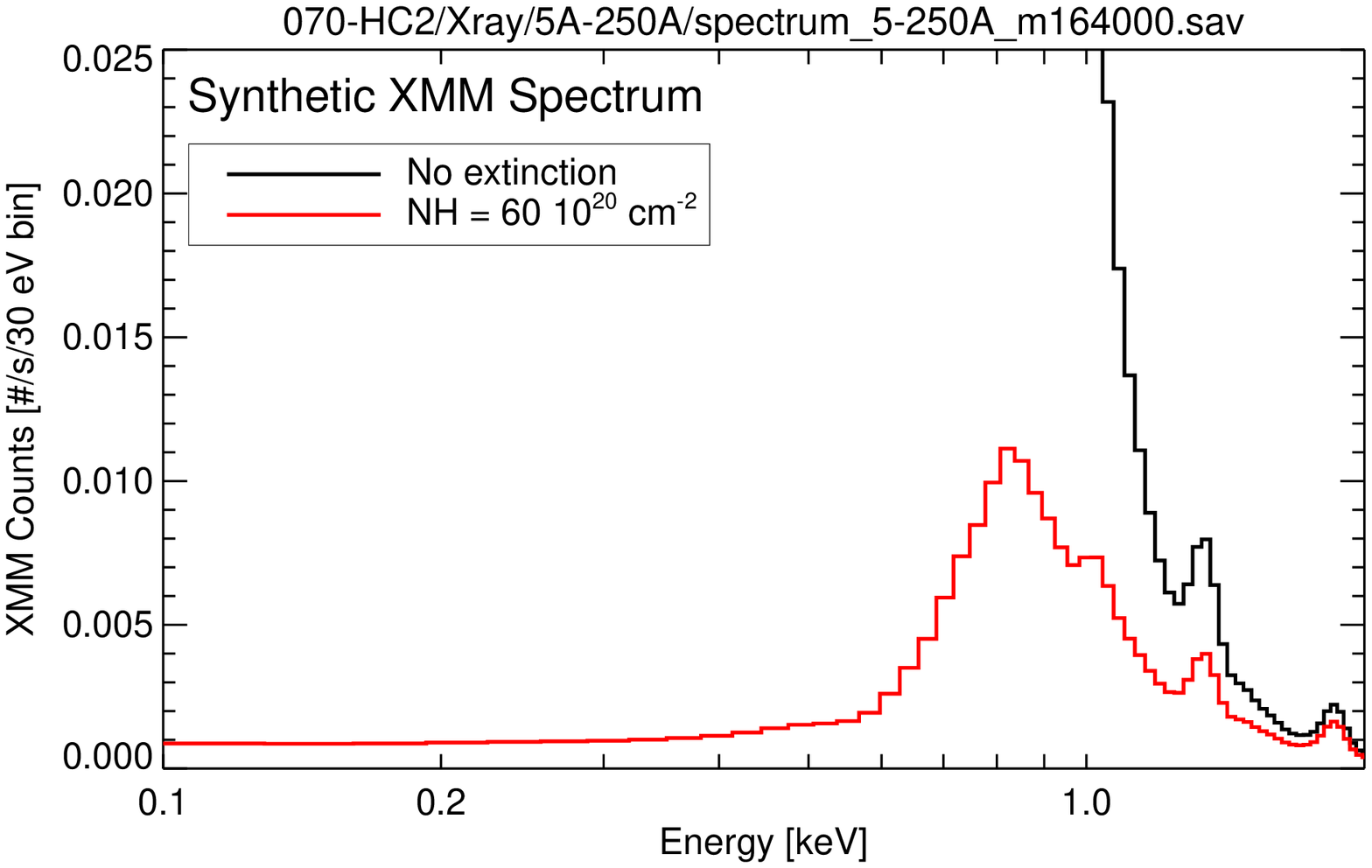}
\caption{\emph{Top}: Spectral flux densities computed from a model selected from
         sequence No.\ 10a-HC2  at a distance of 1~kpc,
	 non-attenuated (black) and attenuated by an intervening hydrogen column
	 density of $N_{\rm H} = 6\times10^{21}~\mbox{cm$^{-2}$}$ (grey).
         The stellar parameters are: $M=0.696$~\Msun, $L=8403~{\rm L}_{\sun},\
	 \teff = 199\,064~{\rm K},\ t=695~{\rm yr}$.
	 \emph{Bottom}: Corresponding simulated count rates per spectral bin of
	 0.03~keV width for the EPIC camera of the XMM-Newton satellite, with (gray) 
         and without (black) absorption.
}
\label{extinction2}
\end{figure}

  A second example is shown in Fig.~\ref{extinction2} where
  the X-ray emission was computed for a model from sequence No.\ 10a-HC2 with
  $M$$\,=\,$$0.696$~\Msun, and $L$$\,\approx\,$$8400$~\Lsun, 
  $T_{\rm eff}$$\,\approx\,$$ 200\,000$~K, representative of, e.g.,
  \object{NGC~7027}.  There is more flux at high energies as compared to the case in
  Fig.~\ref{extinction} because the bubble gas is now hotter
  (cf. Fig.~\ref{comp2.HC}, top).

  According to \citet{Kastetal.01}, the absorption towards
  \object{NGC~7027} is very high, \hbox{$N_{\rm H}
  \approx 6\times10^{21}~\mbox{cm$^{-2}$}$},
  shifting the maximum of the \emph{observed} X-ray emission to higher energies
  (bottom panel of Fig.~\ref{extinction2}).   Our simulated XMM-Newton spectrum
  is very similar to that of \object{NGC~7027} observed by the Chandra X-Ray
  Observatory \citep[cf.][]{Kastetal.01}: the maximum flux occurs 
  between 0.8 and 0.9 keV, probably due to the \ion{Ne}{ix} complex.
  Indeed, by looking at the top panel of Fig. \ref{extinction2} one sees
  a strong line blend around 0.9 keV (\ion{Ne}{ix}) just at the maximum of 
  the reddened flux density distribution. \ion{Ne}{ix} is also strong for
  the model shown in Fig. \ref{extinction}, but far away from the maximum of
  the (reddened) flux density distribution.  
  Because of the larger bubble temperature
  we see also peaks generated by \ion{Mg}{xi} ($\simeq$1.3~keV) and
  \ion{Si}{xiii} ($\simeq$1.9~keV).
  The \ion{O}{vii} complex which is dominant at lower temperatures
  (cf.\ Fig.~\ref{extinction}), is not visible anymore at this spectral 
  resolution \footnote{Note that all the computations presented in this 
  work were performed with the same chemical abundances, viz. those listed 
  in Table~\ref{tab.element}.}.

\begin{figure}[!ht]                 
\includegraphics[bb= 0.3cm 1.0cm 20.3cm 13.0cm, width=\linewidth]{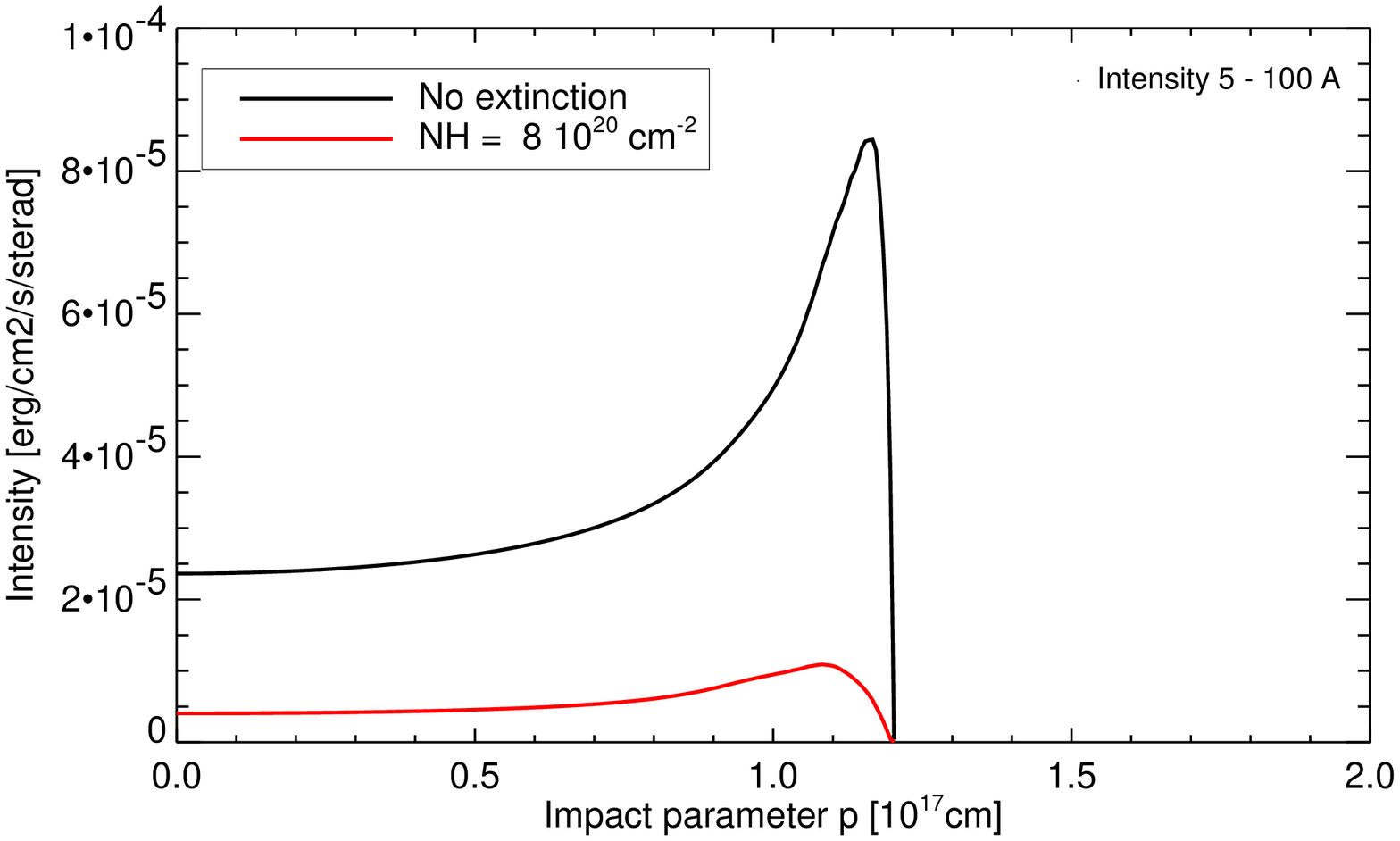}
\includegraphics[bb= 0.3cm 1.0cm 20.3cm 13.0cm, width=\linewidth]{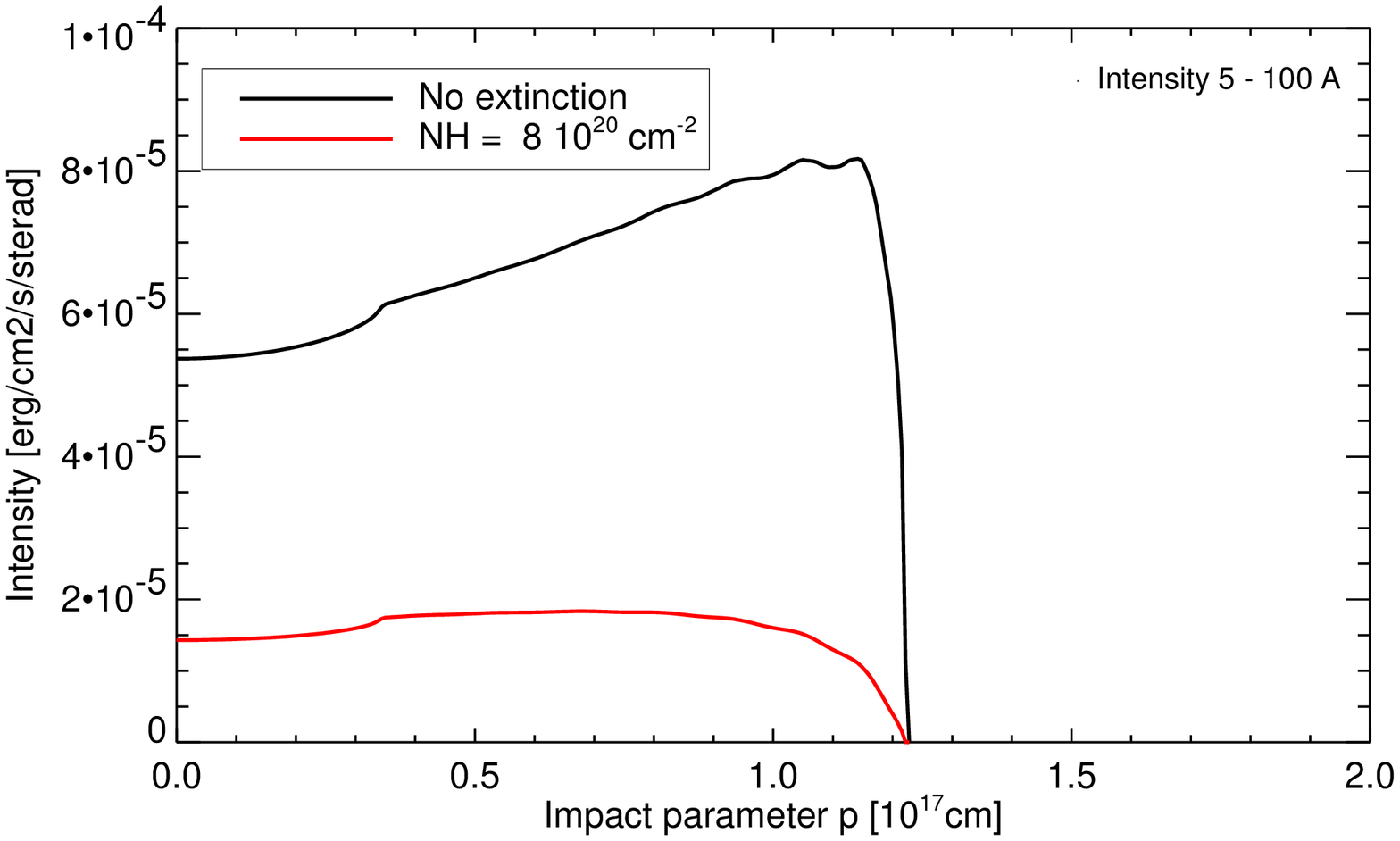}
\caption{\emph{Top}: Intensity profiles from Fig.~\ref{comp.surface},
         middle panel, subject to an absorption corresponding to a hydrogen column
	 density of $N_{\rm H} = 8\times10^{20}~\mbox{cm$^{-2}$}$ (grey).
	 \emph{Bottom}: The same for the bottom panel of Fig.~\ref{comp.surface}.
	 The black lines represent in both panels the unattenuated intensities.  All
	 intensities are integrated from 5--100~\AA, corresponding to 2.5--0.125~keV.
}
\label{595.surf}
\end{figure}
  We see from these examples that the flux detectable from PNe is quite limited to a 
  certain energy range: at high energies, the flux decreases due to the physical 
  structure of the bubble, and at low energies the emission is heavily absorbed by 
  the intervening interstellar matter, and possibly also by dust and neutral gas 
  in and around the PN itself. The range useful for detecting X-rays appears to be 
  between $\approx$\,0.2 and 2~keV (corresponding to the wavelength range 
  $6\dots60$~\AA).
\subsection{Absorbed X-ray surface brightness distributions}
\label{xray.surface}
  We have already discussed the intensity distributions of the X-rays emitted from
  PNe bubbles which are subjected to different treatments of the heat conduction in
  Sect.~\ref{surface}.  Like the spectral energy distribution, also the radial intensity
  distribution is expected to depend on absorption.  The reason is the large variation
  of the absorption with frequency (see Figs.~\ref{extinction} and \ref{extinction2}).
  The emission from the cooler, denser bubble gas is more heavily absorbed than that 
  from the hotter but less dense gas further inwards.  Consequently, the emission is
  weighted towards the hotter, less dense inner regions of the bubble.

  This effect is illustrated in Fig.~\ref{595.surf} where we show examples of radial
  X-ray intensity profiles with and without extinction for our two treatments of
  heat conduction.   In both cases the centre-to-limb variation is considerably
  reduced if interstellar extinction is applied, and the maximum emission is shifted
  inwards.  With {method~2} the X-ray intensity appears to fill the region
  inside the nebular rim nearly homogeneously.

  The radial distribution of the X-ray intensity offers obviously a possibility to
  put constraints on the physics of thermal conduction in very rarefied plasmas.
  Although it is difficult to estimate a definitive amount of center-to-limb
  variations from the existing X-ray maps,
  we note that the recent XMM-Newton observations of \object{NGC~3242} presented by
  \cite{Retal.06} show a smooth distribution of X-ray emission with no apparent limb
  brightening.  Thus we are inclined to conclude that our {method~2}
  (labeled HC2) appears to be more appropriate to describe heat conduction in PNe.
  We will therefore discuss in the following sections only this case.

  Although limb brightening is not really detectable, it is obvious that the 
  X-ray intensity is often not homogeneously distributed across the bubble. 
  In the heat conduction model this would be an indication that the conduction
  efficiency is not the same in all directions, maybe due to the presence of 
  small-scale magnetic fields. It appears more likely, however, that the 
  brightness variations are caused by non-uniform (intra-nebular/interstellar)
  extinction across the nebula 
  \citep[see discussion in][]{Kastetal.02}, as found for many PNe 
  in the visual wavelength range \citep[e.g.][]{Tsamisetal.08,Sandinal.08}.

\begin{table*}[!t]              
\caption{Relevant parameters of objects with detected X-ray emission. 
         The luminosity, $L_{\rm X}$, 
         in the wavelength band 5--28~\AA\ (0.45--2.5~keV), corrected for
         extinction and adjusted according to the distances used here,
         and typical values of temperature, $T_{\rm X}$, and pressure, 
         $P_{\rm X}$, of the X-ray emitting volume as determined by 
         \citet{GCG.05}.
	 The effective temperatures are either spectroscopically derived
	 \citep{Menetal.92} or \ion{He}{ii} Zanstra temperatures 
         (NGC~6543 and NGC~7027: G\'orny, priv. comm.). The distances are 
         either spectroscopic ones (NGC~2392 and NGC~3242: 
         \citealt{Pauletal.04}) or based on expansion parallaxes 
         (NGC~6543: \citealt{Reedetal.99}; NGC~7027: \citealt{Schetal.05a}).
         The distance of NGC~7009 is again from G\'orny (priv. comm.). 
         References for the central-star wind data:
         (1) \citealt{Pauletal.04}; (2) \citealt[][Table 1 therein]{TL.02}; 
         (3) \citealt[][Table 1 therein]{Georgetal.08}.
         All objects listed in this table have central stars with normal, i.e.
         hydrogen-rich, composition.
        }
\label{tab.xray}
\begin{center}
\vskip -2mm
\tabcolsep=1.45mm
\resizebox{\hsize}{!}{
\begin{tabular}{lrcccccccrccc}
\hline\hline\noalign{\smallskip}
  Object   &  $T_{\rm eff}$\ \ ~& Distance  & $L_{\rm star}$ &$\log L_{\rm X}$  & $T_{\rm X}$ 
  &$P_{\rm X}$  &$\log (L_{\rm X}/L_{\rm star})$ & $\log \dot{M}_{\rm wind}$\ \
& $v_{\rm wind}\hspace{3mm}$ &  $L_{\rm wind}$&$\log (L_{\rm X}/L_{\rm wind})$ & Ref. \\[0.6mm]
           &  (K)\ \ \   & (kpc) & (\Lsun)  & (erg\,s$^{-1})$ &(K) & (dyne\,cm$^{-2})$ &
           &(M$_{\sun}\,{\rm yr}^{-1}$)
&(km\,s$^{-1})$ &    (L$_{\sun})$      &  & \\
\noalign{\smallskip}\hline\noalign{\smallskip}
 \object{NGC~2392} & 40\,000 & 1.67 & 5010 &31.20 & $2.0\times10^6$ & $2.6\times10^{-8}$ 
    &$-6.09$ &$-7.74~$ & 420~~ & $0.26$    & $ -1.81$ &(1)\\
 \object{NGC~3242} & 76\,000 & 1.10 & 3120 & 30.90 & $2.2\times10^6$ & $7.6\times10^{-9}$ 
    &$-6.18$ &$-8.40~$ &2400~~ & $1.89$    & $ -2.96$ &(1)\\
 \object{NGC~6543} & 67\,000 & 1.00 & 1590 & 31.00 & $1.7\times10^6$ & $1.8\times10^{-8}$ 
    &$-5.78$ &$-7.73~$ &1340~~ & $2.75$    & $ -3.01$ &(3)\\
 \object{NGC~7009} & 81\,000 & 1.50 & 3600 & 31.34 & $1.8\times10^6$ & $1.4\times10^{-8}$ 
    &$-5.80$ &$-8.55~$ &2770~~ & $1.78$    & $ -2.49$ &(2)\\
 \object{NGC~7027} &200\,000 & 0.80 & 6250 & 31.12 & $7.9\times10^6$ & $1.3\times10^{-7}$
   &$-6.26$ &---   &---\ \ \ &---     &  ---     & \\[0.5mm]
\hline
\end{tabular}
}
\end{center}
\end{table*}
%

\subsection{X-ray luminosities}
\label{xray.lum}
  Now we compare global X-ray properties, in particular the luminosities,
  predicted by our models with the recent observations made by the Chandra and
  XMM-Newton satellites.  We discarded objects with a WC central star because
  wind and hot bubble  have a hydrogen-poor composition which is completely
  different from the (hydrogen-rich) composition assumed in our hydrodynamical
  simulations.   The X-ray properties of the remaining five objects are listed in
  Table~\ref{tab.xray} and are taken from the compilation of \cite{GCG.05}
  in order to have a homogeneous data set to compare with.
  Note that now the X-ray luminosities refer to the energy interval of 
  0.45--2.5~keV (5--28 \AA) only in order to avoid large and uncertain 
  corrections due to interstellar absorption at lower energies.  
  The low energy cut-off used here leads to X-ray luminosities that are, 
  in some cases, considerably \emph{lower} than those quoted in the 
  discovery papers. Additional changes are due to the adjusted distances 
  used in our work.
  The X-ray data from the PNe are supplemented by the corresponding data of 
  the stellar winds if available. 

  The empirical data of Table~\ref{tab.xray} allow already some interesting
  conclusions: Compared with the stellar luminosity, the X-ray contribution
  from the shocked stellar wind (i.e. from the bubble), for the energy band
  considered here, is very small and is only about $10^{-6}\,L_{\rm star}$.
  Also the fraction of the stellar wind power that is converted into X-ray
  emission is quite small: only 1\% to 0.01\%.  Note that these values depend 
  on the definition of the energy band, notably on the
  boundary of the low-energy region where most of the X-rays are emitted (see
  Fig.~\ref{comp.X-ray}).

\begin{figure}[t]                 

\includegraphics*[bb= 2.0cm 1.1cm 20.2cm 12.05cm, width=\columnwidth]{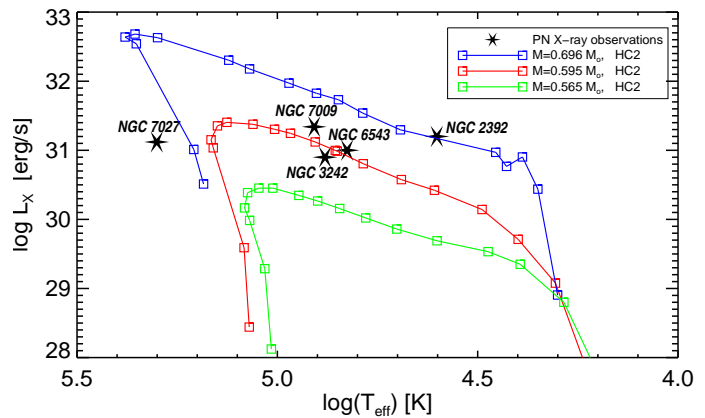}
\caption{X-ray luminosity of the bubble for the energy interval 0.45--2.5~keV
         (5--28~\AA) as a function of the stellar effective temperature as predicted
	 by three model sequences with central star masses 0.565 \Msun, 0.595 \Msun, and
	 0.696 \Msun\ (HC2 models only, see inset).  The star symbols indicate the
	 positions of the objects from Table \ref{tab.xray}.
         The evolution times covered by the model sequences are the same as in
	 Fig.~\ref{X-ray.evol}.
	}
\label{x-ray.HR}
\end{figure}

  We present in the following figures a detailed comparison of the objects listed
  in Table \ref{tab.xray} with the predictions of our hydrodynamical models with
  heat conduction treated according to {method~2}.
  We begin with the ``X-ray Hertzsprung-Russell diagram'' of Fig.\,\ref{x-ray.HR},
  where we present the general evolution of the bubble's X-ray luminosity
  as a function of the stellar effective temperature for sequences with different
  central stars.  The increase of the X-ray luminosities with time seen
  in Fig.~\ref{X-ray.evol} translates into a corresponding increase with effective
  temperature.

  Given the distance uncertainty to individual objects,
  the agreement of our models with the existing observations is very good.
  The X-ray luminosity of all five objects from Table \ref{tab.xray} can be explained by
  sequences with central stars between 0.6 and 0.7 \Msun!  We repeat that the run
  of the X-ray emission with time (or stellar effective temperature) depends on
  the stellar wind properties and the expansion rate of the bubble, which also
  depends indirectly on the wind.  The evolution of the X-ray luminosity as seen in
  Fig.~\ref{x-ray.HR} is the result of the wind model shown in Fig.~\ref{mod.prop}
  (0.595 \Msun).
  The sample of PNe with confirmed soft X-ray emission is, however, too small to
  prove or disprove the wind model used in our simulations.

  Our models without thermal conduction
  fail completely in explaining the observations because their X-ray luminosities
  are too low by about two orders-of-magnitude at the positions of the
  observed objects in Fig.\,\ref{x-ray.HR} (cf. also Fig. \ref{X-ray.evol}).
%
\begin{figure}[t]                 

\includegraphics*[bb= 1.6cm 1.1cm 20.2cm 12.05cm, width=\columnwidth]{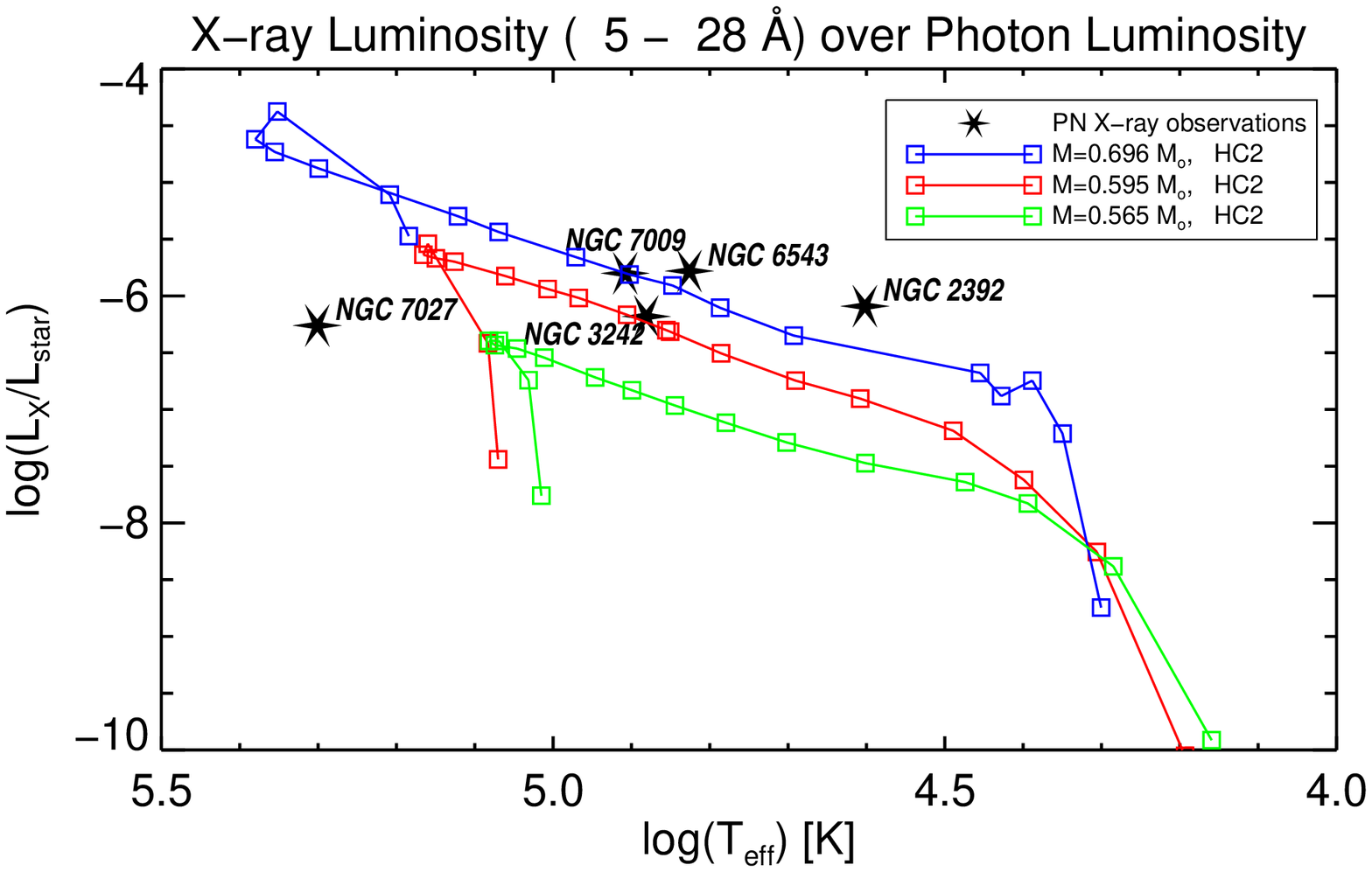}
\vskip 3mm
\includegraphics*[bb= 1.6cm 1.1cm 20.2cm 12.05cm, width=\columnwidth]{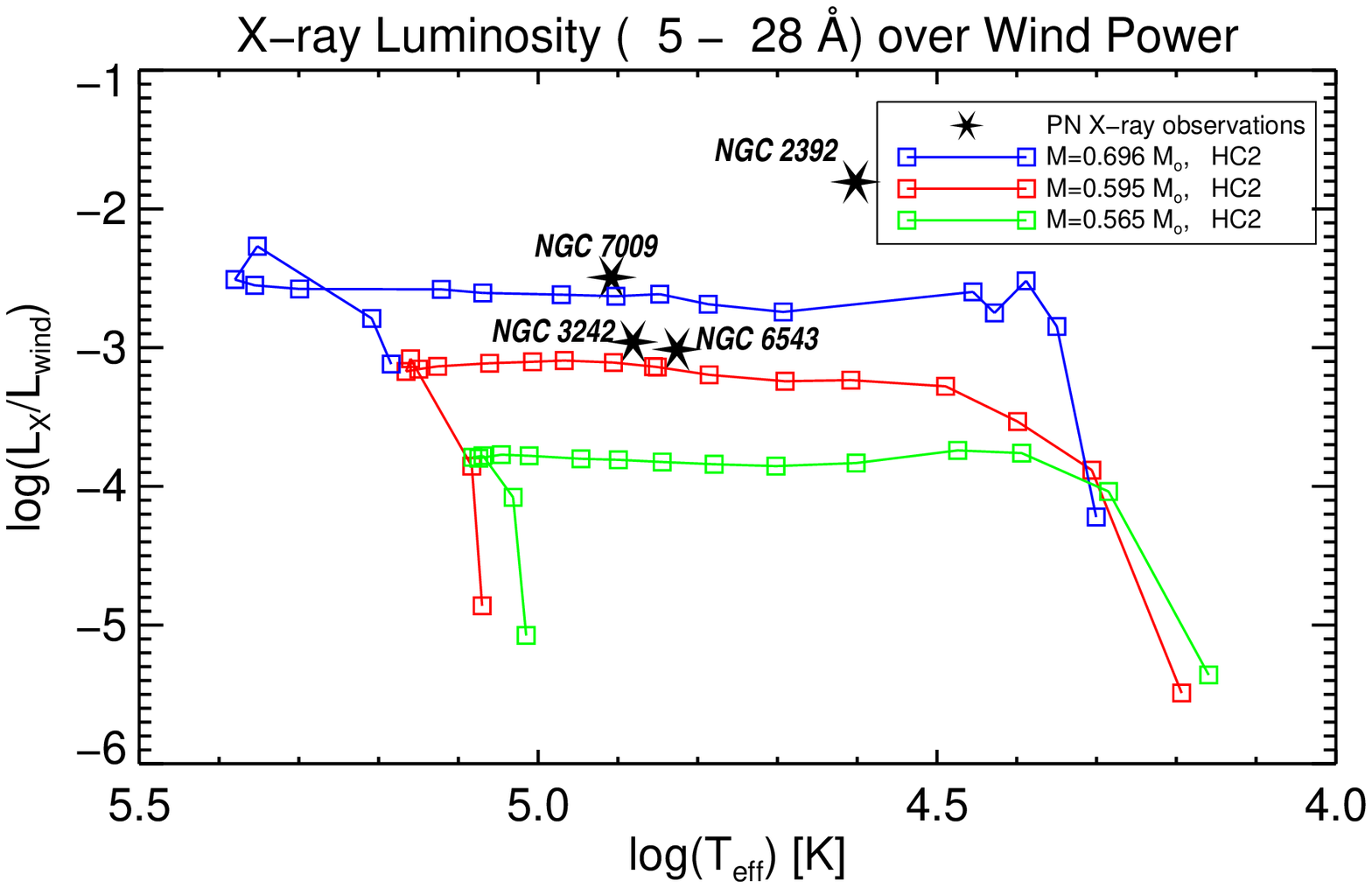}
\caption{\emph{Top}: X-ray luminosity of the bubble (0.45--2.5~keV) over stellar
                     luminosity  vs.  effective temperature.  The model
		     sequences are the same as in Fig.~\ref{x-ray.HR} (see inset).
	 \emph{Bottom}: X-ray luminosity of the bubble (0.45--2.5~keV) over stellar
	               wind power vs. stellar effective temperature.
	The observations from Table~\ref{tab.xray} are again plotted as stars.  Note
	that \object{NGC~7027} cannot be shown in the bottom panel because its stellar
	wind properties are unknown.
	}
\label{x-ray.obs}
\end{figure}

  In order to avoid possible systematic errors caused by 
  distance uncertainties we relate in Fig.\,\ref{x-ray.obs} the X-ray luminosities to 
  both the stellar luminosity, $L_{\rm X}/L_{\rm star}$, and wind luminosity,
  $L_{\rm X}/\Lwind$.  The basic result found from Fig. \ref{x-ray.HR} is
  confirmed in Fig.\,\ref{x-ray.obs} (top panel): our models with thermal conduction
  according to {method~2} (HC2 models) predict soft X-ray emission in
  good agreement with the observations (except for \object{NGC~7027}, see
  discussion below).
  There appears to be a slight preference for our 0.7 \Msun\ models, but 
  considering the uncertainties of $L_{\rm X}$ we don't think that this 
  is a real effect.

  The lower panel of Fig.\,\ref{x-ray.obs} corresponds to that shown in the bottom
  panel of Fig.~\ref{X-ray.evol},
  although the difference between the tracks appears somewhat
  larger.  The reason is the smaller energy range considered:  the temperature
  sensitive emission at the high energy end has now more weight. 
  Again we have a satisfactory agreement between theory and observations. Less
  than 1\,\% of the wind power is converted into X-ray emission. Note, 
  however, that the mass-loss rates which enter into the wind power are 
  notoriously uncertain and may vary by up to a factor of ten between 
  different authors!
 
  As mentioned above, \object{NGC~2392} has an unusually slow wind for its 
  position in the Hertzsprung-Russell diagram compared with the other objects
  listed in Table~\ref{tab.xray}, giving it the lowest wind luminosity and thus
  the highest $L_{\rm X}/\Lwind$ ratio of all objects from the sample.
  Its X-ray luminosity agrees with our models (cf.\ Fig. \ref{x-ray.obs}, 
  top), but its wind power is too small, and hence $L_{\rm X}/\Lwind$ 
  too large (cf.\ Fig. \ref{x-ray.obs}, bottom).  
  It appears possible that fast outflows/jets
  contribute to the X-ray emission of \object{NGC~2392} \citep[see discussion
  in][]{GCGM.05}. 

\begin{figure}[t]                 

\includegraphics*[bb= 1.5cm 1.1cm 20.6cm 12.05cm, width=\columnwidth]{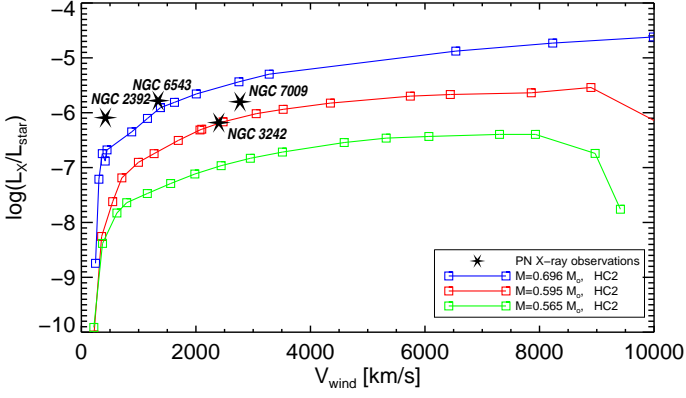}

\caption{$L_{\rm X}/L_{\rm star}$ vs. wind speed, again for our three model sequences
         from Fig.~\ref{x-ray.HR} and the objects from Table~\ref{tab.xray}
         (except for \object{NGC~7027} for which the
	 central star wind is not known).  The X-ray range is again 0.45--2.5 keV.
	}
\label{xray.wind}
\end{figure}

  A further interesting comparison is shown in Fig.~\ref{xray.wind} where
  $L_{\rm X}/L_{\rm star}$ is plotted over $\Vwind$.  This figure demonstrates
  clearly that our models, thanks to the inclusion of heat conduction, give a fully
  consistent description of the \emph{observed} X-ray luminosities also in terms of
  the \emph{observed} large  wind speeds!
  Although the wind of \object{NGC~7027} is not known, it must exist because
  we observe the X-ray emission from the shocked wind gas.  Judging from the
  central star's position close to the white dwarf domain, we estimate a wind speed
  close to the right boundary of Fig.~\ref{xray.wind}, viz. of about
  7000\ldots8000~\kms.

\subsection{X-ray temperatures}
\label{xray.temp}
  A stringent test of our models is the determination of a
  characteristic X-ray emission temperature, $T_\mathrm{X}$, which can be compared with
  the measurements.  We computed $T_\mathrm{X}$ from our models by weighting the electron
  temperature $T_\mathrm{e}(r)$ within the bubble ($r_1$--$r_2$) with the volume 
  emissivity integrated  over the respective energy range, $E_1$--$E_2$ 
 (0.45--2.5 keV or 5--28 \AA):

\begin{equation}
T_\mathrm{X}=\frac{4\,\pi}{L_\mathrm{X}}\,\int_{r_1}^{r_2} r^2\, 
             T_\mathrm{e}(r)\,\eta_\mathrm{X}(r) \, \mathrm{d} r \, ,
\label{e:TX}
\end{equation}
where $L_\mathrm{X}$ is the X-ray luminosity,
\begin{equation}
L_\mathrm{X}=4\,\pi\,\int_{r_1}^{r_2} r^2\, \eta_\mathrm{X}(r) \, \mathrm{d} r \, ,
\label{e:LX}
\end{equation}
and
\begin{equation}
\eta_\mathrm{X}(r)=\int_{E_1}^{E_2} 
\eta (T_\mathrm{e}(r), n_\mathrm{e}(r), E) \, \mathrm{d} E \, ,
\label{e:etaX}
\end{equation}
is the volume emissivity in the energy band ${E_1}$--${E_2}$.

\begin{figure}[t]                 
\includegraphics*[bb= 1.5cm 1.1cm 20.2cm 12.04cm, width=\columnwidth]{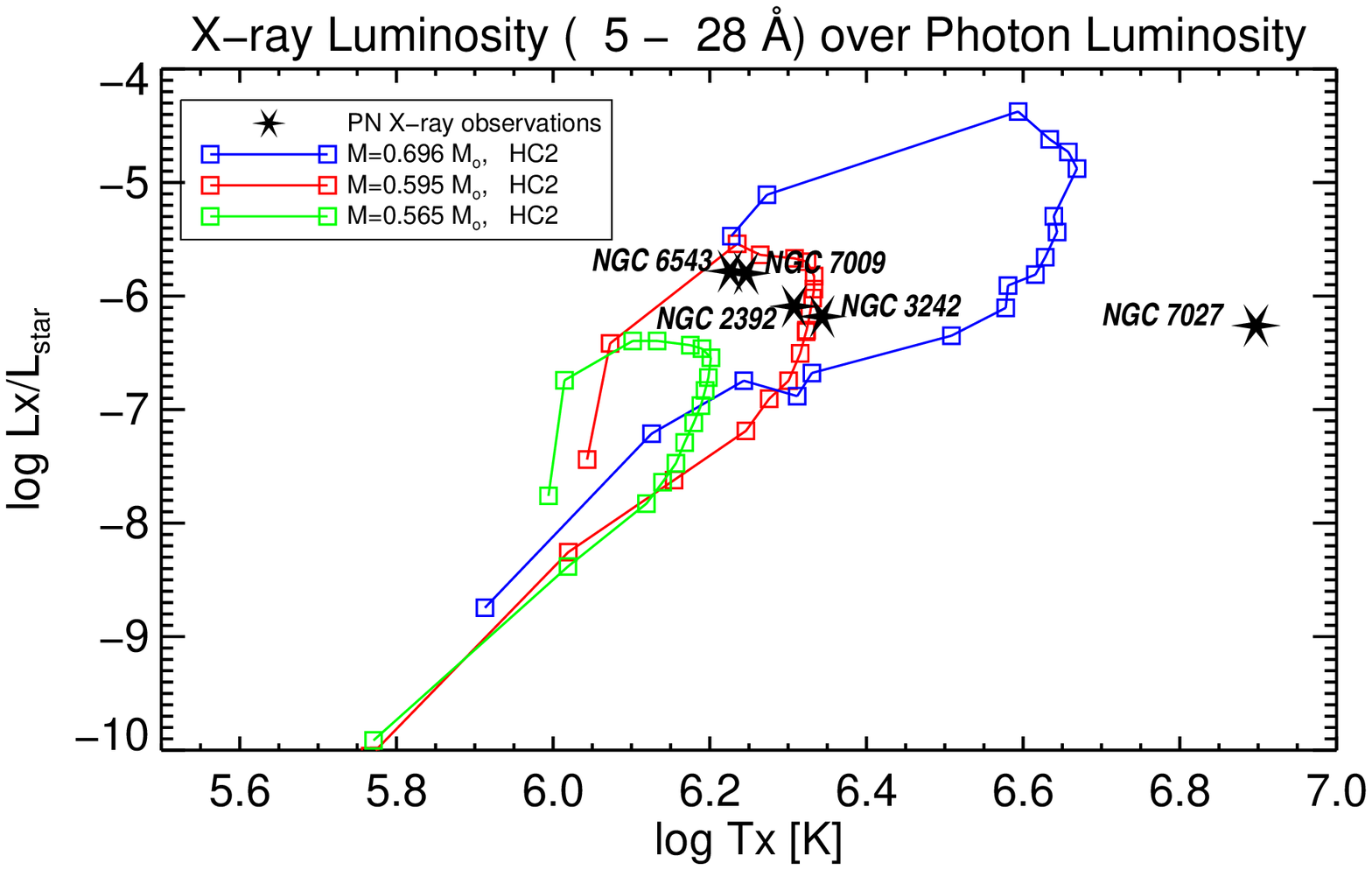}
\vskip 3mm
\includegraphics*[bb= 1.5cm 1.1cm 20.5cm 12.04cm, width=\columnwidth]{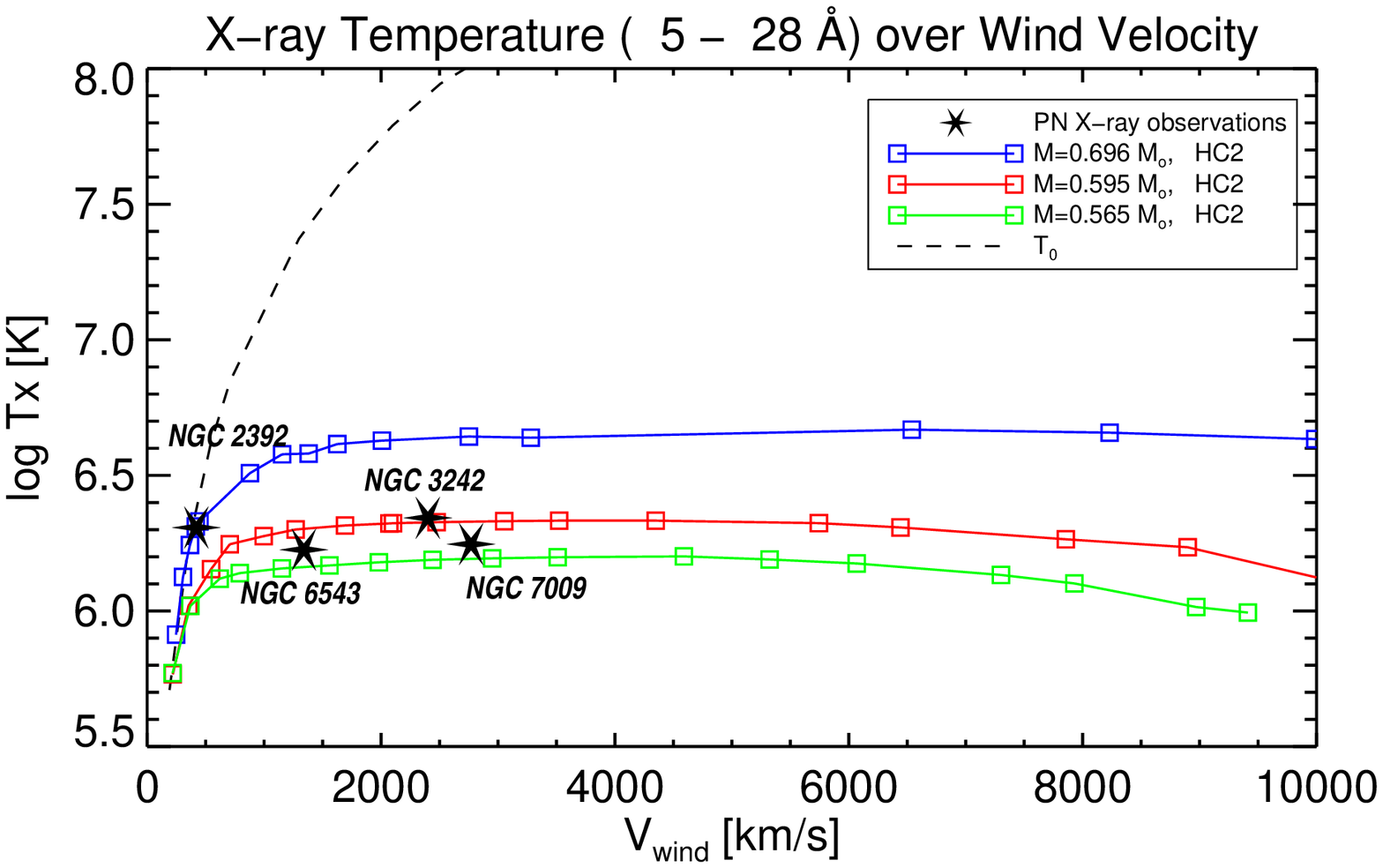}

\caption{\emph{Top:} $L_{\rm X}/L_{\rm star}$ for the 0.45--2.5 keV range as a 
         function of the characteristic temperature of the X-ray emitting 
         region, $T_\mathrm{X}$, again for the three sequences from 
         Fig.~\ref{x-ray.HR}. The individual `tracks' cover the same age 
         spans as shown in Fig.~\ref{X-ray.evol}.
	 Note that the observed values of $T_\mathrm{X}$ 
         (from Table~\ref{tab.xray}) have been derived from the best fit to 
         the spectral energy distributions, while the $T_\mathrm{X}$ for a 
         given hydrodynamical model have been computed according
         to Eq. (\ref{e:TX}). 
         \emph{Bottom:} $T_\mathrm{X}$ as a function of the wind speed,
         $\Vwind$, computed for the three model sequences from  
         Fig.~\ref{x-ray.HR}, and $T_0$ according to Eq.\,(\ref{thb2}) 
         (dashed).
	}
\label{tx}
\end{figure}

  The result is seen in Fig.~\ref{tx} (top panel), covering the
  {whole} evolution from the onset of the bubble formation until the white dwarf
  stage is reached (cf. Fig.~\ref{X-ray.evol}).  In general, $T_\mathrm{X}$ increases
  rapidly to above $10^6$~K while a hot bubble is formed beyond the wind shock.
  Then heat conduction becomes effective, and $T_\mathrm{X}$ increases only slowly
  or remains nearly constant until maximum wind power is reached.
  This refers mainly to the `evaporation' phase during which the the conduction front
  advances outwards (see Sect.~\ref{bubb.struc}).   Afterwards, $T_\mathrm{X}$
  drops in line with the wind power to about $10^6$~K (`condensation' phase).

  The maximum X-ray temperature achieved during the evolution across the HR diagram
  depends on the central star mass:  Thanks to its powerful wind, the sequence with
  the 0.696~\Msun\ model reaches a much larger $T_\mathrm{X}$ than the sequence with 
  the 0.565~\Msun\ star, viz. $4.5\times10^6$~K instead of only $1.6\times10^6$~K.

  Assuming that our computed characteristic X-ray temperatures are reasonable
  representatives of the ones derived from observations, we find excellent
  agreement between theory and observations, except for \object{NGC~7027} whose \Tx\
  ($8\times10^6$~K) is much larger than our models predict (see also below).
  The remaining objects are rather well matched by the 0.595~\Msun\ models with
  heat conduction (sequence No. 6a-HC2).

  The bottom panel of Fig. \ref{tx} illustrates how the temperature of the 
  X-ray emitting region is modified by thermal conduction from a value close 
  to the one predicted by Eq.~(\ref{thb2}). At very low wind speeds, 
  typical for the `early-wind' phase, thermal conduction is unimportant, 
  and \Tx\ equals the post-shock temperature (dashed line in the bottom panel 
  of Fig. \ref{tx}). At faster stellar winds (and larger post-shock 
  temperatures), \Tx\ levels off as a consequence of heat conduction and 
  becomes fairly independent of the wind speed.

\subsection{Correlations with bubble radius}
\label{correlations}
For a given central star mass, our models predict an increase of the
X-ray luminosity with time (bubble radius) during the main phase of
evolution (see Sect.\,\ref{time.evol}, Fig.\,\ref{X-ray.evol}, top).
On the other hand, the same models predict a decrease of $L_\mathrm{X}$
with bubble radius at given central star effective temperature.
This is because, for given $\teff$,  $L_\mathrm{X}$ increases with stellar 
mass (see Sect.\,\ref{xray.lum}, Fig.\,\ref{x-ray.HR}), while the bubble
size decreases due to the shorter evolution time scales of the more massive
stars.
Likewise, in the presence of heat conduction the mean temperature of 
the hot bubble ($\sim$ \Tx) depends on both the wind power and the bubble 
radius according to Eq.\,(\ref{thb4}), but the dependence is weak.
Hence, the expected correlations of $L_\mathrm{X}$ and \Tx\ with
bubble radius $R_{\rm hb}$ depend critically on the mass distribution 
and ages of the observed objects.

In this context we have to discuss the recent findings by \citet{Kastetal.08}
that X-ray temperature and luminosity appear to \emph{decrease} with bubble 
radius. Apart from the fact that these results suffer from uncertainties of 
the distances of the individual objects, which the authors do not
consider at all, we point out that: 
\begin{enumerate}
\item  Figure 4 of \citet{Kastetal.08} contains a mix of objects: 4 PNe have 
       hydrogen-deficient and 3 PNe have hydrogen-rich central stars. The 
       objects from both groups have certainly \emph{per se} distinct physical
       properties and also different evolutionary histories.
\item  The 3 objects with a hydrogen-rich 
       central star (\object{NGC~2392}, \object{NGC~6543}, \object{NGC~7009}) 
       have about the same bubble radius of $\sim\!0.1$ pc. 
\end{enumerate}
The claimed anti-correlations with bubble radius are thus probably purely 
artificial, based entirely on an inappropriate combination of objects with 
different evolutionary background. A much larger, homogeneous sample of 
objects is certainly necessary to construct trustworthy correlations between 
observable quantities that can be compared with theoretical predictions.

\subsection{Individual objects}
\label{indi.obj}
\paragraph{NGC~3242.}
\label{ngc3242}
  This PN is well suited for a closer comparison with our models because its
  bubble is most likely spherical, as judged from the ring-like appearance of the rim.
  The position of \object{NGC~3242} is very close to the 0.595 \Msun\
  track in all the previous Figures (\ref{x-ray.HR}, \ref{x-ray.obs}, \ref{xray.wind}
  and \ref{tx}).  This implies that the evolution of wind power and X-ray emission as
  predicted by our model simulation with heat conduction according to
  {method~2} reflects the real situation in \object{NGC~3242} surprisingly well.
  A detailed comparison between the observed parameters of \object{NGC~3242} and
  two models taken from sequence No. 6a-HC2 which embrace the observed position
  of \object{NGC~3242} in the figures is presented in Table~\ref{param.3242}.
  The density and temperature structure of model 1 is shown in
  Fig.~\ref{comp.HC} (bottom panel).

\begin{table}[t]
\caption{Relevant parameters of two nebular models along the 0.595 \Msun\ track
        (sequence No. 6a-HC2) embracing the position of \object{NGC~3242} in the HR 
         diagram, compared with the observed properties of \object{NGC~3242}. For 
         both observation and models, the X-ray data refer to the 0.45--2.5~keV 
         energy band.
        }
\label{param.3242}
\vskip -1mm
\resizebox{\linewidth}{!}{
\begin{tabular}{lcc|cc}
\hline\hline\noalign{\smallskip}
	& Model 1&Model 2& \object{NGC~3242}     & Ref.\\[0.5mm]
\hline\noalign{\smallskip}
$M/\msun$                 &  0.595 &  0.595   &   0.53, 0.63 & (1), (2)   \\[0.6mm]
$L_\mathrm{star}/\lsun$   & 5\,205 &  5\,051  &  3\,162      & (1)    \\[0.6mm]
$\teff$ (K) & 71\,667& 80\,457  &  75\,000  &  (1)  \\[0.6mm]
$t_{\rm post-agb}$ (yr) &  5\,642  & 6\,121    &  $\simeq\!2\,800$ &   (3)     \\[0.6mm]
$\dotMw$ (\Msun\,yr$^{-1}$)&$9.7\times10^{-9}$ & $8.6\times10^{-9}$&
         $4\times10^{-9}$  &  (1) \\[0.5mm]
$\Vwind$ (\kms) &2\,115 & 2\,490   & 2\,400   &  (1)  \\[0.6mm]
$L_{\rm X}/L_{\rm star}$ & $5.0\times10^{-7}$ & $6.8\times10^{-7}$ & $6.6\times10^{-7}$
                                                        &  Table\,\ref{tab.xray} \\[0.6mm]
$L_{\rm X}/\Lwind$ & $7.2\times10^{-4}$ & $7.7\times10^{-4}$ & $1.1\times10^{-3}$
                                                        &  Table\,\ref{tab.xray} \\[0.6mm]
$T_\mathrm{X}$ (K)& $2.1\times10^6$ & $2.1\times10^6$ & $2.2\times10^6$  &
                                                             (4) \\[0.5mm]
$P_\mathrm{X}$ (dyn\,cm$^{-2}$)&$9.0\times10^{-9}$ & $8.0\times10^{-9}$  & $7.6\times10^{-9}$  &
    (4) \\[0.6mm]
$P_\mathrm{rim}$ (dyn\,cm$^{-2}$)&$\simeq\!8\times10^{-9}$ & $\simeq\!7\times10^{-9}$ & $8.6\times10^{-9}$ &  (4) \\[0.6mm]
$n_{\rm rim}$ (cm$^{-3}$)  &$\simeq$2\,900~~    &$\simeq$2\,400~~  &  2\,600  &  (4) \\[0.6mm]
$n_{\rm shell}$ (cm$^{-3}$)&$\simeq$1\,200~~    &$\simeq$800~~     &  800     &  (4) \\[0.6mm]
$v_\mathrm{rim}$ (\kms)  & 14.0  &  14.5      & 19.5     &  (5) \\[0.6mm]
$v_\mathrm{shell}$ (\kms)& 27.0  &  28.0      & 35.7     &  (5) \\[0.6mm]
\hline\noalign{\smallskip}
\end{tabular}   }
\\[1mm]
(1) \citealt{Pauletal.04};\\
(2) \citealt{KUP.06};\\
(3) Kinematic age from \citealt{Coretal.03},\\
    \mbox{~~~~~} scaled down to the distance of 1.1~kpc used here;\\
(4) \citealt{Retal.06};\\ 
(5) \citealt{Schetal.05}.\\ 
\end{table}

  Given the uncertainty of the observed data and
  the fact that we did not attempt to make any fits to the observations,
  the agreement is very good, especially for the X-ray related data.
  Notice the nearly equal pressures and temperatures of the X-ray emitting
  gas.\footnote{The hot bubble is isobaric despite of the
  radial temperature and density gradients (cf. Fig.\,\ref{comp.HC}).}
  Also mass-loss rates and wind speeds are, within the known uncertainties,
  in reasonable agreement.

  Quantities that depend directly on distance or distance squared differ by larger
  amounts.  Our values for \object{NGC~3242} quoted in Table~\ref{param.3242} are based
  on a distance of 1.1~kpc, according to \citet{Pauletal.04}.  \citet{KUP.06} arrived
  at a larger distance, $d=1.8$~kpc, hence luminosity and kinematic age are larger:
  $L=7760$~\Lsun\ with an age of about 4\,600 years\footnote{%
Kinematic ages based on physical size and expansion velocity are problematic since
they depend strongly on the method used \citep[see][]{Schetal.05a}.  The relatively
large shell expansion speed indicates a significant acceleration during the previous
evolution which lead to an underestimation of the age.%
}.
  These authors give also a larger
  mass-loss rate, $8\times10^{-9}$~\Mdot, from the central star.

  Taken at face values, the numbers listed in Table\,\ref{param.3242} indicate that
  the thermal pressure of the rim, $P_\mathrm{rim}$, exceeds
  that of the bubble by a small margin. In the model we have just the opposite
  situation. The case of \object{NGC~3242} could thus indicate that the wind power
  of the central star has achieved its maximum value already below the present
  effective temperature of 75\,000~K. This would be in line with the more recent
  wind computations conducted by \citet{Pauletal.04} according to which
  central-star mass-loss rates and wind power reach a maximum around stellar 
  temperatures of 50\,000~K, after which the thermal pressure of the bubble
  drops below that of the rim \citep[for more details see][Fig.~6]{StSch.06}. 
  In our present simulations which are based on the older
  \citet{Pauletal.88} recommendations the wind power peaks later, close to maximum
  effective temperature (cf. Fig.\,\ref{mod.prop}, bottom left).

\begin{figure}[t]                 
\includegraphics*[width=0.49\columnwidth]{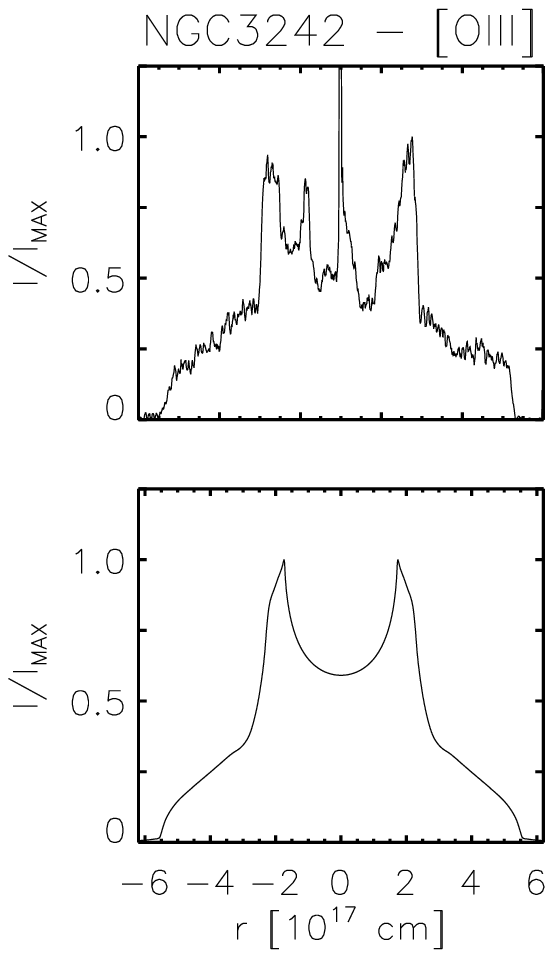}
\includegraphics*[width=0.49\columnwidth]{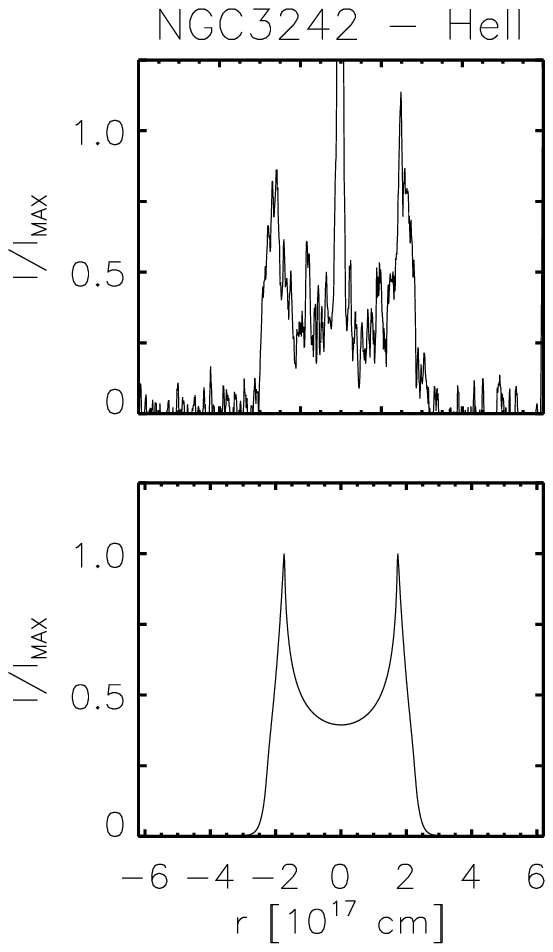}
\caption{\emph{Top}: normalised intensity distributions of \object{NGC~3242}
         in \oiii\ and \heii\ from HST monochromatic images (F502N and F469N, 
         respectively). The cuts are taken along the minor axes and scaled to 
         the model sizes. The spikes at $r=0$ are due to the central star.
	 \emph{Bottom}: normalised intensity distributions of an appropriate
	 hydrodynamical model with a 0.595 \Msun\ central star that matches closely
	 the observations for \object{NGC~3242} (model 2 of Tab.~\ref{param.3242}).
        }
\label{schnitt.3242}
\end{figure}

  One may then ask whether also other structures of \object{NGC~3242} are adequately
  represented by our models.  First of all, the model has a dense rim and a large
  but less dense (attached) shell, and  size and density ratios between rim and shell
  are about two and three, respectively.  These values compare favorably with the 
  observations, and also the (electron) densities are in good agreement: \citet{Retal.06}
  derived electron densities for \object{NGC~3242} of 2600 cm$^{-3}$ in the rim and 
  $800$~cm$^{-3}$ in the shell, respectively (cf. Table \ref{param.3242}).

  Figure \ref{schnitt.3242} gives a further illustration of the close relationship
  between \object{NGC~3242} and our models:  it shows the monochromatic brightness
  distributions in two important emission lines for \object{NGC~3242} and for an
  appropriate model very close to the observed HRD position of \object{NGC~3242}.
  It is remarkable that also the ionization structure in both the real PN
  and the model is such that only the rim is doubly ionized in
  helium.  Yet the model is not perfect:  According to Table~\ref{param.3242} the
  (spectroscopic) expansion velocities of rim and shell are lower than the observed
  ones.  The velocity differences, however, are well matched.

\paragraph{NGC~7027.}
\label{ngc7027}
  We note from Fig.~\ref{x-ray.obs} (top panel) that, although the X-ray luminosity
  of \object{NGC~7027} compares well with those of the other objects shown in this
  figure, it is about 1.5 dex \emph{below} the model prediction.  Also, the
  X-ray emitting region is considerably hotter, viz. with
  \hbox{$8\times10^6$}~K about twice as hot as the maximum reached by our
  0.696~\Msun\ models of sequence No.\ 10-HC2 (see Fig.\,\ref{tx}).
  We have verified that \object{NGC~7027} is embraced by our sequences
  No.\ 10 (no conduction) and No.\ 10a-HC (method~1), so we conclude that 
  thermal conduction may still work but at a lower level than our {method~1} predicts.
  A possible solution would be the presence of a weak magnetic field which
  suppresses thermal conduction to some extent \citep[cf.][]{BBF.90}.

  Recently, \citet{Sabetal.07} reported indeed the detection of polarization across
  \object{NGC~7027} by means of SCUBA observations. The orientation of
  polarization indicates the presence of a toroidal magnetic field
  along the equatorial plane.  Across the central cavity, no clear polarization is
  visible, thus any statements about the orientation of a possible magnetic field 
  inside the bubble is impossible. If this interpretation for the rather low X-ray 
  luminosity is correct, we must infer that heat conduction, in the particular case of
  \object{NGC~7027}, is not fully suppressed by the presence of a (weak) magnetic 
  field. The field geometry is likely to play a role. Indeed, it seems that the
  X-ray emission is suppressed in the equatorial plane \citep{Kastetal.02}.

 Alternatively, the low X-ray luminosity of \object{NGC~7027} might be
 related to the presence of bipolar collimated outflows found 
 by \citet{Coxetal.02}. It is conceivable that a significant amount of
 potentially X-ray emitting matter is lost through multiple openings 
 in the skin of the hot bubble created by the jets.

\paragraph{NGC~2392.}
\label{ngc2392}
 This is a rather peculiar object which poses a problem for our models, as 
 already noted above. Its central star has a wind speed that is much too low 
 for the object's position in the Hertzsprung-Russell diagram at about 
 40\,000 K effective temperature. The wind speed is more typical for the end 
 of the early-wind phase, although the mass-loss rate appears to be rather 
 normal (see Table \ref{tab.xray}). This low wind speed is responsible for a 
 wind luminosity which is about a factor of ten below the wind powers of the 
 other objects of this study (cf. Fig. \ref{wind.models}).

 Despite of this, the X-ray emission from the hot bubble of \object{NGC~2392} 
 compares well with that of the other objects (cf. Fig. \ref{x-ray.obs}) and 
 is much too high for an early-wind with speeds below 500 \kms, as is 
 predicted by our models (Fig. \ref{xray.wind}). It is possible that 
 additional X-ray emission is provided by jets as proposed by \citet{AMS.08}.

\subsection{The UV emission lines}
\label{uv}
  Due to the steep temperature gradient across the bubble/nebula interface, there
  exists only a very narrow region which is suitable for the emission of UV lines
  from highly ionized species. We recall here
  that our code computes the ionization of all elements considered (see
  Table~\ref{tab.element}) time-dependently for the whole computational domain, i.e.
  also in the freely streaming and shocked stellar wind. As a byproduct of our
  simulation, we are thus able to compute also the line emission from the bubble/PN
  interface.

  Recently, \citet{GChuG.04} reported the detection of \ion{O}{vi} $\lambda$\,1032 and
  1038\,\AA\ emission lines in FUSE spectra of \object{NGC~6543}.  Since the central
  star is not very hot, the ionization within the nebular shell is too low as to
  account for O$^{5+}$.  The authors concluded that the
  \ion{O}{vi} lines must originate from the conduction front at temperatures of about
  a few times $10^5$ K.

\begin{figure}[t]             
\includegraphics*[bb= 1.5cm 14.5cm 20.5cm 26.0cm, width=0.99\linewidth]{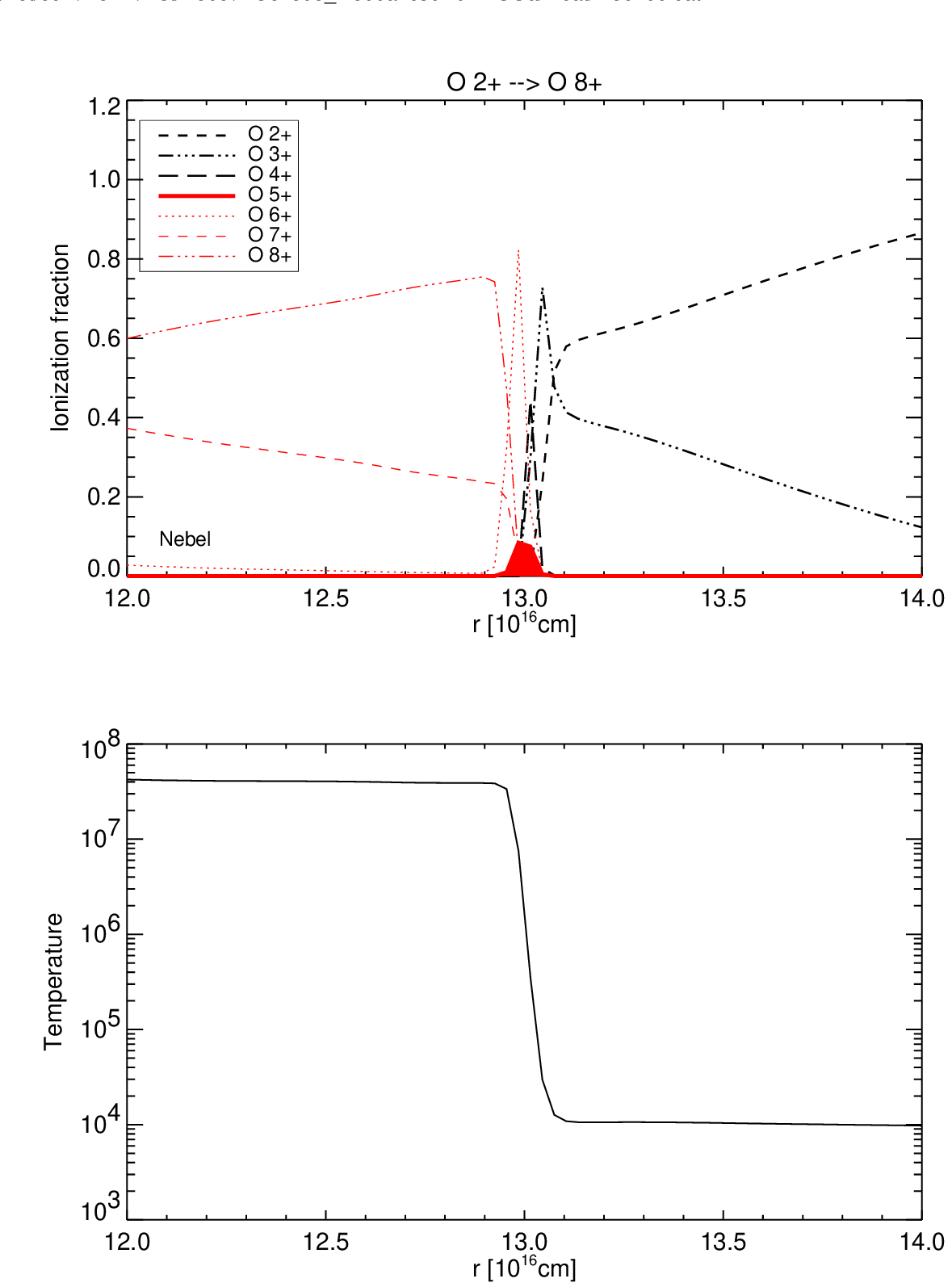}
\includegraphics*[bb= 1.5cm 14.5cm 20.5cm 26.0cm, width=0.99\linewidth]{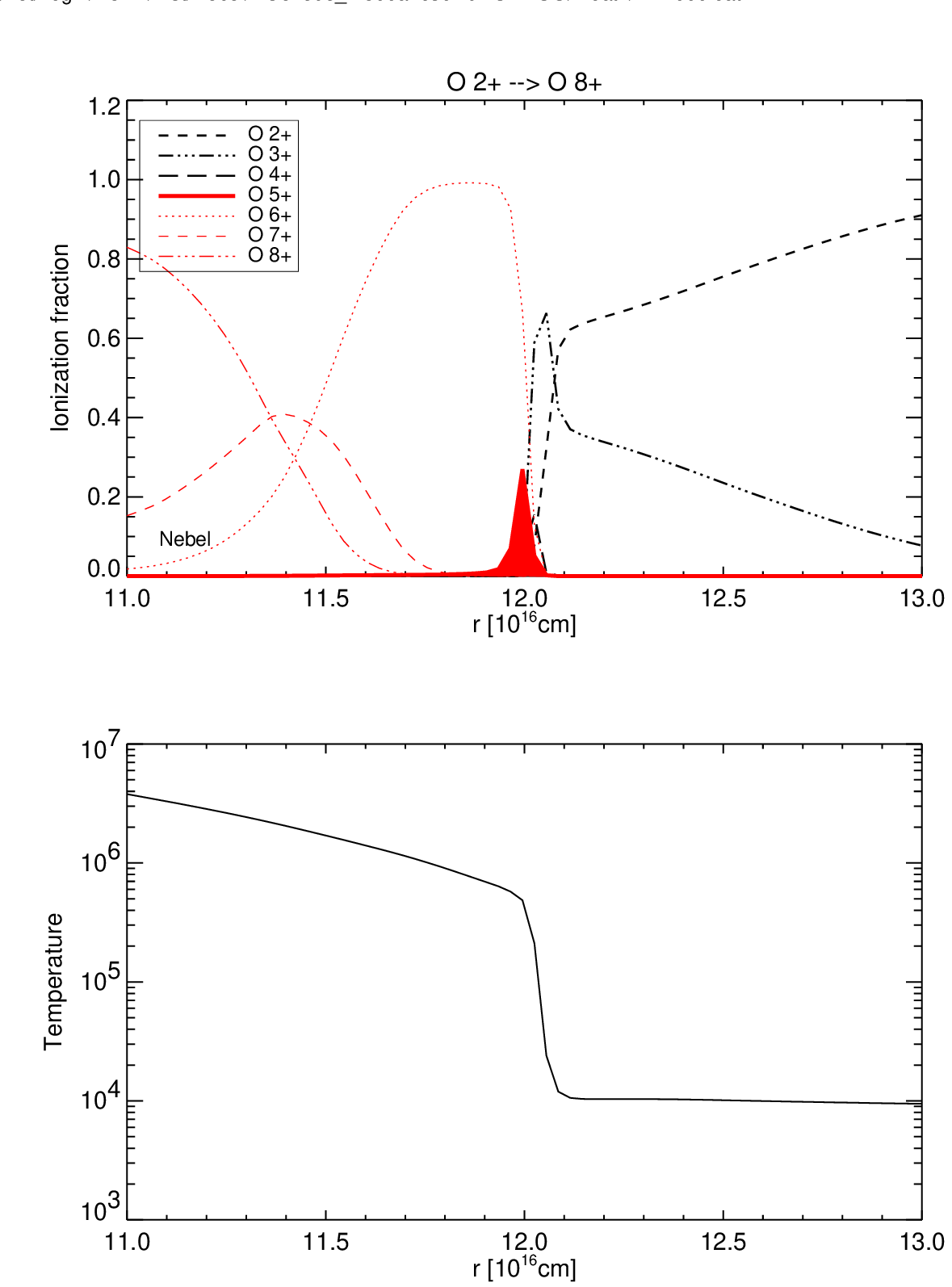}
\includegraphics*[bb= 1.5cm 14.5cm 20.5cm 26.0cm, width=0.99\linewidth]{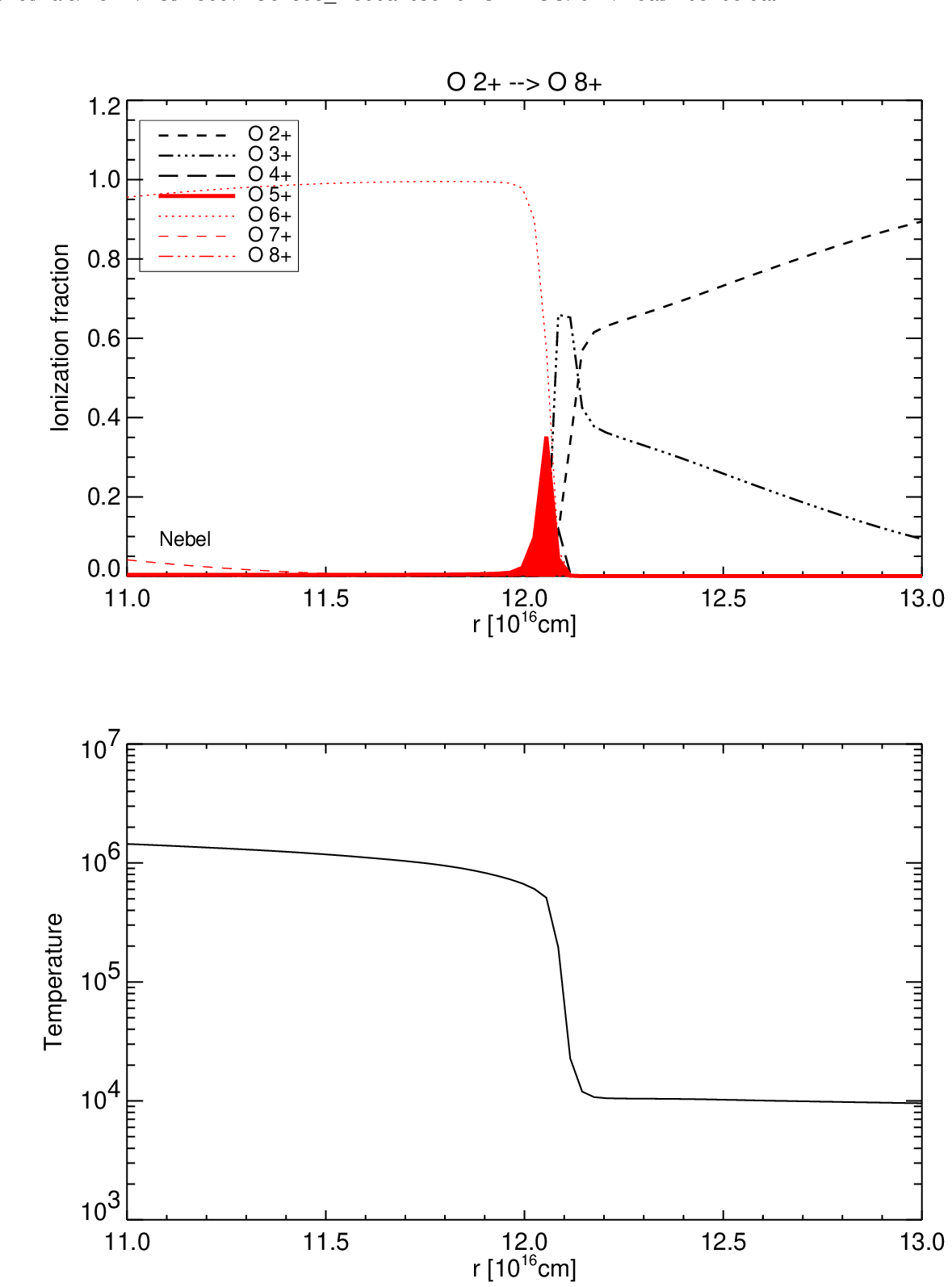}
\caption{Radial profiles of the ionization fractions of oxygen in the vicinity of
         the contact discontinuity/conduction front for different physical treatments:
         \emph{Top}: without thermal conduction; \emph{middle}: with
	 thermal conduction according to {method~1}; \emph{bottom}: with
	 thermal conduction according to {method~2}.
         Thick lines refer to the nebular region, thin ones to the bubble region
	 behind the contact surface/conduction front.
	 The very thin O$^{5+}$ `pocket' is shaded for clarity. Details about the
         three models are given in Table \ref{tab.OVI}.
	 }
\label{ion.ovi}
\end{figure}

  Test calculations showed that the O$^{5+}$ shell or `pocket' is extremely thin,
  only about ${1\times10^{15}}$ cm thick, which is comparable with the spatial 
  resolution of our numerical mesh.  We thus recomputed the sequences Nos. 6a, 
  6a-HC, and 6a-HC2 with a finer mesh (${\Delta r = 3\times10^{14}}$~cm) for 
  ${r\le 1.5\times10^{17}}$~cm in order to achieve a  better resolution of the 
  O$^{5+}$ layers. 

  Figure~\ref{ion.ovi} illustrates how the ionization of oxygen
  varies within the hot bubble, and how the distribution of O$^{+5}$ depends 
  on the physical treatment of this region.
  Note that, since the thermal structure of the conduction front/contact discontinuity
  does not change much during the lifetime of the PN, the ionization fractions remain
  rather stationary with respect to the front/discontinuity. We find that
  the thermal structure close to the conduction front where O$^{5+}$ prevails does not 
  differ much between our two treatments of heat conduction.

  The physical conditions change rapidly across the conduction front:  ahead
  photoionization by the stellar radiation field is the dominant heating and
  ionization mechanism.  For the stellar temperature shown in Fig.~\ref{ion.ovi},
  O$^{2+}$ and O$^{3+}$ are practically the only representatives of oxygen.
  Behind the front the temperatures are so high ($\ga\!10^5$ K) that electron
  collisions determine the ionization state of the gas: now we have a mixture
  of O$^{6+}$, O$^{7+}$, and O$^{8+}$ in proportions ruled by the electron
  temperature.  For instance, in the case without heat conduction the bubble
  temperature is so large that O$^{8+}$ (the bare oxygen nucleus) is the main
  constituent already right behind the contact discontinuity.  Heat conduction
  lowers the temperature gradient behind the front, and consequently O$^{6+}$
  and O$^{7+}$ are the main constituents of oxygen throughout a large fraction of
  the bubble.

\begin{table*}[t]
\centering
\caption{
Dependence of the luminosity of the \ion{O}{vi} lines at $\lambda\, 1032$~\AA\
and  $1038$~\AA\
on the treatment of heat conduction for models similar to those shown in 
Fig.\,\ref{comp.HC}, but computed with higher spatial resolution.
$L$(H$\beta$) and  $L$(\ion{O}{iii} 5007 \AA) refer to the H$\beta$ 
and [\ion{O}{iii}] luminosities of the whole nebula.
}
\label{tab.OVI}
\begin{tabular}{lccclcccc}
\hline\hline\noalign{\smallskip}
No. & $M $    & $L$     & $\teff$ & Thermal    & $L$(H$\beta$) 
    &  $L$(\ion{O}{iii} 5007 \AA)  &  $L$(\ion{O}{vi} 1032 \AA) 
    &  $L$(\ion{O}{vi} 1038 \AA) \\[0.5mm]
    & (\Msun) & (\Lsun) & (K)     & conduction & (\Lsun)        
    & (\Lsun)                     & (\Lsun) 
    &  (\Lsun) \\
\noalign{\smallskip}
\hline\noalign{\smallskip}
6a-HR       & 0.595 & 5\,260 & 67\,978 & no       & $21.7$ & $168.8$  & $4.41\times 10^{-3}$&  $2.21\times 10^{-3}$ \\
6a-HCHR     & 0.595 & 5\,297 & 65\,160 & method~1 & $23.9$ & $174.0$  & $11.3\times 10^{-3}$&  $5.63\times 10^{-3}$ \\
6a-HC2HR    & 0.595 & 5\,292 & 65\,541 & method~2 & $23.6$ & $173.3$  & $10.1\times 10^{-3}$&  $5.04\times 10^{-3}$ \\
\noalign{\smallskip}\hline
\end{tabular}
\end{table*}

 The transition between the photo-heated nebular gas ($T_{\rm e}\simeq 10^4$ K)
  and the shock-heated wind gas ($T_{\rm e}> 10^7$ K)
  occurs very abruptly across the contact discontinuity if there is no thermal
  conduction, and consequently only little room is left for the existence of 
  O$^{5+}$, simply because the temperature is either too low (nebula) or too
  too hot (bubble).
  With thermal conduction included, the temperature increases somewhat more
  gently and thus allows for a larger amount of O$^{5+}$.  This is also reflected in
  the total emission of the \ion{O}{vi} lines listed in 
  Tab.\,\ref{tab.OVI} for the models shown in Fig.~\ref{ion.ovi}. We point out that
  the computed \ion{O}{vi} luminosities are strongly fluctuating from model to 
  model due to the poor numerical resolution of the conduction front. The 
  \ion{O}{vi} line fluxes listed in Tab.\,\ref{tab.OVI} are therefore appropriate
  averages over several contiguous models.

  According to the results for the heat conduction models in Tab.\,\ref{tab.OVI},
  the total flux in the \ion{O}{vi} line at $\lambda\, 1032$~\AA\ emitted by the 
  O$^{5+}$ pocket, assuming a distance of 1 kpc, would be roughly
  $4\times10^{-13}$ erg\,cm$^{-2}$\,s$^{-1}$ .
  This is consistent with the estimates of \citet{GChuG.04} for NGC~6543,
  although their FUSE measurements are based on a slit across
  the central cavity.  The total \ion{O}{vi} $\lambda$\,1032 \AA\ line flux is
  certainly larger than the quoted value of 
  $\simeq\!2\times10^{-13}$ erg\,cm$^{-2}$\,s$^{-1}$.

  We conclude that the luminosity in UV \ion{O}{vi} lines is generated in a
  very thin transition layer which is not very sensitive to the effects of thermal
  conduction. In contrast, the X-ray luminosity depends sensitively on the efficiency
  of thermal conduction. In our models with heat conduction included, the \ion{O}{vi} 
  luminosity is comparable to the X-ray emission coming from more extended parts of
  the bubble (cf. Tables~\ref{tab.xray} and \ref{tab.OVI}).

\section{Summary and conclusions}
\label{diss.concl}

  We presented a detailed numerical approach towards an understanding
  of the diffuse soft X-ray emission from planetary nebulae based on the
  concept of thermal conduction.  
  Since thermal conduction is a physical process 
  inherent to all hydrodynamical systems, becoming important wherever
  the mean free path of the electrons is sufficiently large, we
  included a thermal conduction module into our 1D radiation-hydrodynamics 
  code. We were able to compute the thermal structure of the shocked wind gas 
  inside the nebular cavity self-consistently with the hydrodynamics once 
  the stellar AGB remnant, its initial circumstellar envelope,
  and the post-AGB wind model are specified. By doing so, there is some freedom
  in the treatment of heat conduction in cases where the mean free path of the 
  electrons becomes comparable to the characteristic temperature scale length.
  Magnetic fields that may play a role for the shaping of PNe are not 
  considered.

  Thermal conduction 
  has a twofold effect favoring thermal X-ray emission from the
  shocked wind gas: it lowers the temperature gradient across the
  bubble/nebula interface and heats the cool
  nebular matter, forcing it to `evaporate' inwards.  With time, heat conduction
  accumulates additional matter in the bubble with a characteristic temperature
  of some $10^{6}$ K, which quickly dominates the bubble's mass budget.
  The amount of added matter is controlled by heat transfer across the bubble
  from the inverse wind shock towards the nebula. The bubble mass may increase 
  or decrease with time, depending on the evolution of the stellar wind power.

  We selected several of the hydrodynamic sequences presented in
  \citetalias{Peretal.04} and recomputed them, without changing any of
  the other parameters or boundary conditions, with our heat conduction 
  treatment included.
  The X-ray emission was computed post-facto by employing the CHIANTI code,
  slightly adapted to match our purposes. We were able to find good agreement
  with existing observations, both with respect to X-ray luminosity and surface
  brightness distribution, if {method~2} is used for the treatment of
  heat conduction. At the same time, the stellar wind model employed
  here, based on the theory of radiation-driven winds, implies wind speeds in 
  excess of 1000 \kms\ for typical nebular models (cf. Fig.~\ref{xray.wind}). 
  We conclude that our modeling is fully consistent
  with both the observed wind speeds of more than 1000 \kms\ \emph{and} the
  observational evidence that the soft X-ray emission comes from regions 
  with electron temperatures of about $2\times 10^6$~K (see also 
  Table~\ref{tab.xray}). In contrast, our models without heat conduction fail 
  to reproduce the characteristics of the observed X-ray emission by large 
  amounts if realistic wind parameters are assumed. 

  The basic findings from our modeling can be summarized as follows:
\begin{itemize}
\item  The energy budget of the hot bubble is substantially altered by radiative
       losses at the bubble/nebula interface, both in models with and without 
       thermal conduction.
\item  Heat conduction has a strong influence on the thermal structure of the 
       hot bubble, but leaves the dynamics of the whole system is virtually 
       unchanged, i.e.\ the shaping of the nebular shells and their expansion 
       properties are not affected.
\item  In the absence of heat conduction, the temperature of the hot bubble 
       depends only on the velocity of the central star wind. The resulting 
       X-ray temperature is much too high, and the X-ray luminosity much too 
       low, to explain the existing observations.
\item  If heat conduction is substantial, the bubble temperature is a function of 
       the stellar wind power and the bubble size. The compact nebulae around 
       massive, short-lived central stars are therefore expected to have hotter 
       bubbles than the  nebulae of low-mass, slowly evolving central stars.
\item  The X-ray luminosity is determined by the total emission measure, which 
       decreases by expansion and increases by `evaporation' from the main nebula. 
       The competition between expansion \emph{and} `evaporation' rules the temporal 
       evolution of the X-ray luminosity. 
\item  According to our PN simulations with a time-dependent wind model, the X-ray
       luminosities increase with time during the main phase of evolution when
       the wind power increases with time, in contrast to the case of a constant 
       wind power that would imply a decrease of the X-ray luminosity with time.
\item  For the energy range usually considered for planetary nebulae, the
       X-ray luminosity is below about 1\% of the stellar wind power, and between
       $10^{-4}$ and $10^{-8}$ of the stellar bolometric luminosity. The
       exact numbers depend on the actual evolutionary state and the wind model
       used.
\item  For given effective temperature, the X-ray emission is largest for nebulae 
       around massive central stars even though their small bubbles contain only
       little mass. The X-ray emission measure of their bubbles is nevertheless
       large, primarily because the high electron densities overcompensate the small 
       amount of X-ray emitting gas that massive central stars can accumulate during
       their short lifetimes.
\item  In contrast to the X-ray luminosity, the emission of the UV \ion{O}{vi} lines
       is not very sensitive to the effects of thermal conduction.
\item  Magnetic fields must be absent or extremely weak in all objects with
       diffuse X-ray emission since their presence would strongly depress thermal
       conduction and hence `evaporation'.
       The absence or weakness of magnetic fields implies also that they
       cannot be responsible for shaping these objects.
\end{itemize}
   
   Our numerical treatment as described here leads to X-ray luminosities that are
   substantially below those found by \citet{ZP.98, ZP.96} with their analytical
   approach, although their wind model and central-star evolution are very similar.
   We believe that their analytical approach overestimates the evaporation rate
   because radiative cooling of the gas at the conduction front by line emission 
   is not considered (see also Sect.~\ref{bubble.time}).

   It is likely that the bubble/PN interface becomes dynamically unstable, leading 
   to direct mixing between hot bubble and cool nebular matter \citep[cf.][]{SS.06}.
   The net effect would be similar to heat conduction, i.e.\ a reduction of temperature
   gradients allowing a more efficient X-rays emission, 
   as already mentioned by \citet{Chuetal.97}. The 2D simulations of
   a spherical bubble performed by \citet{SS.06} suggest that the mixing region is
   confined to a rather thin shell at the surface of the bubble.  If so, the
   X-ray emission would be limb brightened, very similar to our models computed
   according to {method~1}. For the time being, the existing observations
   do not seem to be consistent with such a limb brightening.

   \citet{Georgetal.08} found recently that the wind of the central star of 
   \object{NGC~6543} is less depleted in iron compared to the plasma emitting 
   the diffuse X-rays \citep{Georgetal.06}. They concluded that the X-ray 
   emitting plasma ``is derived from nebular gas rather than the stellar 
   wind".  This finding is fully consisting with the heat-conduction models.
   Clearly, more studies of this kind would be very rewarding.

   Finally we want to emphasise that the models introduced in this work are 
   entirely based on normal chemical compositions in the stellar and 
   circumstellar envelopes. The results presented here should therefore 
   \emph{not} be used to interprete X-ray
   emission from objects with hydrogen-deficient central stars 
   such as \object{BD+30\degr 3639}, \object{NGC~40} because their
   evolution is different and not yet understood.
   Additionally, one has to deal with heat conduction and X-ray emission in
   a practically hydrogen-free plasma.


\begin{acknowledgements}
  We are grateful to Dr.\ A. Schwope for introducing us to the secrets of the
  EPIC camera on board XMM-Newton and providing us with the response matrix.
  We are especially thankful to Dr.~Landi who helped us to install the latest
  version of the CHIANTI code. The work of A.W. was supported by DLR under 
  grant No.\ 50 QL 0001.
\end{acknowledgements}


\begin{thebibliography}{}

\bibitem[Akashi \etal(2008)]{AMS.08}
   Akashi, M., Meiron, Y., \& Soker, N. 2008, eprint arXiv:0711.3265

\bibitem[Akashi \etal(2006)]{ASB.06}
   Akashi, M., Soker, N., Behar, E. 2006, \mnras, 368, 1706 

\bibitem[Akashi \etal(2007)]{ASBB.07}
   Akashi, M., Soker, N., Behar, E., \& Blondin, J. 2007, \mnras, 375, 137

\bibitem[Borkowski \etal(1990)]{BBF.90}
   Borkowski, K. J., Balbus, S. A., \& Fristrom, C. C. 1990, \apj, 355, 501

\bibitem[Castor \etal(1975)]{CCW.75}
   Castor, J., McCray, R., \& Weaver, R. 1975, \apj, 200, L107

\bibitem[Chu \etal(1997)]{Chuetal.97}
   Chu, Y.-H., Chang, T. H., \& Conway, G. M. 1997, \apj, 482, 891

\bibitem[Chu \etal(2003)]{Chuetal.03}
   Chu, Y.-H., Guerrero, M. A., \& Gruendl, R. 2003, in Planetary Nebulae: Their
       Evolution and Role in the Universe, ed.\ S. Kwok, M. Dopita, \& R. Sutherland,
       IAU Symp. 209, p. 415

\bibitem[Corradi \etal(2003)]{Coretal.03}
   Corradi, R. L. M., Sch\"onberner, D., Steffen, M., \& Perinotto, M. 2003, \mnras, 340, 417

\bibitem[Cowie \& McKee(1977)]{CMK.77}
   Cowie, L. L., \& McKee, C. F. 1977, \apj, 211, 135

\bibitem[Cox \etal(2002)]{Coxetal.02}
   Cox, P., Huggins, P.J., Maillard, J.-P., Habart, E., Morisset, C., 
   Bachiller, R., \& Forveille, T.\ 2002, \aap, 384, 603 

\bibitem[Dere \etal(1997)]{Deretal.97}
   Dere, K. P., Landi, E., Mason, H. E., Monsignori Fossi, B. C. \& Young, P. R. 1997,
       \aaps, 125, 149

\bibitem[Georgiev \etal(2006)]{Georgetal.06}
   Georgiev, L. N., Richer, M. G., Arrieta, A., \& Zhekov, S. A. 2006, \apj, 639, 185    

\bibitem[Georgiev \etal(2008)]{Georgetal.08}
   Georgiev, L. N., Peimbert, M., Hillier, D. J., Richer, M. G., Arrieta, A., \&
       Peimbert, A. 2008, arXiv:0802.3692

\bibitem[Gruendl \etal(2004)]{GChuG.04}
   Gruendl, R. A., Chu, Y.-H., \& Guerrero, M. A. 2004, \apj, L127

\bibitem[Guerrero(2006)]{G.06}
   Guerrero, M. A. 2006, in Planetary Nebulae in our Galaxy and Beyond,
        ed.\ M. J. Barlow \& R. H. M\'endez, IAU Symp. 234, p. 153

\bibitem[Guerrero \etal(2005a)]{GCG.05}
   Guerrero, M. A., Chu, Y.-H., \& Gruendl, R. 2005a, in Planetary Nebulae as
        Astronomical Tools, ed.\ R. Szczerba, G. Stasi\'nska, \& S. G\'orny,
	AIP Conf. Proc. 804, p. 157

\bibitem[Guerrero \etal(2005b)]{GCGM.05}
   Guerrero, M. A., Chu, Y.-H., Gruendl, R., \& Meixner, M. 2005b, \aap, 430, L69

\bibitem[Kastner(2007)]{Kast.07}
   Kastner, J. H. 2007, in Asymmetrical Planetary Nebulae IV, ed. R. L. M. Corradi
        et al., eprint arXiv:0709.4136

\bibitem[Kastner \etal(2001)]{Kastetal.01}
   Kastner, J. H., Vrtilek, S. D., \& Soker, N. 2001, \apj, 550, L189

\bibitem[Kastner \etal(2002)]{Kastetal.02}
   Kastner, J. H., Jingquiang, L., Vrtilek, S. D., Gatley, I., Merrill, K.M., 
   \& Soker, N. 2002, \apj, 581, 1225

\bibitem[Kastner \etal(2008)]{Kastetal.08}
   Kastner, J. H., Montez, Jr., R., Balick, B., \& De Marco, U. 2008, \apj, 672, 957

\bibitem[Koo \& McKee(1992)]{KMcK.92}
   Koo, B.-C., \& McKee, C. F. 1992, \apj, 388, 103

\bibitem[Kudritzki \etal(2006)]{KUP.06}
   Kudritzki, R.-P., Urbaneja, M. A., \& Puls, J. 2006, in Planetary Nebulae in our
        Galaxy and Beyond, ed.\ M. J. Barlow \& R. H. M\'endez, IAU Symp. 234, p. 119

\bibitem[Landi \& Phillips(2005)]{LP.05}
   Landi, E., \&  Phillips, K. J. H. 2005, \apjs, 160, 286

\bibitem[Marten \& Sczcerba(1997)]{MS.97}
   Marten, H., \& Szczerba, R. 1997, \aap, 248, 590

\bibitem[Mazzotta \etal(1998)]{Mazetal.98}
   Mazzotta, P., Mazzitelli, G., Colafrancesco, S., \& Vittorio, N. 1998, \aaps, 133, 403

\bibitem[Mellema \& Frank(1995)]{MF.95}
   Mellema, G., \& Frank, A. 1995, \mnras, 273, 401

\bibitem[M\'endez \etal(1992)]{Menetal.92}
   M\'endez, R. H., Kudritzki, R. P., \& Herrero, A. 1992, \aap, 260, 329

\bibitem[ Morrison \& McCammon(1983)]{MMcC.83}
   Morrison, R., McCammon, D. 1983, \apj, 270, 119

\bibitem[Pauldrach \etal(1988)]{Pauletal.88}
   Pauldrach, A. W. A., Puls, J., Kudritzki, R. P., M\'endez, R. M., \& Heap, S. H. 1988,
       \aap, 207, 123

\bibitem[Pauldrach \etal(2004)]{Pauletal.04}
   Pauldrach, A. W. A., Hoffmann, T. L., \& M\'endez, R. M. 2004, \aap, 419, 1111

\bibitem[Perinotto \etal(1998)]{Peretal.98}
   Perinotto, M., Kifonidis, K., Sch\"onberner, D., \& Marten, H. 1998,
       \aap, 332, 1044

\bibitem[Perinotto \etal(2004)]{Peretal.04}
   Perinotto, M., Sch\"onberner, D., Steffen, M., \& Calonaci, C. 2004,
       \aap, 414, 993 (Paper I)

\bibitem[Reed \etal(1999)]{Reedetal.99}
   Reed, D. S., Balick, B., Haijan, A. R., \etal\ 1999, \aj, 118, 2430

\bibitem[Reimers(1975)]{Reim.75}
   Reimers, D. 1975, in Problems in Stellar Atmospheres and Envelopes,
       ed.\ B. Baschek W. H. Kegel, \& G. Traving (Berlin: Springer), p. 229

\bibitem[Ruiz \etal(2006)]{Retal.06}
   Ruiz, N., Guerrero, M. A., Chu, Y.-H., Gruendl, R. A., Kwitter, K., \&
       Meixner, M. 2006, in Planetary Nebulae in our Galaxy and Beyond,
       ed.\ M. J. Barlow \& R. H. M\'endez, IAU Symp. 234, p. 497

\bibitem[Sabin \etal(2007)]{Sabetal.07}
   Sabin, L., Zijlstra, A. A., \& Greaves, J. S. 2007, \mnras, 376, 378

\bibitem[Sandin \etal(2008)]{Sandinal.08}
   Sandin, C., Sch{\"o}nberner, D., Roth, M.M., 
   Steffen, M., B{\"o}hm, P., Monreal-Ibero, A. 2008, \aap (in press)

\bibitem[Sch\"onberner \& Steffen(2003)]{SchSt.03}
   Sch\"onberner, D., \& Steffen, M. 2003, in Planetary Nebulae: Their Evolution
       and Role in the Universe, ed.\ S. Kwok, M. Dopita, \& R. Sutherland,
       IAU Symp. 209, p. 147

\bibitem[Sch\"onberner \etal(1997)]{Schetal.97}
   Sch\"onberner, D., \& Steffen, M., Stahlberg, J., Kifonidis, K., \& Bl\"ocker, T.
       1997, in Advances of Stellar Evolution, ed.\ R. T. Rood \& A. Renzini,
       (Cambridge Univ.\ Press), p. 146

\bibitem[Sch\"onberner \etal(2005a)]{Schetal.05a}
   Sch\"onberner, D., Jacob, R., \& Steffen, M. 2005a \aap, 441, 573

\bibitem[Sch\"onberner \etal(2005b)]{Schetal.05}
   Sch\"onberner, D., Jacob, R., Steffen, M., Perinotto, M., Corradi, R. L. M., \&
       Acker, A. 2005b, \aap, 431, 963

\bibitem[Sch\"onberner \etal(2006)]{SSW.06}
   Sch\"onberner, D., Steffen, M., \& Warmuth, A. 2006, in Planetary Nebulae in our
       Galaxy and Beyond, ed.\ M. J. Barlow \& R. H. M\'endez, IAU Symp. 234, p. 161

\bibitem[Soker(1994)]{So.94}
   Soker, N. 1994, \aj, 107, 276

\bibitem[Soker \& Kastner(2003)]{SoKa.03}
   Soker, N., \& Kastner, J. H. 2003, \apj, 583, 368

\bibitem[Spitzer(1962)]{Sp.62}
   Spitzer, L. 1962, \emph{Physics of Fully Ionized Gases}, 
   2nd revised edition, Wiley Interscience Publishers

\bibitem[Steffen \& Sch\"onberner(2006)]{StSch.06}
   Steffen, M., \& Sch\"onberner, D. 2006, in Planetary Nebulae in our Galaxy and
       Beyond, ed.\ M. J. Barlow \& R. H. M\'endez, IAU Symp. 234, p. 285

\bibitem[Steffen \etal(1998)]{Stetal.98}
   Steffen, M., Szczerba, R., \& Sch\"onberner, D. 1998, \aap, 337, 149

\bibitem[Stute \& Sahai(2006)]{SS.06}
   Stute, M., \& Sahai, R. 2006, \apj, 651,882

\bibitem[Tinkler \& Lamers(2002)]{TL.02}
   Tinkler, C. M., \& Lamers, H. J. G. M. 2002, \aap, 384, 987

\bibitem[Tsamis \etal(2008)]{Tsamisetal.08}
   Tsamis, Y. G., Walsh, J. R., P\'equignot, D., \etal\ 2008, \mnras, 386, 22

\bibitem[Volk \& Kwok(1985)]{VK.85}
   Volk, K., \& Kwok, S. 1985, \aap, 153, 79

\bibitem[Weaver \etal(1977)]{Wetal.77}
   Weaver, R., McCray, R., Castor, J., Shapiro, P., \& Moore, R. 1977, \apj, 218, 377

\bibitem[Wrigge \etal(1994)]{WWW.94}
   Wrigge, M., Wendker, H. J., \& Wisotzki, L. 1994, \aap, 286, 219

\bibitem[Zhekov \& Perinotto(1996)]{ZP.96}
   Zhekov, S. A., \& Perinotto, M. 1996, \aap, 309, 648

\bibitem[Zhekov \& Perinotto(1998)]{ZP.98}
   Zhekov, S. A., \& Perinotto, M. 1998, \aap, 334, 239

\end{thebibliography}
\end{document}